\documentclass[11pt,twoside]{article}
\usepackage{axodraw}
\usepackage{graphicx}
\usepackage{colordvi}
\usepackage{latexsym} \usepackage{epsfig}
\begin{document} \thispagestyle{empty}
\begin{center}
{ \Large \bf   BEYOND THE STANDARD MODEL$^\dagger$}
 \vspace{0.6cm}

{\large \bf D. I.~Kazakov } \\[4mm]

{\it Bogoliubov Laboratory of Theoretical Physics, JINR,
 Dubna, Russia \\[0.1cm]
and\\[0.1cm]
Institute for Theoretical and Experimental Physics, Moscow}

%

\end{center}

\begin{abstract}
The present lectures contain an introduction to possible new
physics beyond the Standard Model. Having in mind first of all
accelerator experiments of the nearest future we concentrate on
supersymmetry, a new symmetry that relates bosons and fermions, as
the first target of experimental search.  Since supersymmetry is
widely covered in the literature, we mostly consider novel
developments and applications to hadron colliders. We describe
then the so-called extra dimensional models in less detail and
discuss their possible manifestations.
\end{abstract}\vspace{0.2cm}

 {\bf \Large Preface} \vspace{0.2cm}

When discussing physics beyond the Standard Model one enters terra
incognita and inevitably has to make some choice of the topic.
When doing so I had in mind that most of the audience is working
in one of the LHC collaborations and apparently is looking forward
to discover new physics there. So their main goal will be to find
the Higgs boson and then ... who knows? Search for SUSY is the
main stream and the general belief is that low energy
supersymmetry described by the MSSM is round the corner. So one
has to be prepared. The other widely discussed topic is extra
dimensions. This is a much less motivated subject though is very
intriguing. And if supersymmetry is already elaborated in detail
and may be the subject of precise tests, extra dimensional models
are more speculative and may bring many surprises or ... nothing.

Supersymmetry has already more than 30 years of history and  is
very widely covered in the literature~\cite{reviews} and in the
text books~\cite{WessB}-\cite{Weinberg}. Moreover, I myself gave
lectures on SUSY at the 2000 European School on High Energy
Physics and they are published in the proceedings and are
available in the web~\cite{Lectures}. So I decided not to repeat
the whole subject but keep the main line and to concentrate on the
novel developments. In the year 2000 LEP was still running and
obviously our main expectations to discover supersymmetry were
connected with it. Unfortunately this did not happen. Today we are
looking forward at hadron colliders and this is my main concern in
these lectures. At the same time recent years celebrated
unprecedented development in astroparticle experiments. This is
the new area to look for new physics and in particular for the
manifestation of SUSY. Therefore, I cover partly the motivations
of SUSY in astrophysics and the influence of new astroparticle
data on SUSY models.

When choosing the topic of extra dimensions I am aware of the fact
that this deserves special lectures. At the same time, this
subject is still an actively developing field and many changes in
the ideas and preferences are possible. So I decided to make some
overview without discussing theoretical problems (which are many)
and to present possible experimental signatures since people are
already looking for them. I do not pretend here for any complete
coverage, my aim is to give the flavour of the field.

\noindent
-----------------------------------------------------------

\noindent $^\dagger$ Lectures given at the European School on High
Energy Physics, May-June 2004, \\ Sant Feliu de Guixols, Spain

 \pagebreak
\thispagestyle{empty} {\small \tableofcontents \vglue 0.34cm {\bf
References} \hfill {\bf 49}}

\renewcommand{\thesubsection}{\thesection.\arabic{subsection}}
\renewcommand{\thesubsubsection}{\thesubsection.\arabic{subsubsection}}
\renewcommand{\theequation}{\thesection.\arabic{equation}}

\pagebreak

\section{PART I \ \ SUPERSYMMETRY}
\subsection{Introduction: What is supersymmetry}

{\it Supersymmetry} is a {\it boson-fermion} symmetry that is
aimed to unify all forces in Nature including gravity within a
singe framework. Modern views on supersymmetry in particle physics
are based on string paradigm, though the low energy manifestations
of SUSY can be possibly found at modern colliders and in
non-accelerator experiments.

Supersymmetry emerged from the attempts to generalize the
Poincar\'e algebra to mix representations with different
spin~\cite{super}. It happened to be a problematic task due to the
no-go theorems preventing such generalizations~\cite{theorem}. The
way out was found by introducing the so-called graded Lie
algebras, i.e. adding the anti-commutators to the usual
commutators of the Lorentz algebra. Such a generalization,
described below, appeared to be the only possible one within
relativistic field theory.

If $Q$ is a generator of SUSY algebra, then acting on a boson
state it produces a fermion one and vice versa
 $$ \bar Q| boson> = |fermion> \ \ \ {\rm and} \ \ \
 Q| fermion> = | boson> .$$

Since bosons commute with each other and fermions anticommute, one
immediately finds that SUSY generators should also anticommute,
they must be {\em fermionic}, i.e. they must change the spin by a
half-odd amount and change the statistics. Indeed, the key element
of SUSY algebra is
\begin{equation}\label{ant}
  \{Q_\alpha, \bar{Q}_{\dot \alpha}\}=2\sigma_{\alpha,\dot
\alpha}^\mu P_\mu,
\end{equation}
where $Q$ and $\bar Q$ are SUSY generators and $P_\mu$ is the
generator of translation, the four-momentum.

In what follows we describe SUSY algebra in more detail and
construct its representations which are needed to build a SUSY
generalization of the Standard Model of fundamental interactions.
Such a generalization is based on a softly broken SUSY quantum
filed theory and contains the SM as a low energy theory.

Supersymmetry promises to solve some  problems of the SM and of
Grand Unified Theories. In what follows we describe supersymmetry
as a nearest option for the new physics on a TeV scale.

\subsection{Motivation of SUSY in particle physics}

There are several motivations of introduction of SUSY in particle
physics. Most of them are related to ideas of unification of all
the forces of Nature within the same framework. The incomplete set
is:

  $\bullet$  Unification with gravity

  $\bullet$ Unification  of  gauge couplings

  $\bullet$ Solution of the hierarchy problem

   $\bullet$ Superstring consistency

  $\bullet$  Dark Matter in the Universe

Probably the most challenging is the unification with gravity
which is believed to happen within supergravity which in its turn
is the low energy limit of a string theory. I have considered
these arguments in some detail in my lectures~\cite{Lectures}
 and will not repeat it here. Instead I
will concentrate on the last motivation which became popular in
recent time due to new data coming from astroparticle experiments.

\subsubsection{Astrophysics and Cosmology}

\begin{figure}[ht]\vspace{0.3cm}
\begin{center}
  \leavevmode
 \epsfxsize=6cm \epsfysize=5.5cm
 \hspace*{-6.1cm}\epsffile{solar.eps}\vspace{-5.8cm}

  \epsfxsize=6.3cm
  \hspace*{6cm}
  \epsfysize=5.8cm
 \epsffile{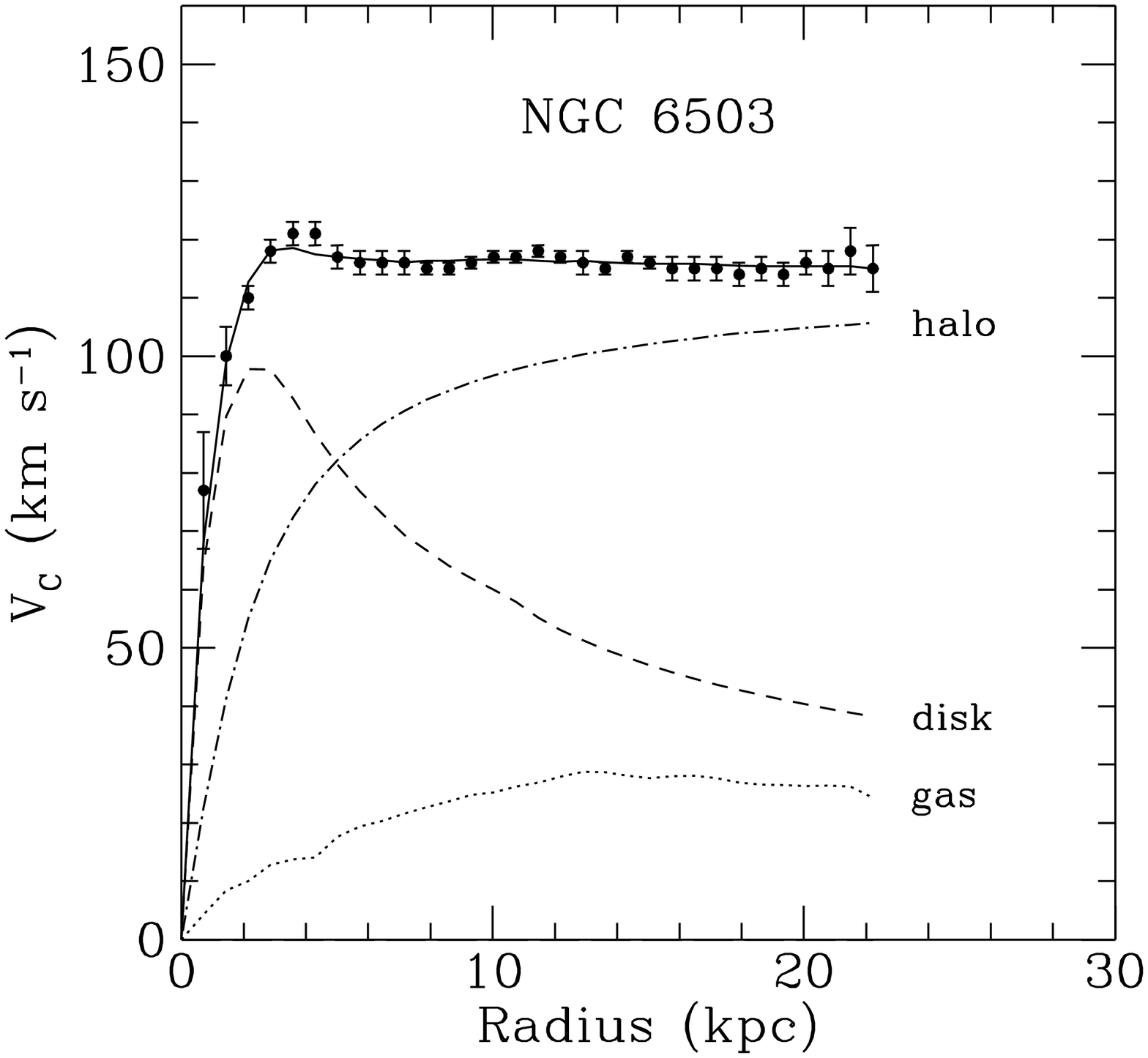}
 \end{center}
 \caption{
Rotation curves for the solar system and galaxy}\label{gal}
\end{figure}

The shining matter is not the only one in the Universe.
Considerable amount consists of the so-called dark matter. The
direct evidence for the presence of the dark matter are the
rotation curves of galaxies (see Fig.\ref{gal})~\cite{rotation}.
What is shown here is the rotation speed of the planets of the
solar system (left) and the stars in some typical spiral galaxy
(right) as a function of a distance from the sun/center of galaxy.
One can see that in the solar system all the planets perfectly fit
the curve obtained from Newton mechanics: centrifugal force is
equal to gravitational force
$$\frac{mv^2}{r}=G\frac{mM}{r^2},\ \ \Rightarrow \ \ v=\sqrt{\frac{GM}{r}}.$$
At the same time, if one looks at stars in the galaxy, one finds A
completely different picture. To explain these curves, one has to
assume the existence of galactic halo made of non shining matter
which takes part in gravitational interaction. The flat rotation
curves of spiral galaxies provide the most direct evidence for the
existence of A large amount of the dark matter. Spiral galaxies
consist of a central bulge and a very thin disc, and are
surrounded by an approximately spherical halo of the dark matter.

According to the latest data~\cite{WMAP}, the matter content of
the Universe is the following:
$$\Omega h^2=1\ \  \Leftrightarrow \ \ \rho=\rho_{crit}$$
$$ \Omega_{vacuum} \approx 73\%, \ \ \Omega_{Dark Matter} \approx
23\%,\  \ \Omega_{Baryon} \approx 4\%$$ Therefore, the amount of
the Dark matter is almost 6 times larger than the usual matter in
the Universe.

There are two possible types of the dark matter: the hot one,
consisting of light relativistic particles and the cold one,
consisting of massive weakly interacting particles (WIMPs). The
hot dark matter might consist of neutrinos; however, this has
problems with galaxy formation. As for the cold dark matter, it
has no candidates within the SM. At the same time, SUSY provides
an excellent candidate for the cold dark matter, namely
neutralino, the lightest superparticle.

\subsection{Basics of supersymmetry}
Sending off the interested reader to~\cite{Lectures} for details
we present here the main ideas and building blocks for
constructing a supersymmetric quantum field theory.

\subsubsection{Algebra of SUSY}

Combined with the usual Poincar\'e and internal symmetry algebra
the Super-Poincar\'e Lie algebra contains additional SUSY
generators $Q_{\alpha}^i$ and $\bar Q_{\dot
\alpha}^i$~\cite{WessB}
 \begin{equation}  \begin{array}{l}
{[} P_{\mu},P_{\nu}{]}  =   0,  \\ {[} P_{\mu},M_{\rho\sigma}{]} =
i(g_{\mu \rho}P_{\sigma} -
 g_{\mu \sigma}P_{\rho}),  \\
{[} M_{\mu \nu} , M_{\rho \sigma} {]}   =    i(g_{\nu \rho}M_{\mu
\sigma} - g_{\nu \sigma}M_{\mu \rho} - g_{\mu \rho}M_{\nu \sigma}
+ g_{ \mu \sigma}M_{\nu \rho}), \\
 {[} B_r , B_s {]}    =    iC_{rs}^t B_{t}, \\
 {[} B_r , P_{\mu} {]}    =    {[} B_r ,M_{\mu \sigma} {]} = 0,  \\
 {[} Q_{\alpha}^i , P_{\mu} {]} =  {[} \bar Q_{\dot \alpha}^i , P_{\mu} {]} = 0, \\
 {[} Q_{\alpha}^i , M_{\mu \nu} {]}  =   \frac{1}{2} (\sigma_{\mu \nu})_{\alpha}
^{\beta}Q_{\beta}^i , \ \ \ \ {[}\bar Q_{\dot \alpha}^i ,M_{\mu
\nu}{]} = - \frac{1}{2} \bar Q_{\dot \beta}^i (\bar \sigma_{\mu
\nu})_{\dot \alpha} ^{\dot \beta} , \\
 {[} Q_{\alpha}^i , B_r {]}  =  (b_r)_{j}^i Q_{\alpha}^j , \ \ \ {[}
 \bar Q_{\dot \alpha}^i , B_r {]} = - \bar Q_{\dot \alpha}^j (b_r)_j^i , \\
 \{ Q_{\alpha}^i , \bar Q_{\dot \beta}^j \}
  =   2 \delta^{ij} (\sigma ^{\mu})_{\alpha \dot \beta }P_{\mu} , \\
 \{ Q_{\alpha}^i , Q_{\beta}^j \}  =   2 \epsilon_{\alpha \beta}Z^{ij} ,
\ \ \ Z_{ij} = a_{ij}^r b_r , \ \ \ \ Z^{ij} = Z_{ij}^+ , \\
 \{ \bar Q_{\dot \alpha}^i , \bar Q_{\dot \beta}^j \}  =  - 2 \epsilon
_{\dot \alpha \dot \beta}Z^{ij} , \ \ \ {[}Z_{ij} , anything {]} =
0 , \\
 \alpha , \dot \alpha  =  1,2 \ \ \ \ i,j = 1,2, \ldots , N .
 \end{array}  \label{group}
 \end{equation}

Here $P_{\mu}$ and $M_{\mu \nu}$  are four-momentum and angular
momentum operators, respectively, $B_r$ are the internal symmetry
generators, $Q^i$ and $\bar Q^i$ are the spinorial SUSY generators
and $Z_{ij}$ are the so-called central charges; $\alpha , \dot
\alpha, \beta , \dot \beta $ are the spinorial indices. In the
simplest case one has one spinor generator $Q_\alpha$ (and the
conjugated one $\bar Q_{\dot{\alpha}}$) that corresponds to an
ordinary or N=1 supersymmetry. When $N>1$ one has an extended
supersymmetry. In what follows we consider the simplest N=1 case
used for phenomenology.

To construct the representations of SUSY algebra (particle states
in SUSY model) we start with the some state labeled by energy and
helicity, i.e. projection of a spin on the direction of momenta
 $$|E,\lambda > $$
and act on it with the SUSY generator $\bar Q$. Then  one obtains
the other state with the same energy (because SUSY generator
commutes with $P_\mu$) but different helicity
\begin{equation}\label{state}
  \bar Q|E,\lambda>= |E,\lambda +1/2>.
\end{equation}
Due to the nilpotent character of SUSY generators (\ref{group}),
the repeated action of the generator $\bar Q$ gives zero. This is
common for N=1 SUSY. One has two states, one bosonic and one
fermionic. This is a generic property of any supersymmetric theory
that the number of bosons equals that of fermions. However, in CPT
invariant theories the number of states is doubled, since CPT
transformation changes the sign of helicity. Hence, in CPT
invariant theories, one has to add the states with opposite
helicity to the above mentioned ones.

Consider some examples. Let us take  $\lambda= 0$. Then one has
the following complete multiplet of SUSY:
 $$\begin{array}{lccccccc} &
\mbox{helicity}&0&1/2& & \mbox{helicity}&0&-1/2 \\
 N=1 \ \ \  \lambda =0 \ \ \  & & & & \stackrel{CPT}{\Longrightarrow} &&  &\\
 & \# \ \mbox{of states}& 1& 1&&\# \ \mbox{of states} &1& 1
\end{array}$$
 which contains one complex scalar and
one spinor with two helicity states.

The other multiplet can be obtained if one starts with
$\lambda=1/2$. Then one has:
$$\begin{array}{lccccccc} &
\mbox{helicity}&1/2&1& & \mbox{helicity}&-1&1/2 \\
 N=1 \ \ \  \lambda =1/2 \ \ \  & & & & \stackrel{CPT}{\Longrightarrow} &&  &\\
 & \# \ \mbox{of states}& 1& 1&&\# \ \mbox{of states} &1& 1
\end{array}$$
This multiplet  contains one spinor field and one massless vector.

Thus, one has two types of supermultiplets: the so-called chiral
multiplet with $\lambda =0$, which contains two physical states
$(\phi,\psi)$ with spin 0 and 1/2, respectively, and the vector
multiplet with $\lambda =1/2$, which also contains two physical
states $(\lambda, A_\mu)$ with spin 1/2 and 1, respectively. These
multiplets are used to describe quarks,  leptons  and vector
bosons in SUSY generalization of the SM.

\subsubsection{Superspace and superfields}

An elegant formulation of supersymmetry transformations and
invariants can be achieved in the framework of superspace
\cite{sspace}.
 Superspace differs from the ordinary Euclidean (Minkowski)
space by adding of two new coordinates, $\theta_{\alpha}$ and
$\bar \theta_{\dot \alpha}$, which are Grassmannian,
 i.e. anti\-com\-muting, variables
 $$\{ \theta_{\alpha}, \theta_{\beta} \} = 0 , \ \ \{\bar
\theta_{\dot \alpha}, \bar \theta_{\dot \beta} \} = 0, \ \
\theta_{\alpha}^2 = 0,\ \ \bar \theta_{\dot \alpha}^2=0, \ \
\alpha,\beta, \dot\alpha, \dot\beta =1,2.$$
 Thus, we go from space to superspace
$$\begin{array}{cc} Space & \ \Rightarrow \ \ Superspace \\
x_{\mu} & \ \ \ \ \ \ \ x_{\mu}, \theta_{\alpha} , \bar \theta_
{\dot \alpha} \end{array}$$

Supersymmetry transformation in superspace looks like an ordinary
translation but in Grassmannian coordinates
 \begin{equation}
 \begin{array}{ccl}
x_{\mu} & \rightarrow & x_{\mu} + i\theta \sigma_{\mu} \bar
\varepsilon
 - i\varepsilon \sigma_{\mu} \bar \theta, \\
\theta & \rightarrow & \theta + \varepsilon , \ \ \bar \theta
\rightarrow  \bar \theta + \bar \varepsilon ,\label{sutr}
\end{array}
 \end{equation}
where $\varepsilon $ and $\bar \varepsilon $ are Grassmannian
transformation parameters. From eq.(\ref{sutr}) one can easily
obtain the representation for the supercharges (\ref{group}), the
generators of supersymmetry, acting on the superspace
 \begin{equation}
Q_\alpha =\frac{\partial }{\partial \theta_\alpha
}-i\sigma^\mu_{\alpha \dot \alpha}\bar{\theta}^{\dot \alpha
}\partial_\mu , \ \ \ \bar{Q}_{\dot \alpha} =-\frac{\partial
}{\partial \bar{\theta}_{\dot \alpha}
}+i\theta_\alpha\sigma^\mu_{\alpha \dot \alpha}\partial_\mu .
\label{q}
 \end{equation}

To define the fields  on a superspace, consider  representations
of the Super-Poincar\'e group (\ref{group}) \cite{WessB}. The
simplest N=1 SUSY multiplets that we discussed earlier are: the
chiral one $\Phi(y,\theta)$ ($y =x + i\theta \sigma \bar \theta $)
and the vector one $V(x,\theta,\bar \theta)$. Being expanded in
Taylor series over  Grassmannian variables $\theta$ and
$\bar\theta$ they give:
 \begin{eqnarray}
\Phi (y, \theta ) & = & A(y) + \sqrt{2} \theta \psi (y) + \theta
 \theta F(y) \label{field} \\
  & \hspace{-0.5cm} = & \hspace{-0.5cm} A(x) +
  i\theta \sigma^{\mu} \bar \theta \partial_{\mu}A(x)
 + \frac{1}{4} \theta \theta \bar \theta \bar \theta \Box A(x) \nonumber\\
 &   + & \sqrt{2} \theta \psi (x) - \frac{i}{\sqrt{2}} \theta
\theta \partial_{\mu} \psi (x) \sigma^{\mu} \bar \theta + \theta
 \theta F(x).\nonumber
 \end{eqnarray}
The coefficients are ordinary functions of $x$ being the usual
fields. They are called the {\em components} of a superfield. In
eq.(\ref{field}) one has 2 bosonic (complex scalar field $A$) and
2 fermionic (Weyl spinor field $\psi$) degrees of freedom. The
component fields $A$ and $\psi$ are called the {\em
superpartners}. The field $F$ is  an {\em auxiliary} field, it has
the "wrong" dimension and has no physical meaning. It is needed to
close the algebra (\ref{group}). One can get rid of the auxiliary
fields with the help of equations of motion.

Thus, a superfield contains an equal number of bosonic and
fermionic degrees of freedom. Under SUSY transformation they
convert  into one another
 \begin{eqnarray}
\delta_\varepsilon A &=& \sqrt 2 \varepsilon \psi, \nonumber \\
\delta_\varepsilon \psi &=& i \sqrt 2 \sigma^\mu \bar \varepsilon
\partial_\mu A + \sqrt 2 \varepsilon F, \label{transf} \\
\delta_\varepsilon F &=&i \sqrt 2  \bar \varepsilon\sigma^\mu\partial_\mu\psi.
\nonumber
 \end{eqnarray}
Notice that the variation of the $F$-component is a total derivative, i.e.
it vanishes when integrated over the space-time.

The vector superfield is real $V = V^+$. It has the following
Grassmannian expansion:
 \begin{eqnarray}
V(x, \theta, \bar \theta) & = & C(x) + i\theta \chi (x) -i\bar
\theta \bar \chi (x)
  + \frac{i}{2} \theta \theta [M(x) + iN(x)] \nonumber \\
  & &\hspace*{-0.9cm} - \ \frac{i}{2} \bar
 \theta \bar \theta [M(x) - iN(x)]
   - \theta \sigma^{\mu}
 \bar \theta v_{\mu}(x) + i \theta \theta
\bar \theta [\lambda (x) + \frac{i}{2}\bar \sigma^{\mu} \partial
_{\mu} \chi (x)] \nonumber \\
 & - & i\bar \theta \bar \theta \theta [\lambda + \frac{i}{2}
\sigma^{\mu} \partial_{\mu} \bar \chi (x)] + \frac{1}{2} \theta
\theta \bar \theta \bar \theta [D(x) + \frac{1}{2}\Box C(x)].
 \label{p}
 \end{eqnarray}
The physical degrees of freedom corresponding to a real vector
superfield $V$ are the vector gauge field $v_{\mu}$ and the
Majorana spinor field $\lambda$. All other components are
unphysical and can be eliminated. Indeed, one can choose a gauge
(the Wess-Zumino gauge) where $C = \chi = M = N =0 $, leaving one
with only physical degrees of freedom except for the auxiliary
field $D$. In this gauge
 \begin{eqnarray}
V & = & - \theta \sigma^{\mu} \bar \theta v_{\mu}(x) + i \theta
\theta \bar \theta \bar \lambda (x) -i\bar \theta \bar \theta
\theta \lambda (x) + \frac{1}{2} \theta \theta \bar \theta \bar
\theta D(x) , \nonumber\\ V^2 & = & - \frac{1}{2} \theta \theta
\bar \theta \bar \theta v_{\mu}(x)v^{\mu}(x) , \nonumber\\ V^3 & =
& 0, \ \ \ etc.
 \end{eqnarray}
 One can define also a field
strength tensor (as analog of $F_{\mu \nu}$ in gauge theories)
 \begin{equation}
W_{\alpha}  =  - \frac{1}{4} \bar D^2 e^V D_{\alpha} e^{-V} , \ \
\ \ \bar W_{\dot \alpha} =  - \frac{1}{4} D^2 e^V \bar D_{\alpha}
e^{-V}, \label{str}
 \end{equation}
 Here $Ds$ are the supercovariant derivatives.
 In the Wess-Zumino gauge the strength tensor is a polynomial
over component fields:
 \begin{equation}
W_\alpha = T^a\left(-i\lambda^a_\alpha +\theta_\alpha D^a-
\frac{i}{2}(\sigma^\mu \bar{\sigma}^\nu \theta)_\alpha F^a_{\mu
\nu }+ \theta^2(\sigma^\mu D_\mu \bar{\lambda}^a)_{\alpha} \right)
,
 \end{equation}
where
 $$F^a_{\mu \nu }=\partial_\mu v^a_\nu - \partial_\nu v^a_\mu
+f^{abc} v^b_\mu v^c_\nu , \ \ \ D_\mu \bar{\lambda }^a=\partial
\bar{\lambda }^a +f^{abc}v^b_\mu \bar{\lambda }^c.$$
 In Abelian case eqs.(\ref{str}) are simplified and take form
$$W_\alpha= - \frac{1}{4} \bar
D^2 D_{\alpha}V , \ \ \ \bar W_{\dot \alpha}= - \frac{1}{4} D^2 \bar
D_{\alpha}V .$$

\subsubsection{Construction of SUSY Lagrangians}

Let us start with the Lagrangian which has no local gauge
invariance. In the superfield notation SUSY invariant Lagrangians
are the polynomials of superfields.  The same way as an ordinary
action is an integral over space-time of Lagrangian density, in
supersymmetric case the action is an integral over the superspace.
The space-time Lagrangian density  is~\cite{WessB,sspace}
 \begin{equation}
{\cal L}  = \int d^2\theta d^2\bar \theta ~\Phi_i^+ \Phi_i+ \int
d^2\theta
 ~[\lambda_i \Phi_i + \frac{1}{2}m_{ij}\Phi_i \Phi_j + \frac{1}{3} y_{ijk}
\Phi_i \Phi_j \Phi_k] +h.c. \label{l}
 \end{equation}
where the first part is a kinetic term and the second one is a
superpotential  ${\cal W}$. We use here integration over the
superspace according to the rules of Grassmannian
integration~\cite{ber}
 $$\int \ d\theta_\alpha =
0 , \ \ \ \ \int \theta _\alpha\ d\theta _\beta=
\delta_{\alpha\beta}. $$
 Performing explicit integration over the Grassmannian parameters,
we get from eq.(\ref{l})
 \begin{eqnarray}
{\cal L} & = & i\partial_{\mu}\bar \psi_i \bar \sigma^{\mu}\psi_i
+ A_i^{\ast} \Box A_i + F_i^{\ast}F_i \label{20} \\
 & + & [\lambda_i F_i + m_{ij}(A_iF_j - \frac{1}{2}\psi_i \psi_j )
+ y_{ijk}(A_iA_jF_k - \psi_i \psi_j A_k ) + h.c. ] . \nonumber
 \end{eqnarray}
The last two terms are the interaction ones. To obtain a familiar
form of the  Lagrangian, we have to solve the constraints
 \begin{eqnarray}
\frac{\partial {\cal L}}{\partial F_k^*} & = & F_k + \lambda_k^* +
m_{ik}^*A_i^* + y_{ijk}^* A_i^*A_j^* = 0, \\ \frac{\partial {\cal
L}}{\partial F_k} & = & F_k^* + \lambda_k + m_{ik}A_i + y_{ijk}
A_iA_j = 0.
\end{eqnarray}
Expressing the auxiliary fields $F$ and $F^*$  from these
equations, one finally gets
 \begin{eqnarray} {\cal L} & = & i\partial_{\mu}\bar \psi_i \bar
\sigma^{\mu}\psi_i + A_i^* \Box A_i - \frac{1}{2}m_{ij}\psi_i
\psi_j - \frac{1}{2}m_{ij}^* \bar \psi_i \bar \psi_j \nonumber \\
& & - y_{ijk}\psi_i \psi_j A_k - y_{ijk}^* \bar \psi_i \bar \psi_j
A_k^* - V(A_i,A_j),  \label{m}
 \end{eqnarray}
where the scalar potential $V = F_k^* F_k $. We will return to the
discussion of the form of the scalar potential in SUSY theories
later.

Consider now the gauge invariant SUSY Lagrangians. They should
contain gauge invariant interaction of the matter fields with the
gauge ones and the kinetic term and the self-interaction of the
gauge fields.

Let us start with the gauge field kinetic terms. In the
Wess-Zumino gauge one has
 \begin{equation} W^{\alpha}W_{\alpha} |_{\theta \theta}= -2i\lambda
\sigma^{\mu}D_{\mu}\bar \lambda -
 \frac{1}{2}F_{\mu \nu}F^{\mu \nu}+\frac{1}{2}D^2
+i \frac{1}{4}F^{\mu \nu}F^{\rho \sigma}\epsilon_{\mu \nu \rho
\sigma },
 \end{equation}
where $D_\mu = \partial_\mu +ig[v_\mu, ]$ is the usual covariant
derivative and the last, the so-called topological $\theta$ term,
is the total derivative. The gauge invariant Lagrangian now has a
familiar form
 \begin{eqnarray}
{\cal L} & = & \frac{1}{4}\int d^2\theta ~W^{\alpha}W_{\alpha}
 + \frac{1}{4} \int d^2 \bar \theta ~\bar W^{\dot \alpha}\bar W_{\dot \alpha}
  \nonumber \\
 & = & \frac{1}{2}D^2 - \frac{1}{4}F_{\mu \nu}F^{\mu \nu} -
 i \lambda \sigma^{\mu}D_{\mu}\bar \lambda. \label{29}
 \end{eqnarray}
To obtain a gauge-invariant interaction with matter chiral
superfields, one has to modify the kinetic term by inserting the
bridge operator
 \begin{equation}
\Phi_i^+  \Phi_i \Rightarrow \Phi_i^+ e^{gV} \Phi_i.
 \end{equation}

A complete SUSY and gauge invariant Lagrangian then looks like
 \begin{eqnarray}
{\cal L}_{SUSY \ YM} & = & \frac{1}{4}\int d^2 \theta
~Tr(W^{\alpha}W_{\alpha})   + \frac{1}{4}\int d^2 \bar{\theta}
~Tr(\bar{W}^{\alpha}\bar{W}_{\alpha})  \label{nonab}\\ &+& \int
d^2 \theta d^2 \bar \theta ~\bar \Phi_{ia}(e^{gV})_b^a\Phi_i^b
+\int d^2 \theta ~{\cal W}(\Phi_i)   +\int d^2 \bar{\theta}
~\bar{{\cal W}}(\bar{\Phi}_i)  , \nonumber
 \end{eqnarray}
where ${\cal W}$ is a  superpotential, which should be invariant
under the group of symmetry of a particular model. In terms of
component fields the above Lagrangian takes the form
 \begin{eqnarray}
{\cal L}_{SUSY \ YM} & = & -\frac{1}{4}F^a_{\mu \nu }F^{a\mu \nu
}-i\lambda^a\sigma^\mu D_\mu \bar{\lambda}^a+\frac{1}{2}D^aD^a
\label{sulag}\\ &&\hspace*{-2.6cm}+\ (\partial_\mu A_i -igv^a_\mu
T^aA_i)^\dagger (\partial_\mu A_i -igv^{a}_\mu T^aA_i)
-i\bar{\psi}_i\bar{\sigma}^\mu (\partial_\mu \psi_i -igv^{a}_\mu
T^a\psi_i) \nonumber \\ && \hspace*{-1.6cm} - \ D^aA^\dagger_i
T^aA_i-i\sqrt{2}A^\dagger_iT^a\lambda^a\psi_i +
i\sqrt{2}\bar{\psi}_iT^aA_i\bar{\lambda}^a+F^\dagger_iF_i
\nonumber \\ &&\hspace*{-1.6cm} +\ \frac{\partial {\cal
W}}{\partial A_i} F_i+ \frac{\partial \bar{{\cal W}}}{\partial
A_i^\dagger}F^\dagger_i -\frac{1}{2}\frac{\partial^2 {\cal
W}}{\partial A_i \partial A_j}\psi_i\psi_j
-\frac{1}{2}\frac{\partial^2 \bar{{\cal W}}}{\partial A_i^\dagger
\partial A_j^\dagger}\bar{\psi}_i\bar{\psi}_j.\nonumber
 \end{eqnarray}
Integrating out the auxiliary fields $D^a$ and $F_i$, one
reproduces the usual Lagrangian.

\subsubsection{The scalar potential}

Contrary to the SM, where the scalar potential is arbitrary and is
defined only by the requirement of the gauge invariance, in
supersymmetric theories it is completely  defined by the
superpotential. It consists of the contributions from the
$D$-terms and $F$-terms. The kinetic energy of the gauge fields
(recall eq.(\ref{29}) yields the $1/2 D^aD^a$ term, and the
matter-gauge interaction (recall eq.(\ref{sulag}) yields the
$gD^aT^a_{ij}A^*_iA_j$ one. Together they give
 \begin{equation}
{\cal L}_D=\frac{1}{2}D^aD^a + gD^aT^a_{ij}A^*_iA_j. \label{d}
 \end{equation}
The equation of motion reads
 \begin{equation}
D^a=-gT^a_{ij}A^*_iA_j. \label{sol}
 \end{equation}
Substituting it back into eq.(\ref{d}) yields the $D$-term part of
the potential
 \begin{equation} {\cal L}_D=-\frac{1}{2}D^aD^a \ \ \ \
\Longrightarrow V_D=\frac{1}{2}D^aD^a,
 \end{equation}
where $D$ is given by eq.(\ref{sol}).

The $F$-term contribution can be derived from the matter field
self-in\-ter\-action eq.(\ref{20}). For a general type
superpotential $W$ one has
 \begin{equation}
{\cal L}_F=F^*_iF_i+(\frac{\partial W}{\partial A_i}F_i + h.c.).
 \end{equation}
Using the equations of motion for the auxiliary field $F_i$
 \begin{equation}
F^*_i=-\frac{\partial W}{\partial A_i} \label{solf}
 \end{equation}
yields
 \begin{equation} {\cal L}_F=-F^*_iF_i \ \ \ \
\Longrightarrow V_F= F^*_iF_i ,
 \end{equation}
where $F$ is given by eq.(\ref{solf}). The full potential is the
sum of the two contributions
 \begin{equation}
V=V_D+V_F.
 \end{equation}

Thus, the form of the Lagrangian  is practically fixed by symmetry
requirements. The only freedom is the field content, the value of
the gauge coupling $g$, Yukawa couplings $y_{ijk}$ and the masses.
Because of the renormalizability constraint $V \leq A^4 $ the
superpotential should be limited by ${\cal W} \leq \Phi^3 $ as in
eq.(\ref{l}). All members of a supermultiplet have the same
masses, i.e. bosons and fermions are degenerate in masses. This
property of SUSY theories contradicts the phenomenology and
requires supersymmetry breaking.

\subsubsection{Spontaneous breaking of SUSY}

Since supersymmetric algebra leads to mass degeneracy in a
supermultiplet, it should be broken  to explain the absence of
superpartners at modern energies. There are several ways of
supersymmetry breaking. It can be broken either explicitly or
spontaneously. Performing SUSY breaking one has to be careful not
to spoil the cancellation of quadratic divergencies which allows
one to solve the hierarchy problem. This is achieved by
spontaneous breaking of SUSY.

Apart from non-supersymmetric theories, in SUSY models the energy
is always nonnegative definite. Indeed, according to quantum
mechanics $$ E = <0| \ H \ |0> $$ and due to SUSY algebra
eq.(\ref{group})
 $\{Q_{\alpha}, \bar Q_{\dot \beta} \} =
2(\sigma^{\mu}) _{\alpha \dot \beta}P_{\mu} ,$ taking into account
that $tr(\sigma^{\mu}P_{\mu}) = 2P_0 ,$
 one gets
$$E = \frac{1}{4} \sum_{\alpha = 1,2}<0| \{Q_{\alpha}, \bar
Q_{\alpha} \} |0> = \frac{1}{4} \sum_{\alpha} |Q_{\alpha}
  |0> |^2 \geq 0 .$$
Hence $$ E = <0| \ H \ |0> \neq 0 \ \ \ \ if \ and \ only \ if \ \
\ Q_{\alpha}|0> \neq 0 .$$

Therefore, supersymmetry is spontaneously broken, i.e. vacuum is
not invariant $(Q_{\alpha} |0> \neq 0 )$, {\em if and only if} the
minimum of the potential is positive $(i.e.\ E > 0)$ .

\begin{figure}[ht]
 \begin{center}
 \leavevmode
  \epsfxsize=12cm
 \epsffile{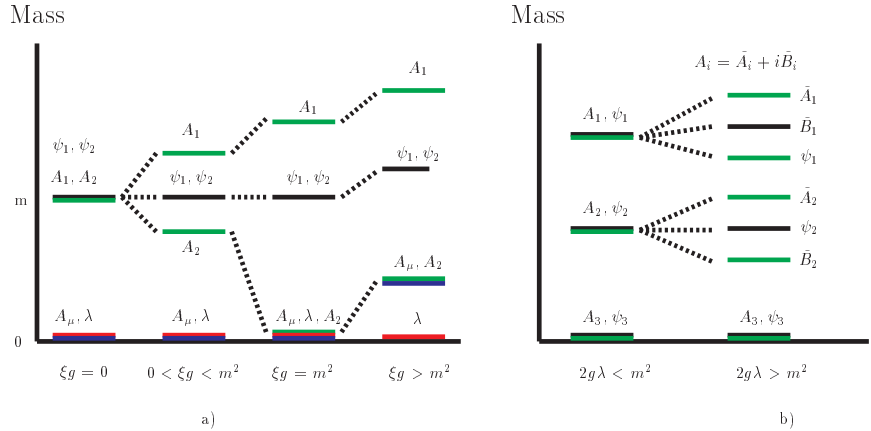}
 \end{center}\vspace{-1cm}
 \caption{Spectrum of spontaneously broken SUSY theories}\label{FI}
 \end{figure}

Spontaneous breaking of supersymmetry is achieved in the same way
as the electroweak symmetry  breaking. One introduces the field
whose vacuum expectation value is nonzero and breaks the symmetry.
However, due to a special character of SUSY, this should be a
superfield whose auxiliary $F$ and $D$ components acquire nonzero
v.e.v.'s. Thus,  among possible spontaneous SUSY breaking
mechanisms one distinguishes the $F$ and $D$ ones.

i) Fayet-Iliopoulos ($D$-term) mechanism \cite{Fayet}. \\
 In this case the, the linear $D$-term is added to the Lagrangian
 \begin{equation} \Delta {\cal
L} = \xi V\vert_{ \theta \theta \bar \theta \bar \theta} = \xi\int
d^4\theta\ V.
\end{equation}
 It is gauge and SUSY invariant by itself; however, it may
lead to spontaneous breaking of both of them depending on the value of
$\xi$. We show in Fig.\ref{FI}a the sample spectrum for two chiral matter
multiplets.
The drawback of this mechanism is the necessity of $U(1)$ gauge
invariance. It can be used in SUSY generalizations of the SM but
not in GUTs.

The mass spectrum also causes some troubles since the following
sum rule is  valid
 \begin{equation}
\sum_{bosonic \ states} m^2_i = \sum_{fermionic \ states} m^2_i ,
\label{sumrule}
 \end{equation}
which is bad for  phenomenology.

ii) O'Raifeartaigh ($F$-term) mechanism \cite{O'R}. \\
 In this case,
several chiral fields are needed and the superpotential should be
chosen in a way that trivial zero v.e.v.s for the auxiliary
$F$-fields be absent.  For instance, choosing the superpotential
to be
 $${\cal W}(\Phi)=
\lambda \Phi_3 +m\Phi_1\Phi_2 +g \Phi_3\Phi_1^2,$$
 one gets the equations for the auxiliary fields
 \begin{eqnarray*}
F^*_1&=&mA_2+2gA_1A_3, \\
F^*_2 &=& mA_1, \\
F^*_3 &=& \lambda +gA^2_1,
 \end{eqnarray*}
which have no solutions with $<F_i> =0$ and SUSY is spontaneously broken.
The sample spectrum is shown in Fig.\ref{FI}b.

The drawbacks of this mechanism is a lot of arbitrariness in the
choice of potential. The sum rule (\ref{sumrule}) is also valid
here.

Unfortunately, none of these mechanisms explicitly works in SUSY
generalizations of the SM. None of the fields of the SM can
develop nonzero v.e.v.s for their $F$ or $D$ components without
breaking  $SU(3)$ or $U(1)$ gauge invariance since they are not
singlets with respect to these groups. This requires the presence
of extra sources of spontaneous SUSY breaking, which we consider
below. They are based, however, on the same  $F$ and $D$
mechanisms.

\subsection{SUSY generalization of the Standard Model. The MSSM}

As has been already mentioned, in SUSY theories the number of
bosonic degrees of freedom equals that of fermionic. At the same
time,  in the SM one has 28 bosonic and 90 (96 with right handed
neutrino) fermionic degrees of freedom. So the SM is to a great
extent non-supersymmetric. Trying to add some new particles to
supersymmetrize the SM, one should take into account the following
observations:

$\bullet$  There are no fermions with quantum numbers of the gauge
bosons;

$\bullet$ Higgs fields have nonzero v.e.v.s; hence they cannot be
 superpartners of quarks and leptons since
this would induce  spontaneous violation of baryon and lepton
numbers;

 $\bullet$ One needs at least two complex chiral Higgs multiplets to
give masses to Up and Down quarks.

The latter is due to the form of a superpotential and chirality of
matter superfields. Indeed, the superpotential should be invariant
under the $SU(3)\times SU(2)\times U(1)$ gauge group. If one looks
at the Yukawa interaction in the Standard Model, one finds that it
is indeed $U(1)$ invariant since the sum of hypercharges in each
vertex equals zero. In the last term this is achieved by taking
the conjugated Higgs doublet $\tilde{H}=i\tau_2H^\dagger$ instead
of $H$. However, in SUSY $H$ is a chiral superfield and hence a
superpotential, which is constructed out of  chiral fields, can
contain only $H$ but not $\tilde H$ which is an antichiral
superfield.

Another reason for the second  Higgs doublet is related to chiral
anomalies. It is known that chiral anomalies spoil the gauge
invariance and, hence, the renormalizability of the theory. They
are canceled in the SM between quarks and leptons in each
generation. However, if one introduces a chiral Higgs superfield,
it contains higgsinos, which are chiral fermions, and contain
anomalies. To cancel them one has to add the second Higgs doublet
with the opposite hypercharge. Therefore, the Higgs sector in SUSY
models is inevitably enlarged, it contains an even number of
doublets.

 {\em Conclusion}: In SUSY models supersymmetry
associates {\em known} bosons with {\em new} fermi\-ons and {\em
known} fermi\-ons with {\em new} bosons.

\subsubsection{The field content}

\begin{figure}[ht]
\begin{center}\vspace{-0.5cm}
 \leavevmode
  \epsfxsize=8cm
 \epsffile{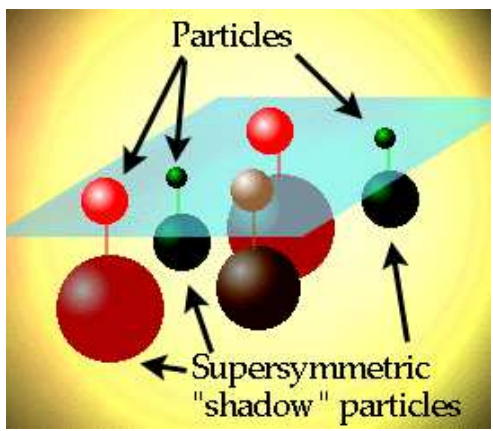}
\end{center}\vspace{-1cm} \caption{The shadow world of SUSY particles~\cite{shadow}
}\label{fig:shadow}
 \end{figure}

Consider the particle content of the Minimal Supersymmetric
Standard Model \cite{MSSM,Lectures}. According to the previous
discussion, in the minimal version  we double the number of
particles (introducing a superpartner to each particle) and add
another Higgs doublet (with its superpartner). Thus, the
characteristic feature of any supersymmetric generalization of the
SM is the presence of superpartners (see Fig.\ref{fig:shadow}). If
supersymmetry is exact, superpartners of ordinary particles should
have the same masses and have to be observed. The absence of them
at modern energies is believed to be explained by the fact that
their masses are very heavy, that means that supersymmetry should
be broken.

The particle content of the MSSM then appears as (tilde denotes a
superpartner of an ordinary particle).
\begin{center}
{\bf Particle Content of the MSSM}\vglue 0.6cm
\nopagebreak[4]
\renewcommand{\tabcolsep}{0.03cm}
\begin{tabular}{lllccc}
Superfield & \ \ \ \ \ \ \ Bosons & \ \ \ \ \ \ \ Fermions &
$SU_c(3)$& $SU_L(2)$ & $U_Y(1)$ \\ \hline \hline Gauge  &&&&& \\
${\bf G^a}$   & gluon \ \ \ \ \ \ \ \ \ \ \ \ \ \ \  $g^a$ &
gluino$ \ \ \ \ \ \ \ \ \ \ \ \ \tilde{g}^a$ & 8 & 1 & 0 \\ ${\bf
V^k}$ & Weak \ \ \ $W^k$ \ $(W^\pm, Z)$ & wino, zino \
$\tilde{w}^k$ \ $(\tilde{w}^\pm, \tilde{z})$ & 1 & 3& 0 \\ ${\bf
V'}$   & Hypercharge  \ \ \ $B\ (\gamma)$ & bino \ \ \ \ \ \ \ \ \
\ \ $\tilde{b}(\tilde{\gamma })$ & 1 & 1& 0 \\ \hline Matter &&&&
\\ $\begin{array}{c} {\bf L_i} \\ {\bf E_i}\end{array}$ & sleptons
\ $\left\{
\begin{array}{l} \tilde{L}_i=(\tilde{\nu},\tilde e)_L \\ \tilde{E}_i =\tilde
e_R \end{array} \right. $ & leptons \ $\left\{ \begin{array}{l}
L_i=(\nu,e)_L
\\ E_i=e_R \end{array} \right.$ & $\begin{array}{l} 1 \\ 1 \end{array} $  &
$\begin{array}{l} 2 \\ 1 \end{array} $ & $\begin{array}{r} -1 \\ 2
\end{array} $ \\ $\begin{array}{c} {\bf Q_i} \\ {\bf U_i} \\ {\bf D_i}
\end{array}$ & squarks \ $\left\{ \begin{array}{l}
\tilde{Q}_i=(\tilde{u},\tilde d)_L \\ \tilde{U}_i =\tilde u_R \\
\tilde{D}_i =\tilde d_R\end{array}  \right. $ & quarks \ $\left\{
\begin{array}{l} Q_i=(u,d)_L \\ U_i=u_R^c \\ D_i=d_R^c \end{array}
\right.$ & $\begin{array}{l} 3
\\ 3^* \\ 3^* \end{array} $  & $\begin{array}{l} 2 \\ 1 \\ 1 \end{array} $ &
$\begin{array}{r} 1/3 \\ -4/3 \\ 2/3 \end{array} $ \\ \hline Higgs
&&&& \\ $\begin{array}{c} {\bf H_1} \\ {\bf H_2}\end{array}$ &
Higgses \ $\left\{
\begin{array}{l} H_1 \\ H_2 \end{array}  \right. $ & higgsinos \ $\left\{
 \begin{array}{l} \tilde{H}_1 \\ \tilde{H}_2 \end{array} \right.$ &
$\begin{array}{l} 1 \\ 1 \end{array} $  & $\begin{array}{l} 2 \\ 2
\end{array} $ &
$\begin{array}{r} -1 \\ 1
\end{array} $
 \\ \hline \hline
\end{tabular}
\end{center}
\vglue .5cm

The presence of an extra Higgs doublet in SUSY model is a novel
feature of the theory. In the MSSM  one has two doublets with the
quantum numbers (1,2,-1) and (1,2,1), respectively:
$$H_1=\left(\begin{array}{c} H^0_1 \\ H_1^- \end{array}\right) =
\left( \begin{array}{c} v_1 +\frac{
S_1+iP_1}{\sqrt{2}} \\ H^-_1 \end{array}\right), \ H_2=\left(
\begin{array}{c} H^+_2 \\ H_2^0 \end{array}  \right) = \left(
\begin{array}{c} H^+_2 \\ v_2 +\frac{
S_2+iP_2}{\sqrt{2}}
\end{array} \right),$$
where  $v_i$ are the vacuum expectation values of the neutral
components.

Hence, one has 8=4+4=5+3 degrees of freedom. As in the case of the
SM, 3 degrees of freedom can be gauged away, and one is left with
5 physical states compared to 1 in the SM. Thus, in the MSSM, as
actually in any of two Higgs doublet models,
 one has  five  physical Higgs bosons: two CP-even neutral,
one CP-odd neutral  and two charged. We consider the mass
eigenstates below.

\subsubsection{Lagrangian of the MSSM}

 The Lagrangian of the MSSM consists of two parts; the
first part is SUSY generalization of the Standard Model, while the
second one represents the SUSY breaking as mentioned above.
 \begin{equation}
 {\cal L}={\cal L}_{SUSY}+{\cal L}_{Breaking},
 \end{equation}
where
 \begin{eqnarray}
{\cal L}_{SUSY} &= &\sum_{SU(3),SU(2),U(1)}^{}\frac{1}{4}
\left(\int d^2\theta \ Tr W^\alpha W_\alpha + \int d^2\bar\theta \
Tr \bar W^{\dot \alpha}\bar W_{\dot \alpha} \right) \nonumber \\
 && \hspace*{-2cm} +\sum_{Matter}^{}\int d^2\theta d^2 \bar\theta \
\Phi^\dagger_ie^{\displaystyle g_3\hat V_3 + g_2\hat V_2 + g_1\hat
V_1}\Phi_i +\int d^2\theta \ ({\cal W}_R+{\cal W}_{NR}) + h.c .
 \end{eqnarray}
The index $R$ in a superpotential refers to the so-called
$R$-parity~\cite{r-symmetry}. The first part of ${\cal W}$ is
R-symmetric
 \begin{equation}
W_{R} = \epsilon_{ij}(y^U_{ab}Q_a^j U^c_bH_2^i +
y^D_{ab}Q_a^jD^c_bH_1^i
       +  y^L_{ab}L_a^jE^c_bH_1^i + \mu H_1^iH_2^j), \label{R}
 \end{equation}
where $i,j=1,2,3$ are the $SU(2)$ and $a,b=1,2,3$ are the
generation indices; colour indices are suppressed. This part of
the Lagrangian almost exactly repeats that of the SM except that
the fields are now the superfields rather than the ordinary fields
of the SM. The only difference is the last term which describes
the Higgs mixing. It is absent in the SM since there is only one
Higgs field there.

The second part is R-nonsymmetric
 \begin{equation}
W_{NR} =  \epsilon_{ij}(\lambda^L_{abd}L_a^i L_b^jE_d^c +
\lambda^{L\prime}_{abd}L_a^iQ_b^jD_d^c +\mu'_aL^i_aH_2^j)
+\lambda^B_{abd}U_a^cD_b^cD_d^c. \label{NR}
 \end{equation}
These terms are absent in the SM. The reason is very simple: one
can not replace the superfields in eq.(\ref{NR}) by the ordinary
fields like in eq.(\ref{R}) because of the Lorentz invariance.
These terms have a different property, they violate either lepton
(the first 3 terms  in eq.(\ref{NR})) or baryon number (the last
term). Since both effects are not observed in Nature, these terms
must be suppressed or be excluded. One can avoid such terms if one
introduces  special symmetry called the $R$-symmetry. This is the
global $U(1)_R$ invariance
 \begin{equation}
U(1)_R: \ \ \theta \to e^{i\alpha} \theta ,\ \  \Phi \to
e^{in\alpha}\Phi , \label{RS}
 \end{equation}
which is reduced  to the discrete group $Z_2$, called the
$R$-parity. The $R$-parity quantum number
 is given by
$R=(-1)^{3(B-L)+2S}$ for particles with spin $S$. Thus, all the
ordinary particles have the $R$-parity quantum number equal to
$R=+1$, while all the superpartners have $R$-parity quantum number
equal to $R=-1$. The $R$-parity obviously forbids the $W_{NR}$
terms. However, it may well be that these terms are present,
though experimental limits on the couplings are very
severe~\cite{rex}
$$\lambda^L_{abc}, \ \ \lambda^{L\prime}_{abc} < 10^{-4}, \ \ \ \
\ \lambda^B_{abc} < 10^{-9}.$$

\subsubsection{Properties of interactions}

If one assumes that the $R$-parity is preserved, then the
 interactions of superpartners are essentially the same as in the
SM, but two of three particles involved into an interaction at any
vertex are replaced by superpartners. The reason for it is the
$R$-parity. Conservation of the $R$-parity has two consequences

$\bullet$ the superpartners are created in pairs;

$\bullet$  the lightest superparticle (LSP) is stable. Usually it
is  photino $\tilde \gamma $, the superpartner of a photon with
some admixture of neutral higgsino.

\begin{figure}[ht]\vspace{-0.5cm}
\begin{center}
 \leavevmode
  \epsfxsize=13cm \hspace*{-1cm}
 \epsffile{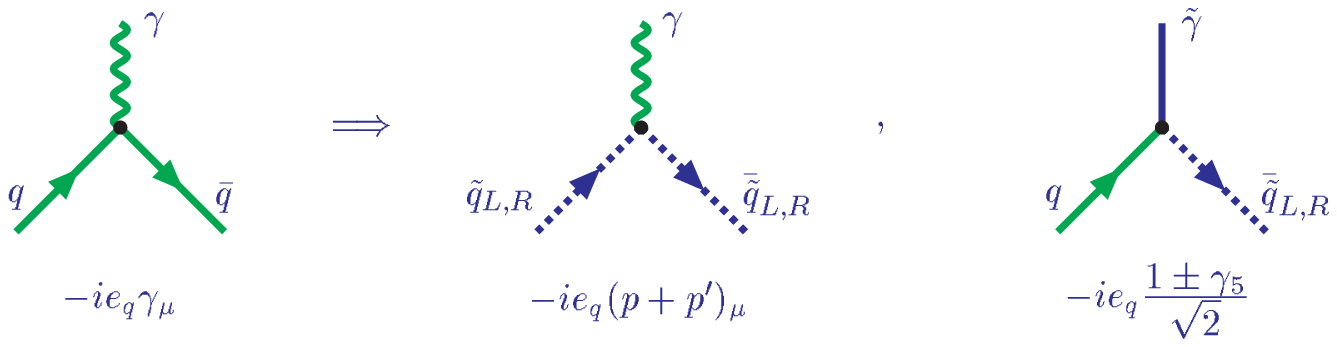}\vspace{-0.2cm}
 \leavevmode
  \epsfxsize=13cm \hspace*{-1cm}
 \epsffile{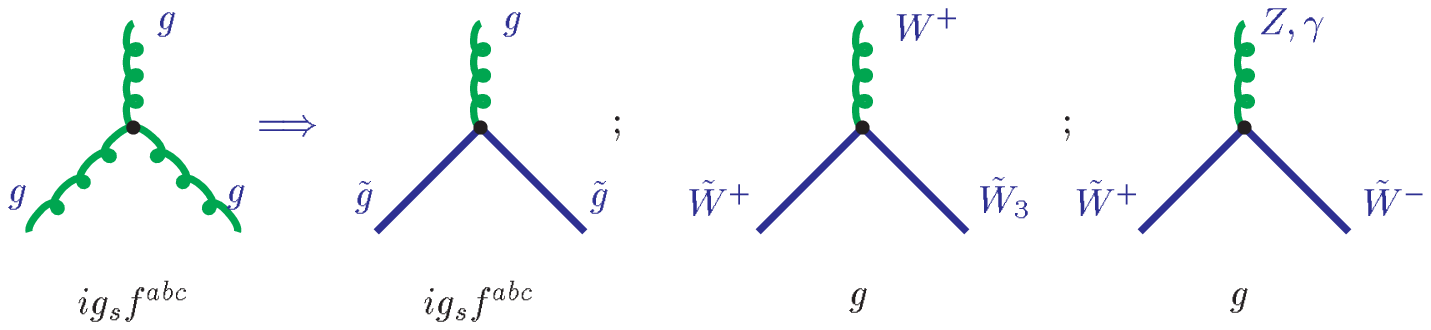}
 \leavevmode
  \epsfxsize=13cm \hspace*{-1cm}\vspace*{-0.9cm}

 \epsffile{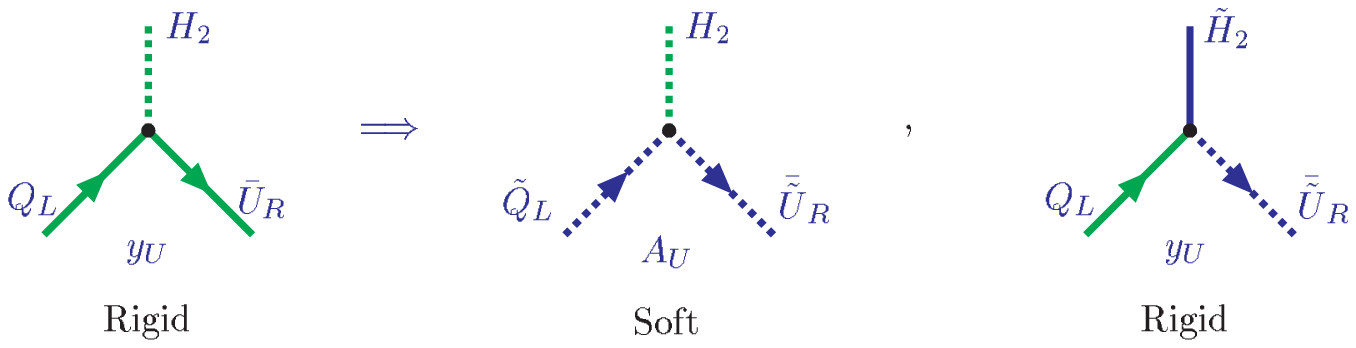}
 \end{center}
\vspace{-1.5cm}
 \caption{Gauge-matter interaction, gauge self-interaction and Yukawa-type interaction}
 \label{yukint}
 \end{figure}

Typical vertices are  shown in Figs.\ref{yukint}. The tilde above
a letter denotes the corresponding superpartner. Note that the
coupling is the same in all the vertices involving superpartners.

\subsubsection{Creation and decay of superpartners}

\begin{figure}[ht]\vspace{-0.6cm}
 \begin{center}
 \leavevmode
  \epsfxsize=11cm
 \epsffile{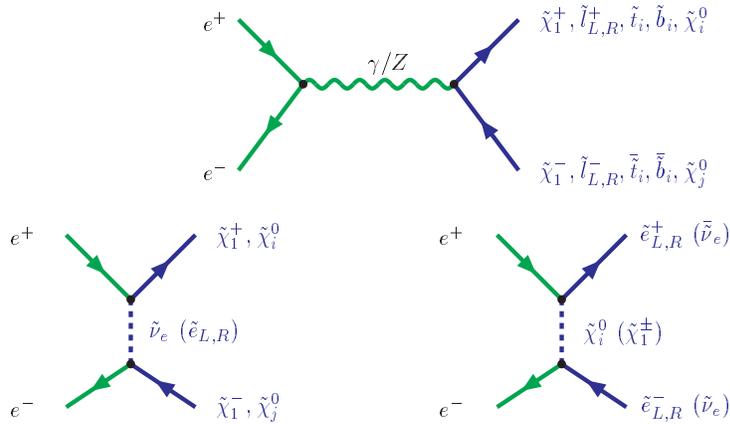}
 \end{center}\vspace{-1cm}
 \caption{Creation of superpartners}\label{creation}
 \end{figure}

The above-mentioned rule   together with the Feynman rules for the
SM enables one to draw the diagrams describing creation of
superpartners. One of the most promising processes is the $e^+e^-$
annihilation (see Fig.\ref{creation}).

\begin{figure}[htb]\vspace{-0.7cm}
 \begin{center}
 \leavevmode
  \epsfxsize=12cm \epsfysize=6cm
 \epsffile{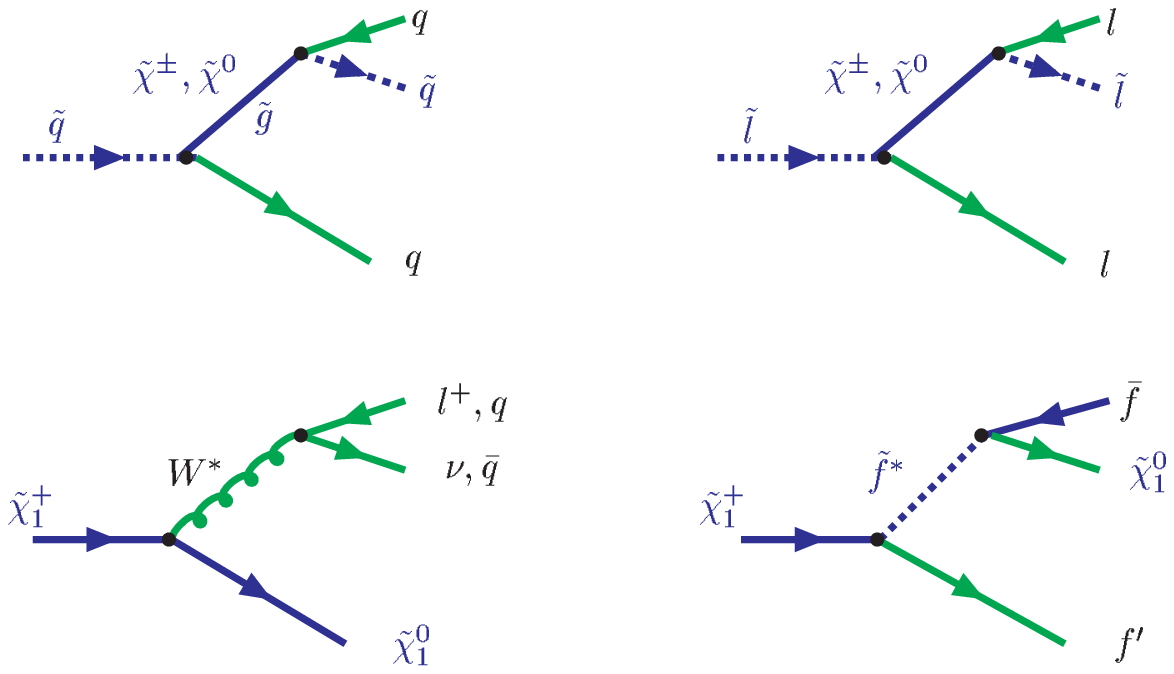}
 \end{center}\vspace{-1cm}
 \caption{Decay of superpartners}\label{decay}
 \end{figure}
The usual kinematic restriction is given by the c.m. energy
$m^{max}_{sparticle} \leq \frac{\sqrt s}{2}.\ $ Similar processes
take place at hadron colliders with electrons and positrons being
replaced by quarks and gluons.

Creation of superpartners can be  accompanied by creation of
ordinary particles as well. We consider various experimental
signatures for $e^+e^-$ and hadron colliders below. They crucially
depend on SUSY breaking pattern and on the mass spectrum of
superpartners.

The decay properties of superpartners also depend on their masses. For
the quark and lepton superpartners the main processes are shown in
Fig.\ref{decay}.

When the $R$-parity is conserved, new particles will eventually
end up giving neutralinos (the lightest superparticle) whose
interactions are comparable to those of neutrinos and they leave
undetected. Therefore, their signature would be missing energy and
transverse momentum.  Thus, if supersymmetry exists in Nature and
if it is broken somewhere below 1 TeV, then it will be possible to
detect it in the nearest future.

\subsection{Breaking of SUSY in the MSSM}

Since none of the fields of the MSSM can develop non-zero v.e.v.
to break SUSY without spoiling the gauge invariance, it is
supposed that spontaneous supersymmetry breaking takes place via
some other fields. The most common scenario for producing
low-energy supersymmetry breaking is called the {\em hidden
sector} one~\cite{hidden}. According to this scenario, there exist
two sectors: the usual matter belongs to the "visible" one, while
the second, "hidden" sector, contains fields which lead to
breaking of supersymmetry. These two sectors interact with each
other by exchange of some fields called {\em messengers}, which
mediate SUSY breaking from the hidden to the visible sector. There
might be various types of messenger fields: gravity, gauge, etc.
The hidden sector is the weakest part of the MSSM. It contains a
lot of ambiguities and leads to uncertainties of the MSSM
predictions considered below.

So far there are known four main mechanisms to mediate SUSY breaking from a
hidden to a visible sector:\vspace{-0.2cm}

\begin{itemize}
\item Gravity mediation (SUGRA)~\cite{gravmed};\\[-0.7cm]
\item Gauge mediation~\cite{gaugemed};\\[-0.7cm]
\item Anomaly mediation~\cite{anommed};\\[-0.7cm]
\item Gaugino mediation~\cite{gauginomed}.
\end{itemize}

\begin{figure}[ht]\vspace{-1cm}
 \begin{center}
\leavevmode
  \epsfxsize=9cm \epsfysize=8cm
 \epsffile{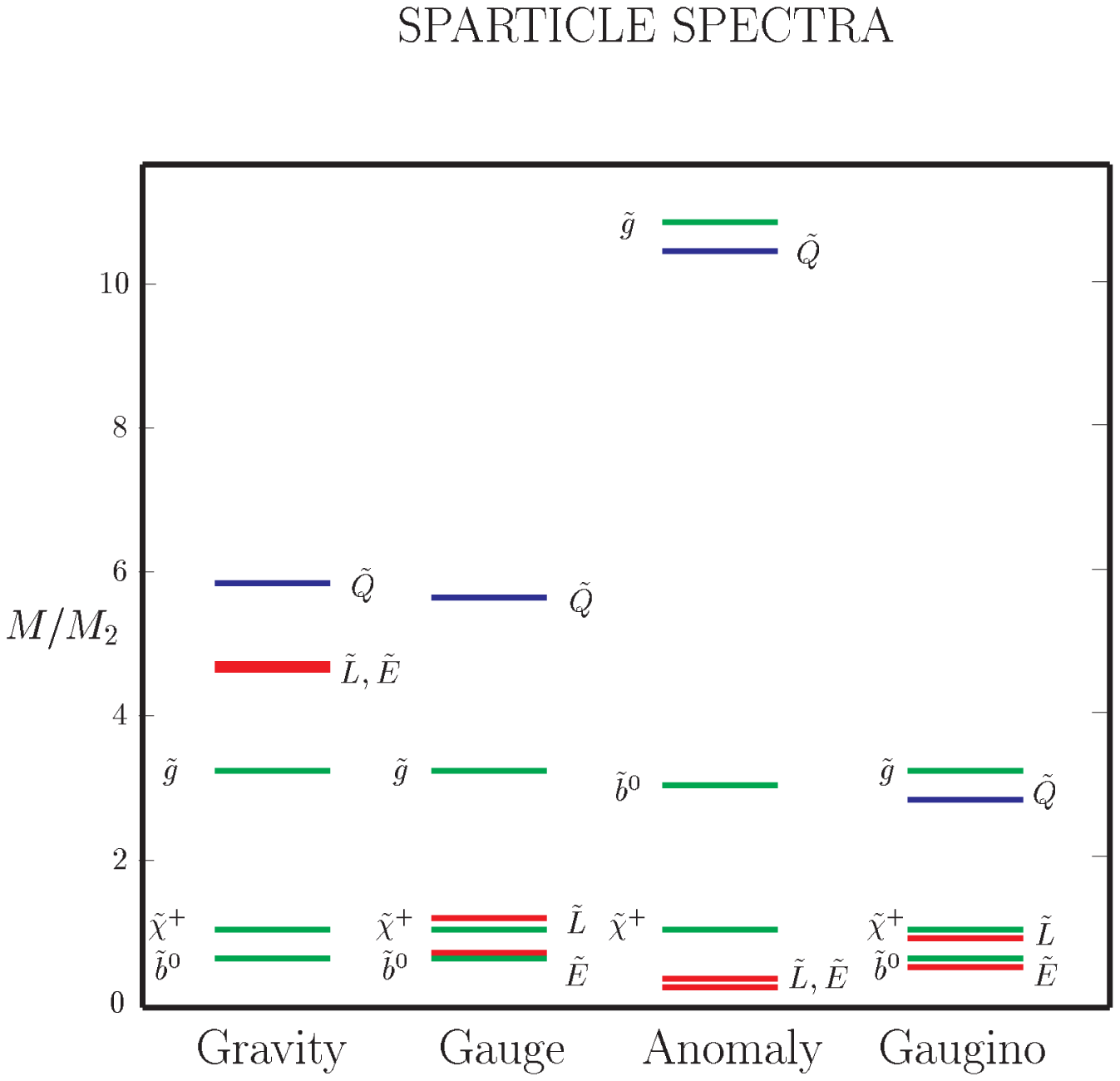}
\end{center}\vspace{-1cm} \caption{Superparticle spectra for various
mediation mechanisms}\label{spectra} \end{figure}

All four mechanisms of soft SUSY breaking are different in details
but are  common in results. Predictions for the sparticle spectrum
depend on the mechanism of SUSY breaking. For comparison of the
four above-mentioned mechanisms we show  in Fig.\ref{spectra} the
sample spectra as the ratio to the gaugino mass
$M_2$~\cite{Peskin}.

In what follows, to calculate the mass spectrum of superpartners,
we need an explicit form of SUSY breaking terms. For the
 MSSM and without  the $R$-parity violation one has
\begin{eqnarray}
-{\cal L}_{Breaking} & = & \sum_{i}^{}m^2_{0i}|\varphi_i|^2+\left( \frac 12
\sum_{\alpha}^{}M_\alpha \tilde \lambda_\alpha\tilde \lambda_\alpha+
BH_1H_2\right. \label{soft}\\
 & + & \left. A^U_{ab}\tilde Q_a\tilde U^c_bH_2+A^D_{ab}\tilde Q_a
 \tilde D^c_bH_1+ A^L_{ab}\tilde L_a\tilde E^c_bH_1
  +h.c.\right) , \nonumber
\end{eqnarray}
where we have suppressed the $SU(2)$ indices.  Here $\varphi_i$
are all scalar fields, $\tilde \lambda_\alpha $  are the gaugino
fields, $\tilde Q, \tilde U, \tilde D$ and $\tilde L, \tilde E$
are the squark and slepton  fields, respectively, and $H_{1,2}$
are the SU(2) doublet Higgs fields.

Eq.(\ref{soft}) contains a vast number of free parameters which
spoils the prediction power of the model. To reduce their number,
we adopt the so-called {\em universality} hypothesis, i.e., we
assume the universality or equality of various soft parameters at
a high energy scale. Namely, following the so-called mSUGRA SUSY
breaking scenario, we put all the spin 0 particle masses to be
equal to the universal value $m_0$, all the spin 1/2 particle
(gaugino) masses to be equal to $m_{1/2}$ and all the cubic and
quadratic terms, proportional to $A$ and $B$, to repeat the
structure of the Yukawa superpotential (\ref{R}). This is an
additional requirement motivated by the supergravity mechanism of
SUSY breaking. Universality is not a necessary requirement and one
may consider nonuniversal soft terms as well. However, it will not
change the qualitative picture presented below; so for simplicity,
in what follows we consider the universal boundary conditions. In
this case, eq.(\ref{soft}) takes the form
\begin{eqnarray}
-{\cal L}_{Breaking} & = & m_0^2\sum_{i}^{}|\varphi_i|^2+\left( \frac 12
m_{1/2}\sum_{\alpha}^{} \tilde \lambda_\alpha\tilde \lambda_\alpha\right.
\label{soft2}\\
 &  & \left.\hspace*{-2.2cm}  +\ A[y^U_{ab}\tilde Q_a\tilde U^c_bH_2+y^D_{ab}\tilde Q_a
 \tilde D^c_bH_1+ y^L_{ab}\tilde L_a\tilde E^c_bH_1]
 +  B[\mu H_1H_2] +h.c.\right) , \nonumber
\end{eqnarray}

The soft terms explicitly break supersymmetry. As will be shown
later, they lead to the mass spectrum of superpartners different
from that of  ordinary particles. Remind that the masses of quarks
and leptons remain zero until $SU(2)$ invariance is spontaneously
broken.

\subsubsection{The soft terms and the mass formulas}

There are two main sources of the mass terms in the Lagrangian:
the $D$ terms and soft ones. With given values of
$m_0,m_{1/2},\mu,Y_t,Y_b,Y_\tau, A$, and $B$ one can construct the
mass matrices for all the particles. Knowing them at the GUT
scale, one can solve the corresponding RG equations, thus linking
the values at the GUT and electroweak scales. Substituting these
parameters into the mass matrices, one can predict the mass
spectrum of superpartners \cite{spectrum,BEK}.

{\bf Gaugino-higgsino mass terms} The mass matrix for  gauginos,
the superpartners of the gauge bosons, and for higgsinos, the
superpartners of the Higgs bosons, is nondiagonal, thus leading to
their mixing. The mass terms look like
\begin{equation}
{\cal L}_{Gaugino-Higgsino}=
 -\frac{1}{2}M_3\bar{\lambda}_a\lambda_a
 -\frac{1}{2}\bar{\chi}M^{(0)}\chi -(\bar{\psi}M^{(c)}\psi + h.c.),
\end{equation}
where $\lambda_a , a=1,2,\ldots ,8,$ are the Majorana gluino
fields and
\begin{equation}
\chi = \left(\begin{array}{c}\tilde{B}^0 \\ \tilde{W}^3 \\
\tilde{H}^0_1 \\ \tilde{H}^0_2
\end{array}\right), \ \ \ \psi = \left( \begin{array}{c}
\tilde{W}^{+} \\ \tilde{H}^{+}
\end{array}\right)
\end{equation}
are, respectively, the Majorana neutralino and Dirac chargino
fields.

The neutralino mass matrix is
\begin{equation}
M^{(0)}=\left(
\renewcommand{\tabcolsep}{0.04cm}
\begin{tabular}{cccc}
$M_1$ & $0$ & -$M_Z\cos\beta \sin_W$ & $M_Z\sin\beta \sin_W$ \\
$0$ &
$M_2$ & $M_Z\cos\beta \cos_W$   & -$M_Z\sin\beta \cos_W$  \\
-$M_Z\cos\beta \sin_W$ & $M_Z\cos\beta \cos_W$  & $0$ & -$\mu$ \\
$M_Z\sin\beta \sin_W$ & -$M_Z\sin\beta \cos_W$  & -$\mu$ & $0$
\end{tabular} \right),\label{neut)}
\end{equation}
where $\tan\beta = v_2/v_1$ is the ratio of two Higgs v.e.v.s and
$\sin_W= \sin\theta_W$ is the usual sinus of the weak mixing
angle. The physical neutralino masses  $M_{\tilde{\chi}_i^0}$ are
obtained as eigenvalues of this matrix after diagonalization.

For charginos one has
\begin{equation}
M^{(c)}=\left(
\begin{array}{cc}
M_2 & \sqrt{2}M_W\sin\beta \\ \sqrt{2}M_W\cos\beta & \mu
\end{array} \right).\label{char}
\end{equation}
This matrix has two chargino eigenstates
$\tilde{\chi}_{1,2}^{\pm}$ with mass eigenvalues
\begin{eqnarray}
M^2_{1,2}&=&\frac{1}{2}\left[M^2_2+\mu^2+2M^2_W \right.\\ &&
\left.\hspace*{-1.5cm} \mp\
\sqrt{(M^2_2-\mu^2)^2+4M^4_W\cos^22\beta
+4M^2_W(M^2_2+\mu^2+2M_2\mu \sin 2\beta )}\right].\nonumber
\end{eqnarray}

{\bf Squark and slepton masses} Non-negligible Yukawa couplings
cause a mixing between the electroweak eigenstates and the mass
eigenstates of the third generation particles.  The mixing
matrices for $\tilde{m}^{2}_t,\tilde{m}^{2}_b$ and
$\tilde{m}^{2}_\tau$ are
\begin{equation} \label{stopmat}
\left(\begin{array}{cc} \tilde m_{tL}^2& m_t(A_t-\mu\cot \beta )
\\ m_t(A_t-\mu\cot \beta ) & \tilde m_{tR}^2 \end{array}  \right),
\nonumber
\end{equation}
\begin{equation} \label{sbotmat}
\left(\begin{array}{cc} \tilde  m_{bL}^2& m_b(A_b-\mu\tan \beta )
\\ m_b(A_b-\mu\tan \beta ) & \tilde  m_{bR}^2 \end{array}
\right), \nonumber
\end{equation}
\begin{equation} \label{staumat} \left(\begin{array}{cc}
\tilde  m_{\tau L}^2& m_{\tau}(A_{\tau}-\mu\tan \beta ) \\
m_{\tau}(A_{\tau}-\mu\tan \beta ) & \tilde m_{\tau R}^2
\end{array}  \right)               \nonumber
\end{equation}
with
\begin{eqnarray*}
  \tilde m_{tL}^2&=&\tilde{m}_Q^2+m_t^2+\frac{1}{6}(4M_W^2-M_Z^2)\cos
  2\beta ,\\
  \tilde m_{tR}^2&=&\tilde{m}_U^2+m_t^2-\frac{2}{3}(M_W^2-M_Z^2)\cos
  2\beta ,\\
  \tilde m_{bL}^2&=&\tilde{m}_Q^2+m_b^2-\frac{1}{6}(2M_W^2+M_Z^2)\cos
  2\beta ,\\
  \tilde m_{bR}^2&=&\tilde{m}_D^2+m_b^2+\frac{1}{3}(M_W^2-M_Z^2)\cos
  2\beta ,\\
 \tilde m_{\tau L}^2&=&\tilde{m}_L^2+m_{\tau}^2-\frac{1}{2}(2M_W^2-M_Z^2)\cos
2\beta ,\\ \tilde m_{\tau
R}^2&=&\tilde{m}_E^2+m_{\tau}^2+(M_W^2-M_Z^2)\cos
  2\beta
\end{eqnarray*}
and the  mass eigenstates are  the eigenvalues of these mass matrices. For
the light generations the mixing is negligible.

The first terms here ($\tilde{m}^2$) are the soft ones, which are
calculated using the RG equations starting from their values at
the GUT (Planck) scale. The second ones are the usual masses of
quarks and leptons and the last ones are the $D$-terms of the
potential.

\subsubsection{The Higgs potential}

As has already been mentioned, the Higgs potential in the MSSM is
totally defined by superpotential ${\cal W}$ and the soft terms.
Due to the structure of ${\cal W}$ the Higgs self-interaction is
given by the $D$-terms while the $F$-terms contribute only to the
mass matrix. The tree level potential is
\begin{eqnarray}
V_{tree}(H_1,H_2)&=&m^2_1|H_1|^2+m^2_2|H_2|^2-m^2_3(H_1H_2+h.c.)
\label{Higpot} \nonumber\\ &+&
\frac{g^2+g^{'2}}{8}(|H_1|^2-|H_2|^2)^2 +
\frac{g^2}{2}|H_1^+H_2|^2,
\end{eqnarray}
where $m_1^2=m^2_{H_1}+\mu^2, m_2^2=m^2_{H_2}+\mu^2$. At the GUT
scale $m_1^2=m^2_2=m_0^2+\mu^2_0, \ m^2_3=-B\mu_0$. Notice that
the Higgs self-interaction coupling in eq.(\ref{Higpot}) is  fixed
and  defined by the gauge interactions as opposed to the SM.

The potential (\ref{Higpot}), in accordance with supersymmetry, is
positive definite and stable. It has no nontrivial minimum
different from zero.  Indeed, let us write the minimization
condition for  the potential (\ref{Higpot})
\begin{eqnarray} \frac 12\frac{\delta V}{\delta
H_1}&=&m_1^2v_1 -m^2_3v_2+ \frac{g^2+g'^2}4(v_1^2-v_2^2)v_1=0,
\label{min1}
\\ \frac 12\frac{\delta V}{\delta H_2}&=&m_2^2v_2-m^2_3v_1+ \frac{
g^2+g'^2}4(v_1^2-v_2^2)v_2=0, \label{min2}
\end{eqnarray}
where we have  introduced the notation $$<H_1>\equiv v_1= v
\cos\beta , \ \  <H_2>\equiv v_2= v \sin\beta, \ \ v^2=
v_1^2+v_2^2,\ \  \tan\beta \equiv \frac{v_2}{v_1}.$$ Solution of
eqs.(\ref{min1}),(\ref{min2}) can be expressed in terms of $v^2$
and $\sin 2\beta$
\begin{equation} v^2=\frac{\displaystyle  4(m^2_1-m^2_2\tan^2 \beta
)}{\displaystyle (g^2+ g'^2)(\tan^2\beta -1)},\ \ \ \sin2\beta
=\frac{\displaystyle 2m^2_3}{\displaystyle m^2_1+m^2_2}.
\label{min}
\end{equation}
One can easily see from eq.(\ref{min}) that if
$m_1^2=m_2^2=m_0^2+\mu_0^2$, $v^2$ happens to be negative, i.e.
the minimum does not exist.  In fact, real positive solutions to
eqs.(\ref{min1}),(\ref{min2}) exist only if the following
conditions are satisfied:
\begin{equation}
m_1^2+m_2^2 > 2 m_3^2, \ \ \ \  m_1^2m_2^2 < m_3^4 , \label{cond}
\end{equation}
which is not the case at the GUT scale. This means that
spontaneous breaking of the $SU(2)$  gauge invariance, which is
needed in the SM to give masses for all the particles, does not
take place in the MSSM.

This strong statement is valid, however, only at the GUT scale.
Indeed, going down with energy, the parameters of the potential
(\ref{Higpot}) are renormalized.  They become the "running"
parameters with the energy scale dependence given by the RG
equations. The running of the parameters leads to a remarkable
phenomenon known as  {\em radiative spontaneous symmetry breaking}
to be discussed below.

Provided conditions (\ref{cond}) are satisfied, the mass matrices
at the tree level are \\ CP-odd components
 $P_1$ and $P_2$ :
\begin{equation}
{\cal M}^{odd} = \left.\frac{\partial^2 V}{\partial P_i \partial
P_j} \right |_{H_i=v_i} = \left( \begin{array}{cc}  \tan\beta &1
\\1& \cot\beta \end{array}\right) m_3^2,
\end{equation}
 CP-even neutral components $S_1$ and $S_2$:
\begin{equation}
{\cal M}^{ev} = \left.\frac{\partial^2 V}{\partial S_i \partial
S_j} \right| = \left(
\begin{array}{cc} \tan\beta & -1
\\-1& \cot\beta \end{array}\right) m_3^2 +\left( \begin{array}{cc}
\cot\beta & -1 \\-1& \tan\beta \end{array}\right) M_Z
\frac{\sin2\beta}{2},
\end{equation}
Charged components $H^-$ and $H^+$:
\begin{equation}
{\cal M}^{ch} =\left.\frac{\partial^2 V}{\partial H^+_i
\partial H^-_j} \right|_{H_i=v_i} = \left( \begin{array}{cc}
\tan\beta &1 \\1& \cot\beta \end{array}\right)
 (m_3^2+M_W\cos\beta\sin\beta).
\end{equation} Diagonalizing the mass matrices, one gets the mass
eigenstates:
 $$\begin{array}{l} \left\{
\begin{array}{lllr} G^0 \ &=& -\cos\beta P_1+\sin \beta P_2 , & \ \ \
Goldstone \ boson \ \to Z_0, \\ A \ &=& \sin\beta P_1+\cos \beta P_2 , & \
\ \ \ \ \ \ \ Neutral \ CP=-1 \ Higgs, \end{array}\right.\\  \\ \left\{
\begin{array}{lllr} G^+ &=& -\cos\beta (H^-_1)^*+\sin \beta H^+_2 , &\ \
Goldstone \ boson \ \to W^+, \\ H^+ &=& \sin\beta (H^-_1)^*+\cos \beta
H^+_2 , &\ \ \ Charged \ Higgs, \end{array}\right.\\ \\ \left\{
\begin{array}{lllr} h \ &=& -\sin\alpha S_1+\cos\alpha S_2 , & \ \ \ \ \ \
\ SM \ Higgs \ boson \ CP=1, \\ H \ &=& \cos\alpha S_1+\sin\alpha S_2 , &
\ \ \ \ \ \ \ Extra \ heavy \ Higgs \ boson , \end{array}\right.
\end{array}$$ where the mixing angle $\alpha $ is
given by equation: \ \ $ \tan 2\alpha = \tan 2\beta
\left(\frac{m^2_A+M^2_Z}{m^2_A-M^2_Z}\right).$

The physical Higgs bosons acquire the following masses
\cite{MSSM}:
 \begin{eqnarray} \mbox{CP-odd
neutral Higgs} \ \ A: && \ \ \ \ \ \ m^2_A = m^2_1+m^2_2,
\nonumber \\ \mbox{Charge Higgses} \ \ H^{\pm}: &&
 \ \ \ \ \ m^2_{H^{\pm}}=m^2_A+M^2_W ,
 \end{eqnarray}
CP-even neutral Higgses \ \ H, h:
\begin{equation} m^2_{H,h}=
\frac{1}{2}\left[m^2_A+M^2_Z \pm
\sqrt{(m^2_A+M_Z^2)^2-4m^2_AM_Z^2\cos^22\beta}\right],
\end{equation}
where, as usual,
 $$ M^2_W=\frac{g^2}{2}v^2, \ \
M^2_Z=\frac{g^2+g'^2}{2}v^2 .$$ This leads to the once celebrated
SUSY mass relations
\begin{equation}\begin{array}{c} m_{H^{\pm}} \geq M_W, \ \
m_h \leq m_A \leq M_H, \\[0.2cm] m_h \leq M_Z |\cos 2\beta| \leq
M_Z,\ \  m_h^2+m_H^2=m_A^2+M_Z^2.\end{array}\label{bound}
\end{equation}

Thus, the lightest neutral Higgs boson happens to be lighter than
the $Z$ boson, which clearly distinguishes it from the SM one.
Though we do not know the mass of the Higgs boson in the SM, there
are several indirect constraints leading to the lower boundary of
$m_h^{SM} \geq 135 $ GeV~\cite{bound}. After including the
radiative corrections,  the mass of the lightest Higgs boson in
the MSSM, $m_h$, however, increases. We consider it in more detail
below.

\subsubsection{Renormalization group analysis}

To calculate the low energy values of the soft terms, we use the
corresponding RG equations. The one-loop RG equations for the
rigid MSSM couplings are \cite{Ibanez}
\begin{eqnarray}
\frac{d\tilde{\alpha}_i}{dt}&=&b_i \tilde{\alpha}_i^2, \ \ \ \ t\equiv \log
Q^2/M_{GUT}^2 \nonumber\\ \frac{dY_U}{dt} & = &
-Y_L\left(\frac{16}{3}\tilde{\alpha}_3 + 3\tilde{\alpha}_2 +
\frac{13}{15}\tilde{\alpha}_1-6Y_U-Y_D\right) , \nonumber \\ \frac{dY_D}{dt}
& = & -Y_D\left(\frac{16}{3}\tilde{\alpha}_3 + 3\tilde{\alpha}_2 +
\frac{7}{15}\tilde{\alpha}_1-Y_U-6Y_D-Y_L\right), \nonumber
\\ \frac{dY_L}{dt} & = &- Y_L\left( 3\tilde{\alpha}_2 +
  \frac{9}{5}\tilde{\alpha}_1-3Y_D-4Y_L\right), \label{eq}
\end{eqnarray}
where we use the notation $\tilde \alpha = \alpha/4\pi= g^2/16\pi^2,\ Y
=y^2/16\pi^2$.

 For the soft terms one finds
\begin{eqnarray}
\frac{dM_i}{dt} & = & b_i \tilde{\alpha}_iM_i . \nonumber\\ \frac{dA_U}{dt}
& = & \frac{16}{3}\tilde{\alpha}_3 M_3 + 3\tilde{\alpha}_2 M_2 +
\frac{13}{15}\tilde{\alpha}_1 M_1+6Y_UA_U+Y_DA_D, \nonumber
\\
 \frac{dA_D}{dt} & = & \frac{16}{3}\tilde{\alpha}_3 M_3 +
3\tilde{\alpha}_2 M_2 + \frac{7}{15}\tilde{\alpha}_1
M_1+6Y_DA_D+Y_UA_U+Y_LA_L, \nonumber \\ \frac{dA_L}{dt} & = &
3\tilde{\alpha}_2 M_2 + \frac{9}{5}\tilde{\alpha}_1 M_1+3Y_DA_D+4Y_LA_L,
\nonumber
\\ \frac{dB}{dt} & = & 3\tilde{\alpha}_2 M_2 +
\frac{3}{5}\tilde{\alpha}_1 M_1+3Y_UA_U+3Y_DA_D+Y_LA_L. \nonumber
\\
 \frac{d\tilde{m}^2_Q}{dt} & =&- \left[
(\frac{16}{3}\tilde{\alpha}_3M^2_3 + 3\tilde{\alpha}_2M^2_2 +
\frac{1}{15}\tilde{\alpha}_1M^2_1)\right. \nonumber\\
 &-&\left.
Y_U(\tilde{m}^2_Q+\tilde{m}^2_U+m^2_{H_2}+A^2_U)
  -Y_D(\tilde{m}^2_Q+\tilde{m}^2_D+m^2_{H_1}+A^2_D)\right], \nonumber\\
\frac{d\tilde{m}^2_U}{dt} & = &- \left[(\frac{16}{3}\tilde{\alpha}_3M^2_3
+\frac{16}{15}\tilde{\alpha}_1M^2_1)
-2Y_U(\tilde{m}^2_Q+\tilde{m}^2_U+m^2_{H_2}+A^2_U)\right] , \nonumber\\
\frac{d\tilde{m}^2_D}{dt} & =
 &- \left[(\frac{16}{3}\tilde{\alpha}_3M^2_3+ \frac{4}{15}\tilde{\alpha}_1M^2_1)
-2Y_D(\tilde{m}^2_Q+\tilde{m}^2_D+m^2_{H_1}+A^2_D)\right],
\nonumber\\
  \frac{d\tilde{m}^2_L}{dt} & = &
-\left[3(
 \tilde{\alpha}_2M^2_2 + \frac{1}{5}\tilde{\alpha}_1M^2_1)
-Y_L(\tilde{m}^2_L+\tilde{m}^2_E+m^2_{H_1}+A^2_L)\right],
\nonumber
\\ \frac{d\tilde{m}^2_E}{dt} & = &-\left[ (
 \frac{12}{5}\tilde{\alpha}_1M^2_1)-2Y_L(
\tilde{m}^2_L+\tilde{m}^2_E+m^2_{H_1}+A^2_L)\right], \nonumber
\\
\frac{d\mu^2}{dt}&=&-\mu^2\left[3(\tilde{\alpha}_2+
\frac{1}{5}\tilde{\alpha}_1)-(3Y_U+3Y_D+Y_L)\right],\label{eq2}\\
\frac{dm^2_{H_1}}{dt} & = & -\left[3(\tilde{\alpha}_2M^2_2
+\frac{1}{5}\tilde{\alpha}_1M^2_1)
-3Y_D(\tilde{m}^2_Q+\tilde{m}^2_D+m^2_{H_1}+A^2_D)\right.
\nonumber
\\ &&\left. - Y_L(\tilde{m}^2_L+\tilde{m}^2_E+m^2_{H_1}+A^2_L)\right] ,
\nonumber\\ \frac{dm^2_{H_2}}{dt} & = &-\left[ 3(\tilde{\alpha}_2
M^2_2 +\frac{1}{5}\tilde{\alpha}_1M^2_1)
-3Y_U(\tilde{m}^2_Q+\tilde{m}^2_U+m^2_{H_2}+A^2_U)\right].
\nonumber
\end{eqnarray}

\begin{figure}[hbt]
\begin{center}
\leavevmode \epsfxsize=6cm \epsffile{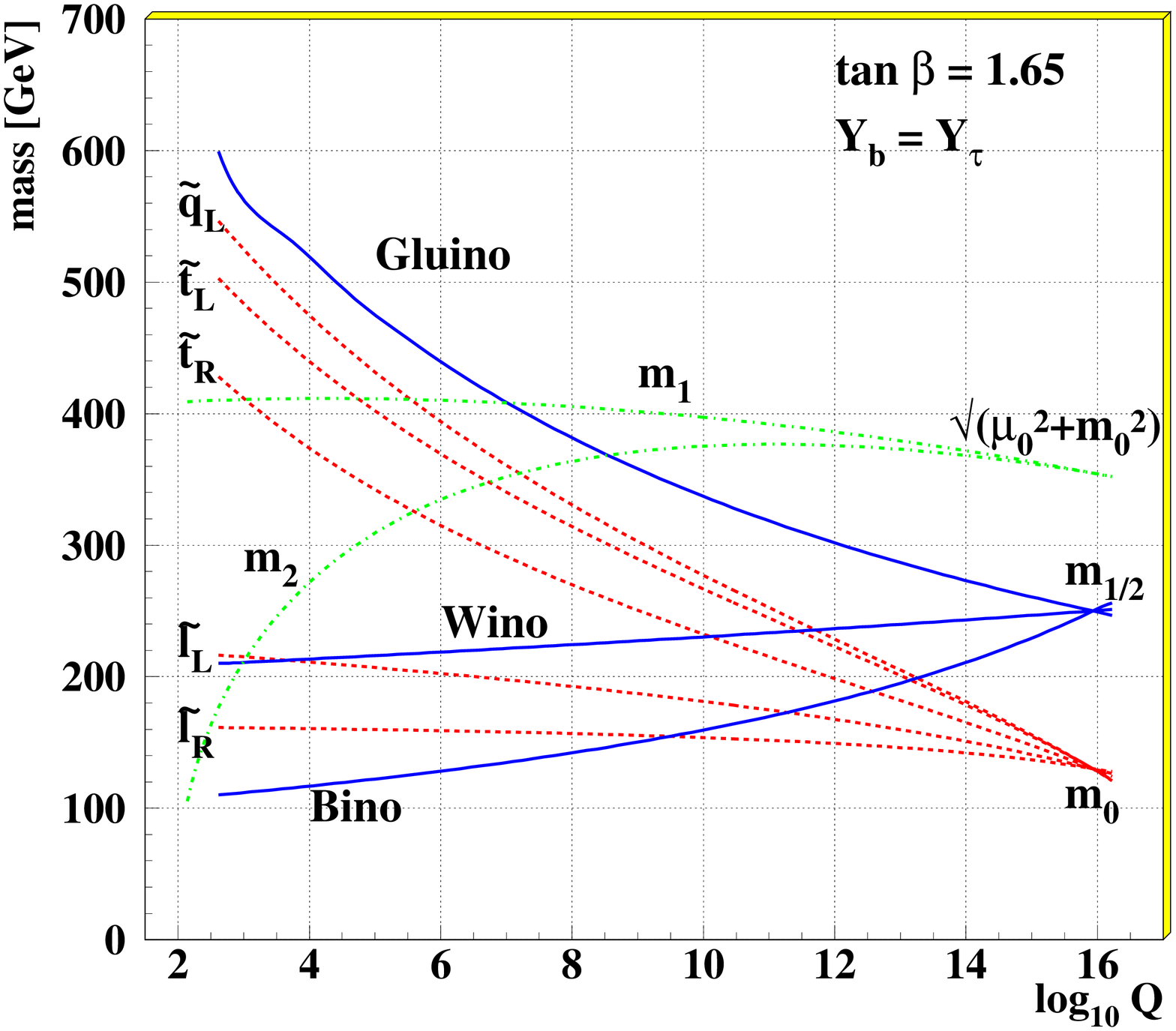}
 \leavevmode
\epsfxsize=6cm \epsffile{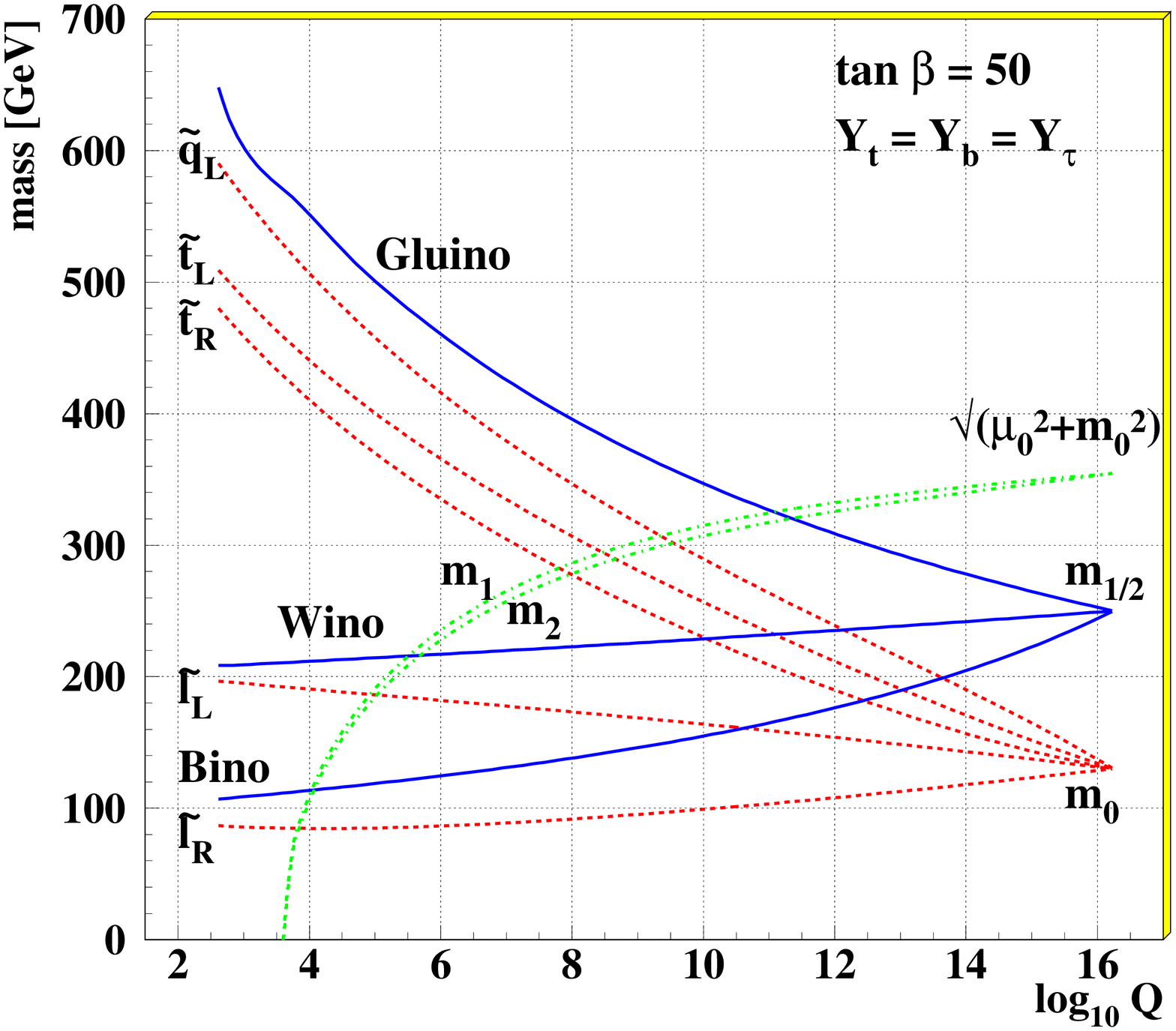}\vspace{-0.5cm}
\caption{An example of evolution of sparticle masses and soft
supersymmetry breaking parameters $m_1^2=m^2_{H_1}+\mu^2$ and
$m_2^2=m^2_{H_2}+\mu^2$ for low (left) and high (right) values of
$\tan\beta$ } \label{16}
\end{center}
\end{figure}

Having all the RG equations, one can now find the RG flow for the
soft terms.  Taking the initial values of the soft masses at the
GUT scale in the interval between $10^2\div 10^3$ GeV consistent
with the SUSY scale suggested by unification of the gauge
couplings~\cite{ABF,Lectures} leads to the  RG flow of the soft
terms shown in Fig.\ref{16}.~\cite{spectrum,BEK}
%
%

One should mention the following general features common to any
choice of initial conditions:

 i) The gaugino masses follow the running of the gauge couplings
 and split at low energies. The gluino mass is running faster
 than the others  and is usually the heaviest due to the strong interaction.

 ii) The squark and slepton masses also split at low energies, the
 stops (and sbottoms) being the lightest due to relatively big Yukawa couplings
 of the third generation.

  iii) The Higgs masses (or at least one of them) are running down
 very quickly and may even become negative.

\begin{figure}[htb]
\begin{center}
 \leavevmode
  \epsfxsize=5.5cm
 \epsffile{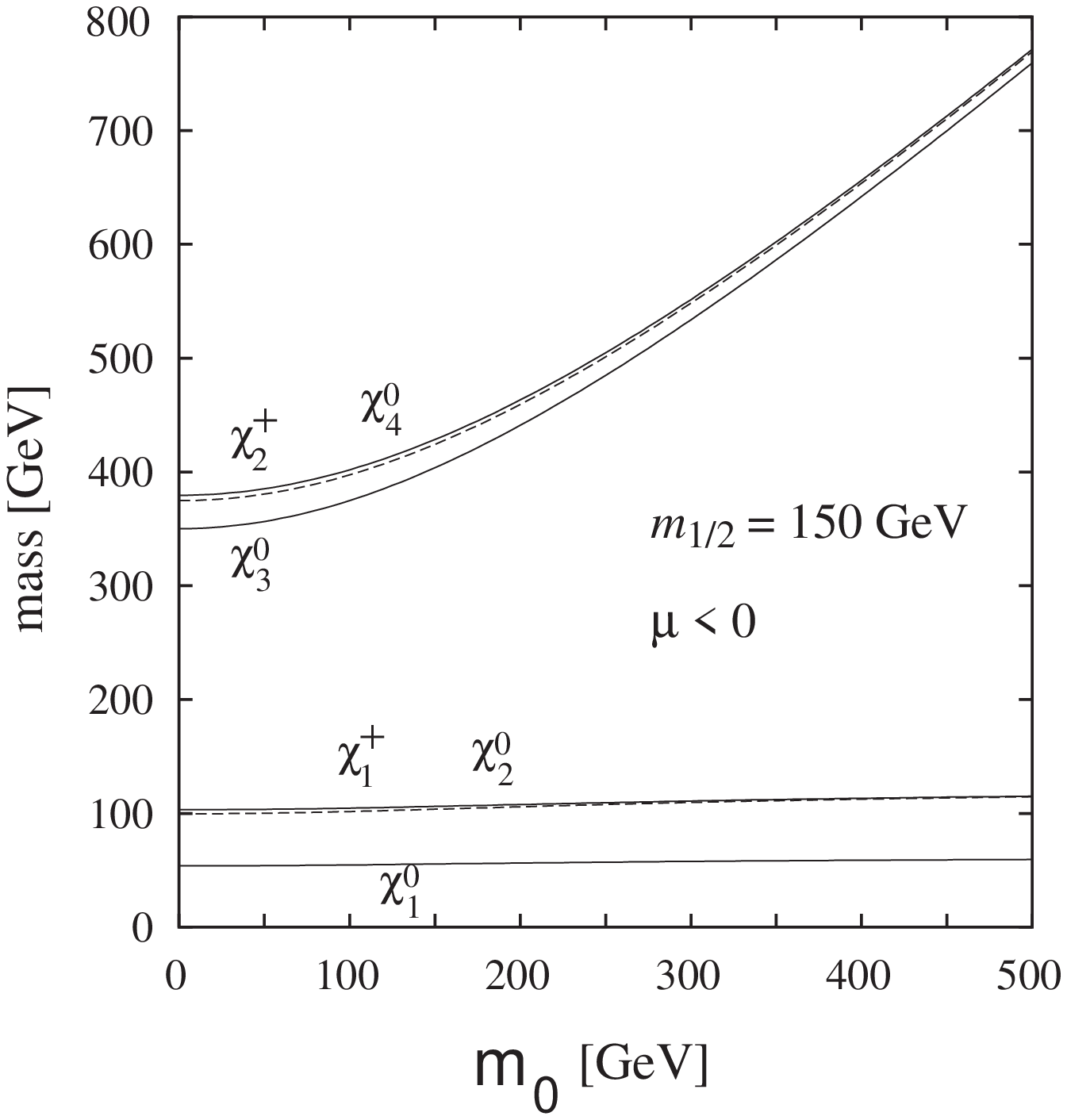}
 \epsfxsize=5.9cm\epsffile{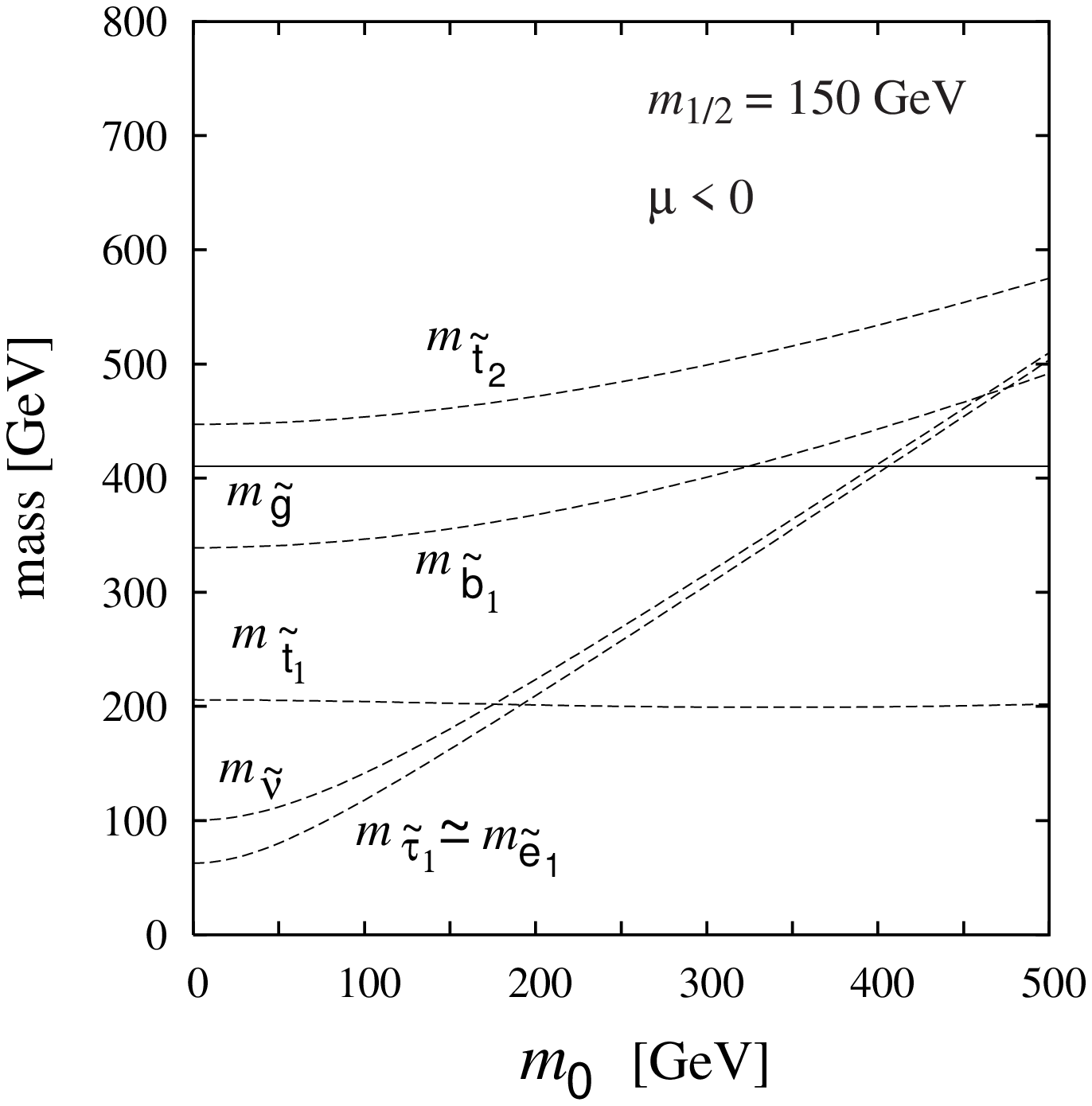}
\end{center}\vspace{-0.5cm}
\caption{The masses of sparticles as functions of the initial
value $m_0$} \label{fig:barger}
\end{figure}
Typical dependence of the mass spectra on the initial conditions
($m_0$) is also shown in Fig.\ref{fig:barger} ~\cite{Barger}. For
a given value of $m_{1/2}$ the masses of the lightest particles
are practically independent of $m_0$, while the heavier ones
increase with it monotonically. One can see that the lightest
neutralinos and charginos as well as the stop squark may be rather
light.


The running of the Higgs masses leads to the phenomenon known as
{\em radiative electroweak symmetry breaking}. Indeed, one can see
in Fig.\ref{16}  that $m_2^2$ (or both $m_1^2$ and $m_2^2$)
decreases when going down from the GUT scale to the $M_Z$ scale
and  can even become negative.  As a result, at some value of
$Q^2$  the conditions (\ref{cond}) are satisfied, so that the
nontrivial minimum appears. This triggers spontaneous breaking of
the $SU(2)$ gauge invariance. The vacuum expectations of the Higgs
fields acquire nonzero values and provide masses to quarks,
leptons and $SU(2)$ gauge bosons, and additional masses to their
superpartners.

In this way one also obtains the explanation of why the two scales
are so much different. Due to the logarithmic running of the
parameters, one needs  a long "running time" to get $m_2^2$ (or
both $m_1^2$ and $m_2^2$) to be negative when starting from a
positive value of the order of $M_{SUSY}\sim 10^2 \div 10^3$ GeV
at the GUT scale.

\subsection{Constrained MSSM}

\subsubsection{Parameter space of the MSSM}

The Minimal Supersymmetric Standard Model has the following free
parameters:
\begin{enumerate}
\item[i)] three gauge couplings $\alpha_i$;\\[-0.6cm]
\item[ii)] three matrices of the Yukawa couplings $y^i_{ab}$, where
$i = L,U,D$;\\[-0.6cm]
\item[iii)] the Higgs field mixing parameter  $\mu $;\\[-0.6cm]
\item[iv)] the soft supersymmetry breaking parameters.
\end{enumerate}
Compared to
the SM there is an additional Higgs mixing parameter, but the
Higgs self-coupling, which is arbitrary in the SM,  is fixed by
supersymmetry. The main uncertainty comes from the unknown soft
terms.

With the universality hypothesis one is left with the following
set of 5 free parameters defining the mass scales
 $$ \mu, \ m_0, \ m_{1/2}, \ A \ \mbox{and}\
 B \leftrightarrow \tan\beta = \frac{v_2}{v_1}. $$
While choosing parameters and making predictions, one has two
possible ways to proceed:

 i) take the low-energy parameters like superparticle masses
  $\tilde{m}_{t1},\tilde{m}_{t2}, m_A$,
$\tan\beta$, mixings $X_{stop},\mu$, etc. as input  and calculate
cross-sections as functions of these parameters.

 ii) take the high-energy parameters like the above mentioned 5
soft parameters as input, run the RG equations and find the
low-energy values. Now the calculations can be carried out in
terms of the initial parameters. The experimental constraints are
sufficient to determine these parameters, albeit with large
uncertainties.

Both the ways are used in a phenomenological analysis. We show
below how it works in practice.

\subsubsection{The choice of constraints}

When subjecting constraints on the MSSM, perhaps, the most
remarkable fact is that all of them can be fulfilled
simultaneously. In our analysis we impose the following
constraints on the parameter space of the MSSM:

$\bullet$ Gauge coupling constant unification; \\ This is one of
the most  restrictive constraints which we have discussed
in~\cite{Lectures}. It fixes the scale of SUSY breaking of an
order of 1 TeV.

$\bullet$ $M_Z$ from electroweak symmetry breaking;\\
Radiative EW symmetry breaking  (see eq.(\ref{min})) defines the
mass of the Z-boson
\begin{equation}
\label{defmz} M_Z^2=2\frac{m_1^2-m_2^2\tan^2\beta}{\tan^2\beta-1}
.
\end{equation} This condition determines the value of $\mu^2$ for given
values of $m_0$ and $m_{1/2}$.

\begin{figure}[ht]
\begin{center}
 \leavevmode
  \epsfxsize=7.5cm \epsfysize=7.5cm
 \epsffile{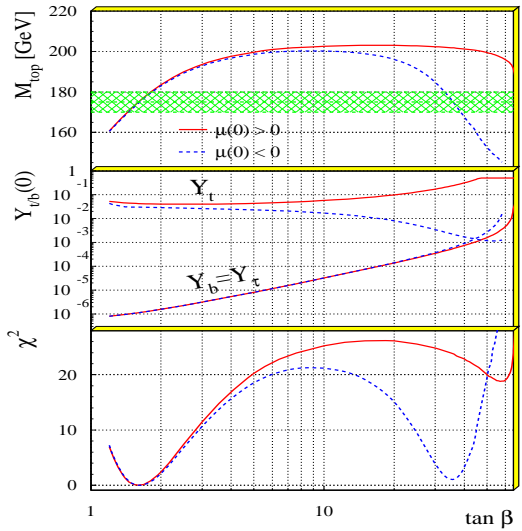}
\end{center}\vspace{-1cm}
\caption{The upper part shows the top quark mass as a function of
$\tan\beta$ for $m_0= 600$ GeV, $m_{1/2}=400$ GeV. The middle part
shows the corresponding values of the Yukawa couplings at the GUT
scale and the lower part of the $\chi^2$ values.} \label{fig:tb}
\end{figure}

$\bullet$ Yukawa coupling constant unification;\\ The masses of
top, bottom and $\tau$ can be obtained from the low energy values
of the running Yukawa couplings via
 \begin{equation} m_t=y_t\
v\sin\beta, \ \ m_b=y_b\ v\cos\beta, \ \ m_\tau=y_\tau \
v\cos\beta . \label{yuk}
\end{equation}
They can be translated to the pole masses with account taken of
the radiative corrections. The requirement of bottom-tau Yukawa
coupling unification, i.e. equality of $b$-quark and $\tau$-lepton
masses at the GUT scale, strongly restricts the possible solutions
in $m_t$ versus $\tan\beta$ plane~\cite{bbog} as it can be seen
from Fig.\ref{fig:tb}.  Releasing this constraint one may use
intermediate values of $\tan\beta$.

$\bullet$ Precision measurement of decay rates;\\
We take the branching ratio $BR(b\to s \gamma)$ which has been
measured by the CLEO~\cite{CLEO} collaboration and later by
ALEPH~\cite{ALBSG} and yields the world average of $BR(b\to s
\gamma)=(3.14\pm0.48)\cdot 10^{-4}$. The Standard Model
contribution to this process gives slightly lower result, thus
leaving window for SUSY. This requirement imposes severe
restrictions on the parameter space, especially for the case of
large $\tan\beta$.

$\bullet$ Anomalous magnetic moment of muon.\\
Recent measurement of the anomalous magnetic moment indicates
small deviation from the SM of the order of 2 $\sigma$. The
deficiency may be easily filled with SUSY contribution, which is
proportional to $\mu$. This requires positive sign of $\mu$ that
kills a half of the parameter space of the MSSM~\cite{Anom}.

$\bullet$ Experimental lower limits on SUSY masses; \\ SUSY
particles have not been found so far and from the searches at LEP
one knows  the lower limit on the charged lepton and chargino
masses of about  half of the centre of mass energy~\cite{LEPSUSY}.
The lower limit on the neutralino masses  is smaller. There exist
also limits on squark and gluino masses from the hadron
colliders~\cite{TEVSUSY}.  These limits restrict the  minimal
values for the SUSY mass parameters.

$\bullet$ Dark Matter constraint; \\
In the early Universe all particles were produced abundantly and
were in thermal equilibrium through annihilation and production
processes. The time evolution of the number density of the
particles is given by Boltzmann equation and can be evaluated
knowing the thermally averaged total annihilation cross section.
The WIMP's fall out of the equilibrium at a temperature of about
$m_\chi /22$~\cite{equi} and a relic cosmic abundance remains. At
the present, the mass density in units of the critical density is
given by~\cite{abun}
\begin{equation}\label{abun}
  \Omega_\chi h^2=\frac{m_\chi n_\chi}{\rho_c}\approx \left(
  \frac{2\cdot 10^{-27} cm^3s^{-1}}{<\sigma v>}\right).
\end{equation}
The amount of neutralinos should not be too big to overclose the
Universe and, at the same time, it should be enough to produce the
right amount of the Dark matter. Taking the value of the Hubble
parameter to be $h_0>0.4$ one finds that the contribution of each
relic particle species $\chi$ has to obey conservative bounds
$\Omega_\chi h^2_0 \sim 0.1\div 0.3$. This serves as a very severe
bound on SUSY parameters~\cite{relictst}. We show below that
recent very precise data from WMAP collaboration, which measured
thermal fluctuations of Cosmic Microwave Background radiation and
restricted the amount of the Dark matter in the Universe up to
$23\pm4\%$, leave a very narrow band of allowed region in
parameter space.

Having in mind the above mentioned constraints one can  find the
most probable region of the parameter space by minimizing the
$\chi^2$ function~\cite{BEK}. We first choose the value of the
Higgs mixing parameter $\mu$ from the requirement of radiative EW
symmetry breaking, then we take the values of $\tan\beta$ from the
requirement of Yukawa coupling unification (see Fig.\ref{fig:tb}).
One finds two possible solutions: low $\tan\beta$ solution
corresponding to $\tan\beta \approx 1.7$ and high $\tan\beta$
solution corresponding to $\tan\beta \approx 30\div 60$.

The low $\tan\beta$ solution which predicts light particles was
very popular at the time of LEP. Unfortunately, LEP  found neither
superpartners nor the light Higgs boson. A modern limit on the
value of $\tan\beta$ comes from non-observation of the Higgs boson
up to 114 GeV and  restricts $\tan\beta \ge 3\div 4$. Moreover,
since most of the SUSY radiative corrections are proportional to
$\tan\beta$, large values of $\tan\beta$ are preferable.

What is left are the values of the soft parameters $A,\ m_0$ and
$m_{1/2}$. However, the role of the trilinear coupling $A$ is not
essential. In what follows, we consider the plane $m_0,m_{1/2}$
and find the allowed region in this plane. Each point at this
plane corresponds to a fixed set of parameters and allows one to
calculate the spectrum, the cross-sections, etc.

\begin{figure}[htb]
\begin{center}\vspace*{-0.5cm}
  \leavevmode
  \epsfxsize=5.3cm
 \epsffile{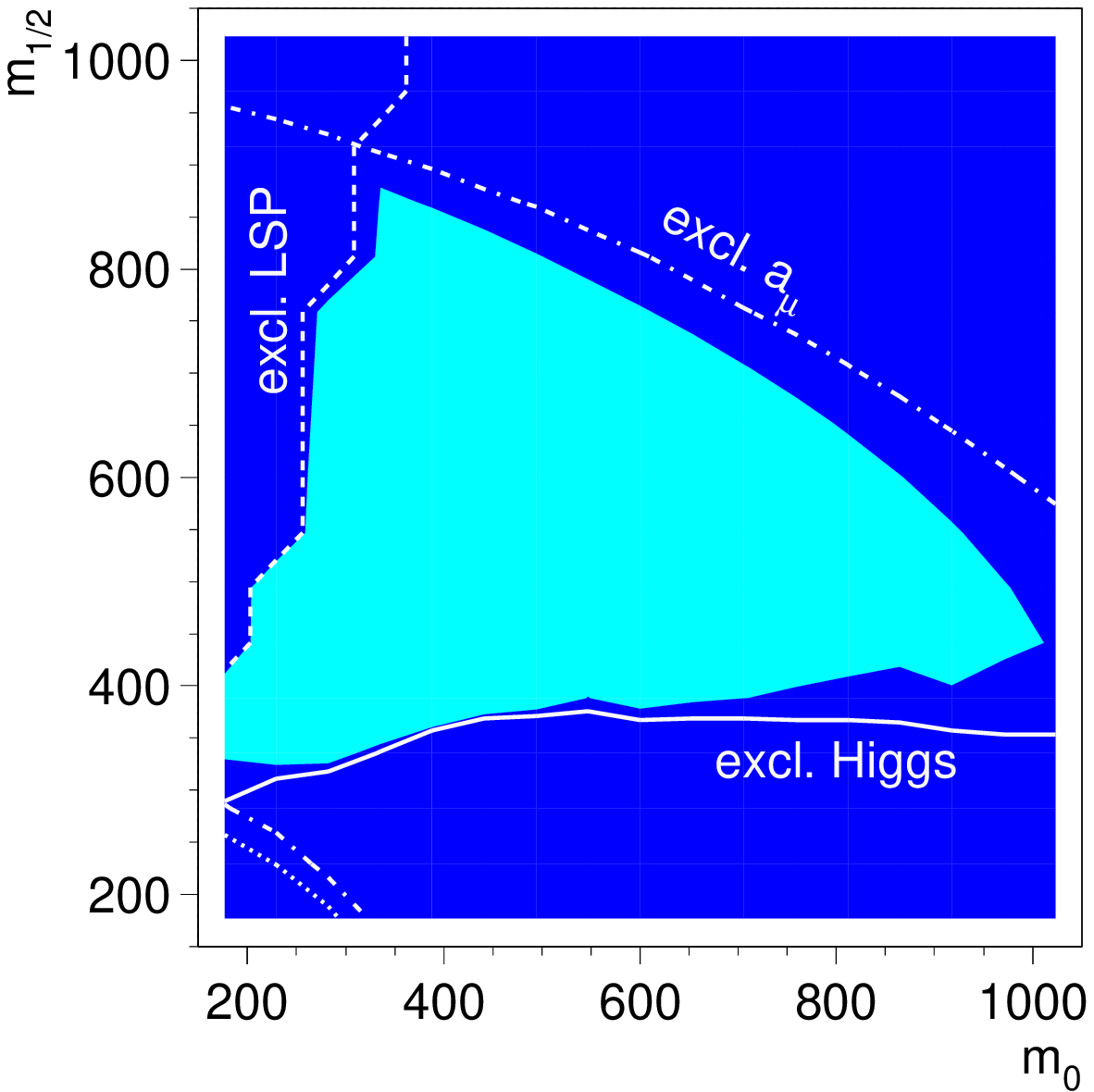}
 \epsfxsize=5.3cm
 \epsffile{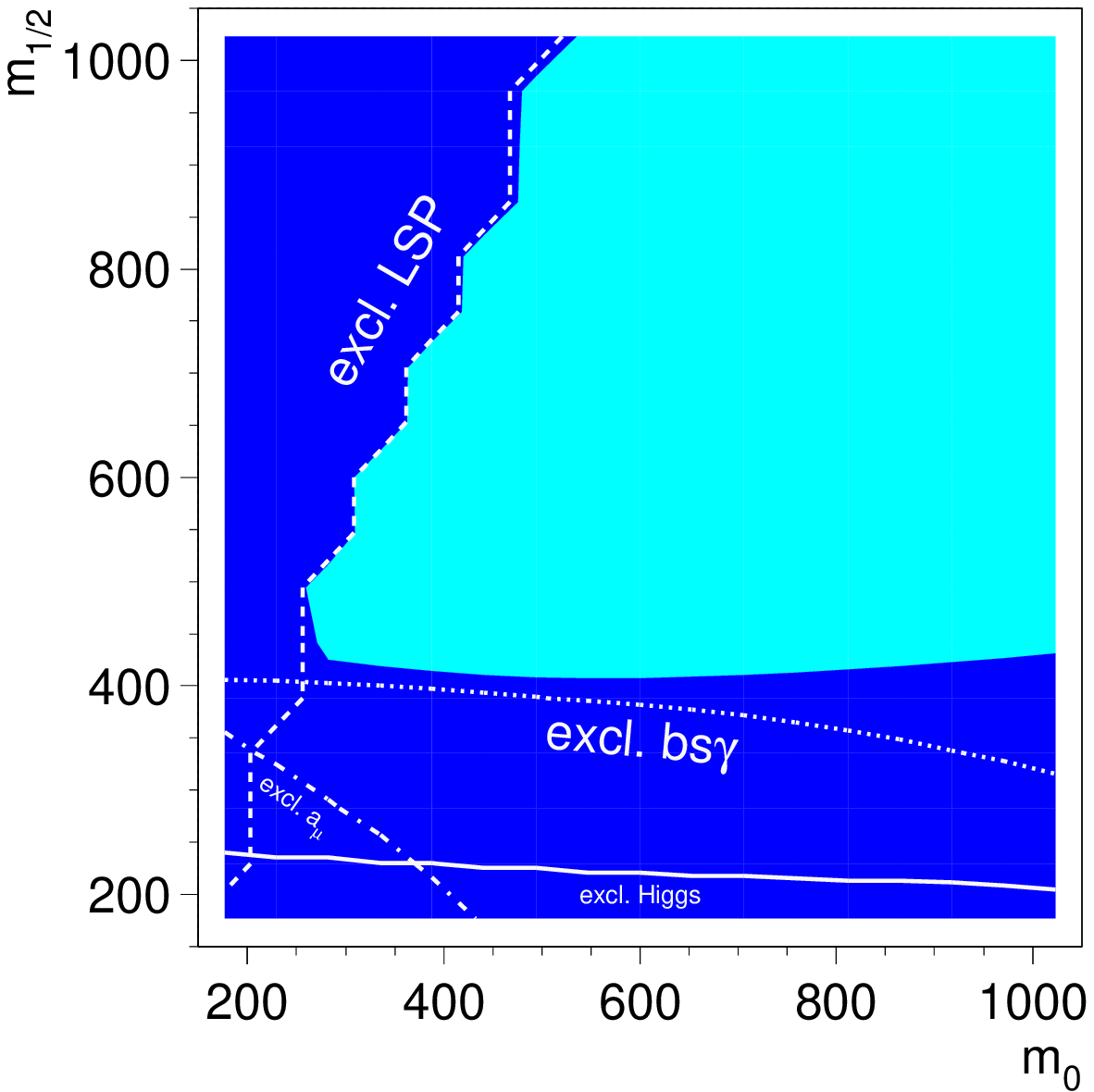}\vspace{-0.6cm}
 \end{center}
\hspace*{5cm} $\tan\beta=35$ \hspace{3cm} $\tan\beta=50$
 \caption{\label{chi}
Allowed regions of parameter space for high $\tan\beta$ scenario.}
\end{figure}

\begin{figure}[ht]
\begin{center}\vspace*{-0.1cm}
 \epsfxsize=5.3cm
 \epsffile{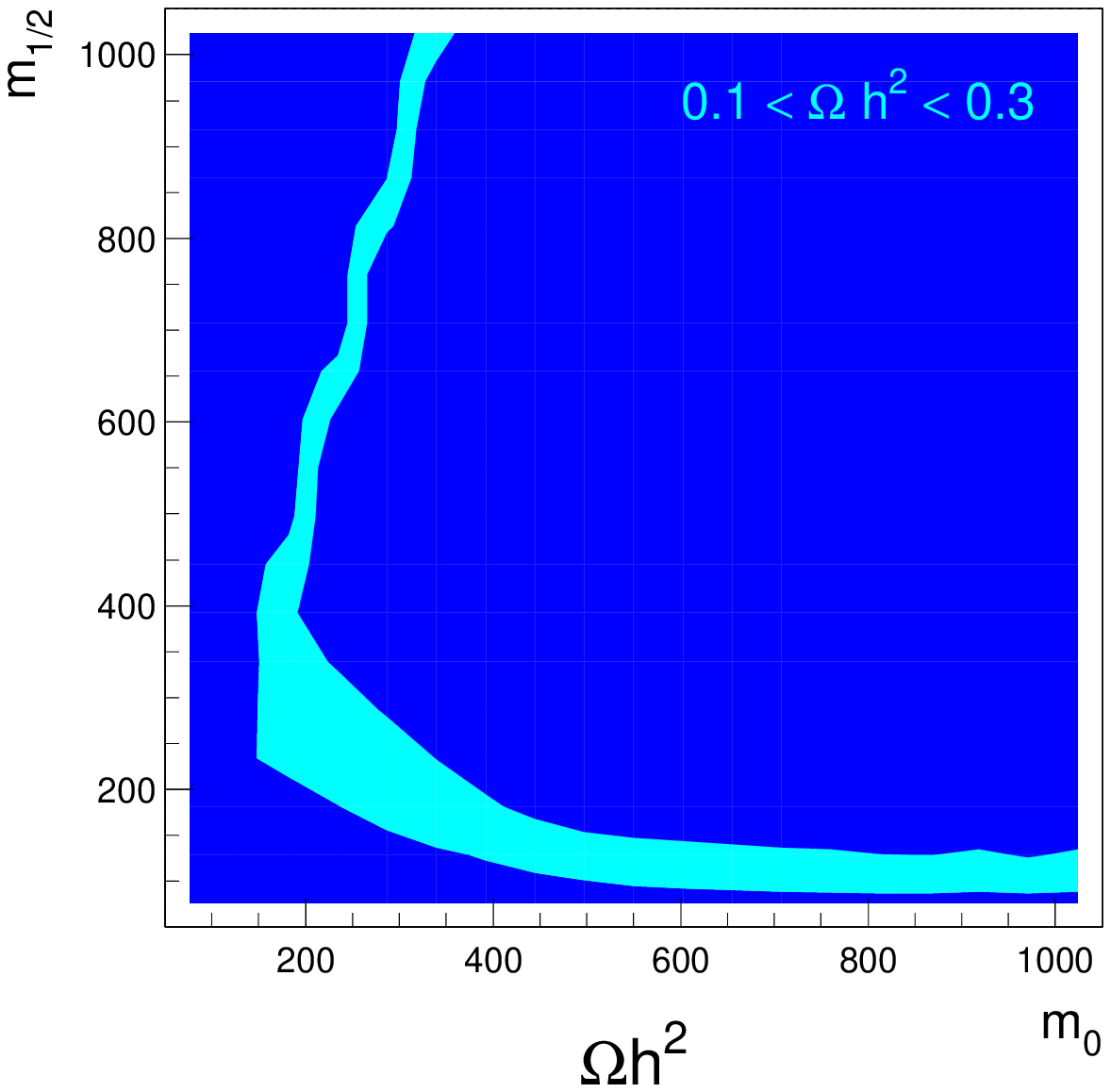}
 \epsfxsize=5.3cm
\epsffile{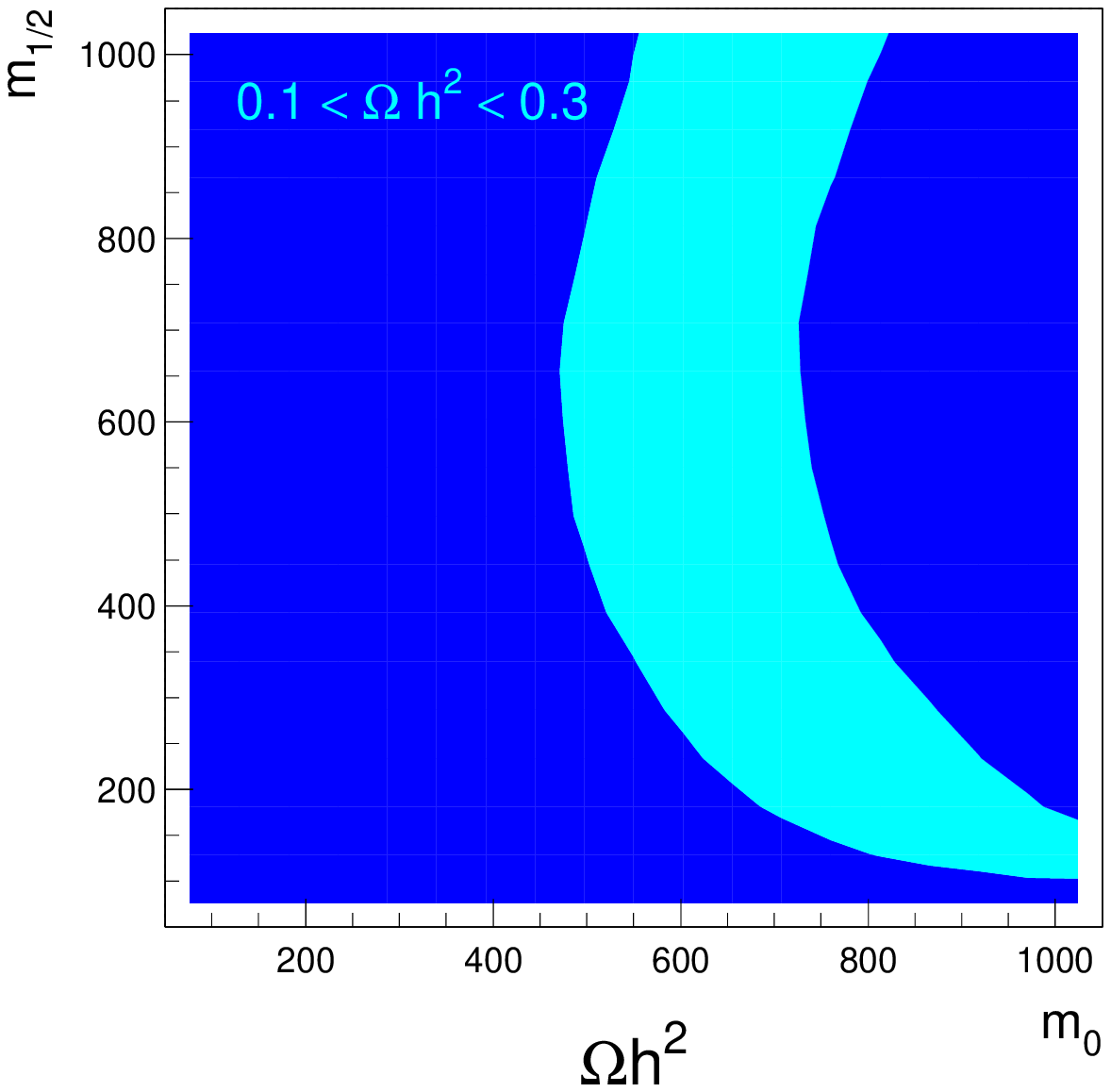}
 \end{center}\vspace{-0.3cm}
\hspace*{5cm} $\tan\beta=35$ \hspace{3cm} $\tan\beta=50$
 \caption{\label{chi2}
  Restrictions on parameter space from the requirement of the
   right amount of the Dark Matter.}
\end{figure}

We present the allowed regions of the parameter space for two
typical values of $\tan\beta$  in Fig.\ref{chi}. This plot
demonstrates the role of various constraints in the $\chi^2$
function. The contours enclose domains by the particular
constraints used in the analysis \cite{BS}. Fig.\ref{chi2} shows
the role of the Dark Matter constraint (before WMAP).

\begin{figure}[htb]
\begin{center}\vspace*{-0.3cm}

 \epsfxsize=5.5cm
 \epsffile{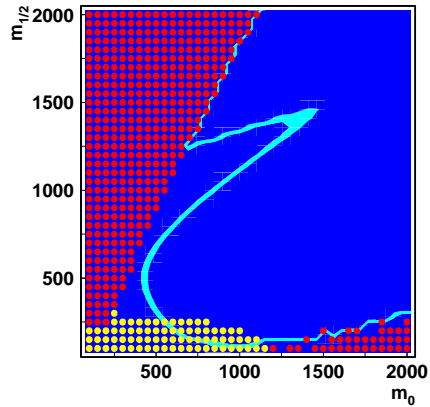}
 \end{center}\vspace{-0.3cm}
 \caption{\label{WMAP} The light shaded (blue) line is the region
 allowed by WMAP for $\tan\beta=51$, $\mu>0$ and $A_0=0.5m_0$. The
 excluded regions where the stau would be the LSP (red, left upper corner)
  or EWSB fails (red, right corner) or the Higgs boson is too light
  (yellow, left low corner) are indicated by
the dots.}
\end{figure}

Taking into account the WMAP data puts even more severe constrains
due to very high precision of measurement. This constraint is
shown in Fig.\ref{WMAP} as a narrow light blue
band~\cite{deB,Wband}. We have taken here a twice wider region in
the $m_0,m_{1/2}$ plane, thus allowing higher masses of
superpartners.

\clearpage
\subsubsection{The mass spectrum of superpartners}

When the parameter set is fixed, one can calculate the mass
spectrum of superpartners. Below we show  the typical mass
spectrum~\cite{deB} for large $\tan\beta$ solution. At the top we
show  the fitted values of the soft SUSY breaking parameters and
at the bottom of the table on can see also the values of some
observables used as constraints and fitted by the choice of
parameters.

\begin{table}[htb]
\renewcommand{\arraystretch}{1.10}
\begin{center}
\normalsize \vspace{0.5cm}
\begin{tabular}{|c|c|c|}
\hline
 Parameter & Value & Value \\
\hline
 $m_0$ & 500 GeV & 500 GeV \\
 $m_{1/2}$ & 350 GeV & 550 GeV \\
 $\tan\beta$ & 50 & 52 \\
 $A_0$ & $0\cdot m_0$ & $0\cdot m_0$ \\
 sign $\mu$ & + & + \\
\hline
 Particle & Mass [GeV] & Mass [GeV] \\  \hline
 & & \\
 $\tilde \chi^0_{1,2,3,4}$ & 144, 259, 447, 462 & 230, 420, 665, 676 \\
 $\tilde \chi^\pm_{1,2}$ & 259, 463 & 420, 677 \\
 $\tilde g$ & 803 & 1231 \\
 $\tilde t_{1,2}$ & 618, 769 & 899, 1066 \\
 $\tilde b_{1,2}$ & 679, 758 & 960, 1052 \\
 $\tilde u_{1,2}$ & 864, 889 & 1185, 1230 \\
 $\tilde d_{1,2}$ & 862, 892 & 1180, 1233 \\
 $\tilde \tau_{1,2}$ & 318, 496 & 289, 565 \\
 $\tilde l_{1,2}$ & 519, 556 & 544, 626 \\
 $\tilde \nu_\tau$ & 475 & 538 \\
 $\tilde \nu$ & 550 & 621 \\
 $h$ & 115.0 & 118.0 \\
 $H$ & 375.4 & 493.6  \\
 $A$ & 375.7 & 496.0 \\
 $H^\pm$ & 386.7 & 505.0 \\
\hline
 Observable & Value & Value \\
\hline
 $Br(b\to X_s\gamma)$ & $1.63 \cdot 10^{-4}$ & $2.68 \cdot
 10^{-4}$\\
 $Br(B_s\to \mu^+\mu^-)$ & $\sim 5\cdot 10^{-8}$ & $\sim 2\cdot
 10^{-8}$ \\
 $a_\mu$ & $363\cdot 10^{-11}$ & $224\cdot 10^{-11}$ \\
 $\Omega h^2$ & 0.117 & 0.113 \\
 \hline
\end{tabular} \end{center}
\caption[]{\label{t2} mSUGRA parameters and the corresponding mass
spectrum of superpartners.}
\end{table}

 Notice the
low values of the masses of the lightest Higgs boson and of the
lightest neutralino which is the LSP. They happen to be very
sensitive to the value of $\tan\beta$ and increase with increase
of the latter.

\subsubsection{Experimental signatures at $e^+e^-$ colliders}
Experiments are finally beginning to push into a significant
region of supersymmetry parameter space. We know the sparticles
and their couplings, but we do not know their masses and mixings.
Given the mass spectrum one can calculate the cross-sections and
consider the possibilities of observing  new particles at modern
accelerators. Otherwise, one can get  restrictions on unknown
parameters.

 We start with $e^+e^-$ colliders. In the leading order creation
of superpartners is given by the diagrams shown in
Fig.\ref{creation} above. For a given center of mass energy the
cross-sections depend on the mass of created particles and vanish
at the kinematic  boundary. Experimental signatures are defined by
the decay modes which vary with the mass spectrum. The main ones
are summarized below. A characteristic feature of all possible
signatures is the missing energy and transverse momenta, which is
a trade mark of a new physics.

$$\begin{array}{lll}
\mbox{\underline{Production}}&\mbox{\underline{Key Decay
Modes}}&~~ \mbox{\underline{Signatures}} \\ && \\ \bullet
~~~~~\tilde{l}_{L,R}\tilde{l}_{L,R}~~ &\tilde{l}^\pm_R \to l^\pm
\tilde{\chi}^0_i \searrow  \mbox{cascade}~~~~~ & \mbox{acomplanar
pair of}
\\ &  \tilde{l}^\pm_L \to l^\pm \tilde{\chi}^0_i \nearrow \mbox{decays} &
\mbox{charged leptons} + \Big/ \hspace{-0.3cm E_T} \\ \bullet
~~~~~\tilde{\nu}\tilde{\nu}& \tilde{\nu}\to l^\pm \tilde{\chi}^0_1
 & \Big/ \hspace{-0.3cm E_T}\\
\bullet  ~~~~~\tilde{\chi}^\pm_1\tilde{\chi}^\pm_1 &\tilde{\chi}^\pm_1 \to
 \tilde{\chi}^0_1 l^\pm \nu, \ \tilde{\chi}^0_1q \bar q' &
 \mbox{isol lept + 2 jets} + \Big/ \hspace{-0.3cm E_T} \\
&  \tilde{\chi}^\pm_1 \to \tilde{\chi}^0_2 f \bar f' &\mbox{pair
of acomplanar}\\
 &  \tilde{\chi}^\pm_1 \to l \tilde{\nu}_l \to
l\nu_l\tilde{\chi}^0_1 &\mbox{leptons} + \Big/ \hspace{-0.3cm E_T}
\\ & \tilde{\chi}^\pm_1 \to \nu_l \tilde{l} \to \nu_l
l\tilde{\chi}^0_1&\mbox{4 jets} + \Big/ \hspace{-0.3cm E_T}
\\
 \bullet  ~~~~~\tilde{\chi}^0_i\tilde{\chi}^0_j &
\tilde{\chi}^0_i \to \tilde{\chi}^0_1 X, \tilde{\chi}^0_j  \to
\tilde{\chi}^0_1 X' & X=\nu_l \bar \nu_l \ \mbox{invisible} \\ &&
~~= \gamma,2l,\mbox{2 jets} \\ && 2l + \Big/ \hspace{-0.3cm E_T},
l+2j + \Big/ \hspace{-0.3cm E_T} \\
\bullet ~~~~~\tilde{t}_i\tilde{t}_j & \tilde{t}_1 \to c
\tilde{\chi}^0_1 & \mbox{2 jets}+ \Big/ \hspace{-0.3cm E_T} \\ &
\tilde{t}_1 \to b \tilde{\chi}^\pm_1 \to b f\bar
f'\tilde{\chi}^0_1 & \mbox{2 b jets}+ \mbox{2 leptons} + \Big/
\hspace{-0.3cm E_T} \\ && \mbox{2 b jets}+\mbox{lepton} + \Big/
\hspace{-0.3cm E_T}
\\
\bullet  ~~~~~\tilde{b}_i\tilde{b}_j & \tilde{b}_i \to b
\tilde{\chi}^0_1 & \mbox{2 b jets}+ \Big/ \hspace{-0.3cm E_T} \\ &
\tilde{b}_i \to b \tilde{\chi}^0_2 \to b f\bar f'\tilde{\chi}^0_1
& \mbox{2 b jets}+ \mbox{2 leptons} + \Big/ \hspace{-0.3cm E_T} \\
&& \mbox{2 b jets}+ \mbox{2 jets}+ \Big/ \hspace{-0.3cm E_T}
 \end{array} $$

\begin{figure}[ht]
 \leavevmode
  \epsfxsize=7.5cm
 \epsffile{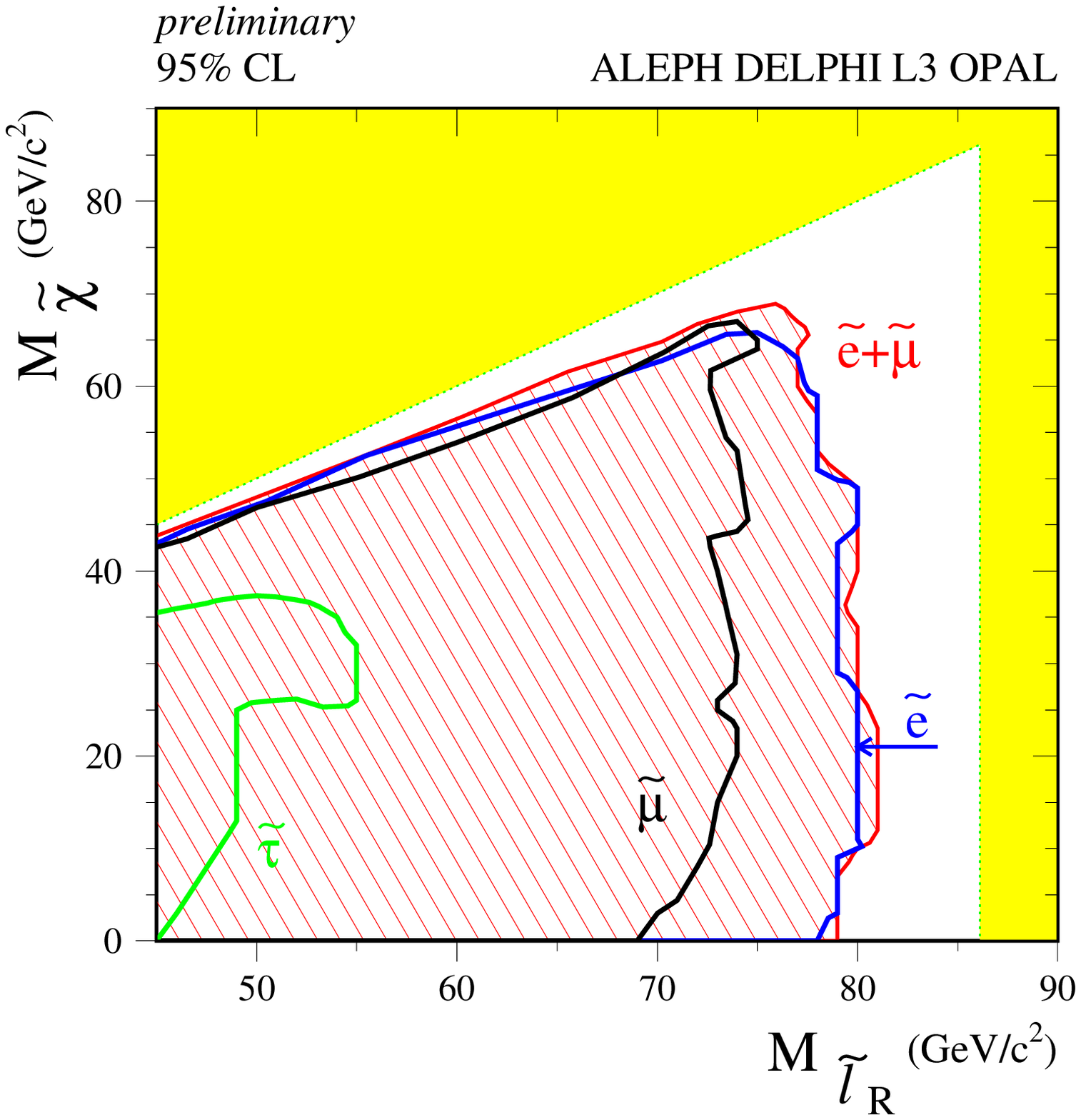}
 \vspace{-9.3cm}

 \epsfxsize=6.0cm
 \hspace*{8cm}
 \epsffile{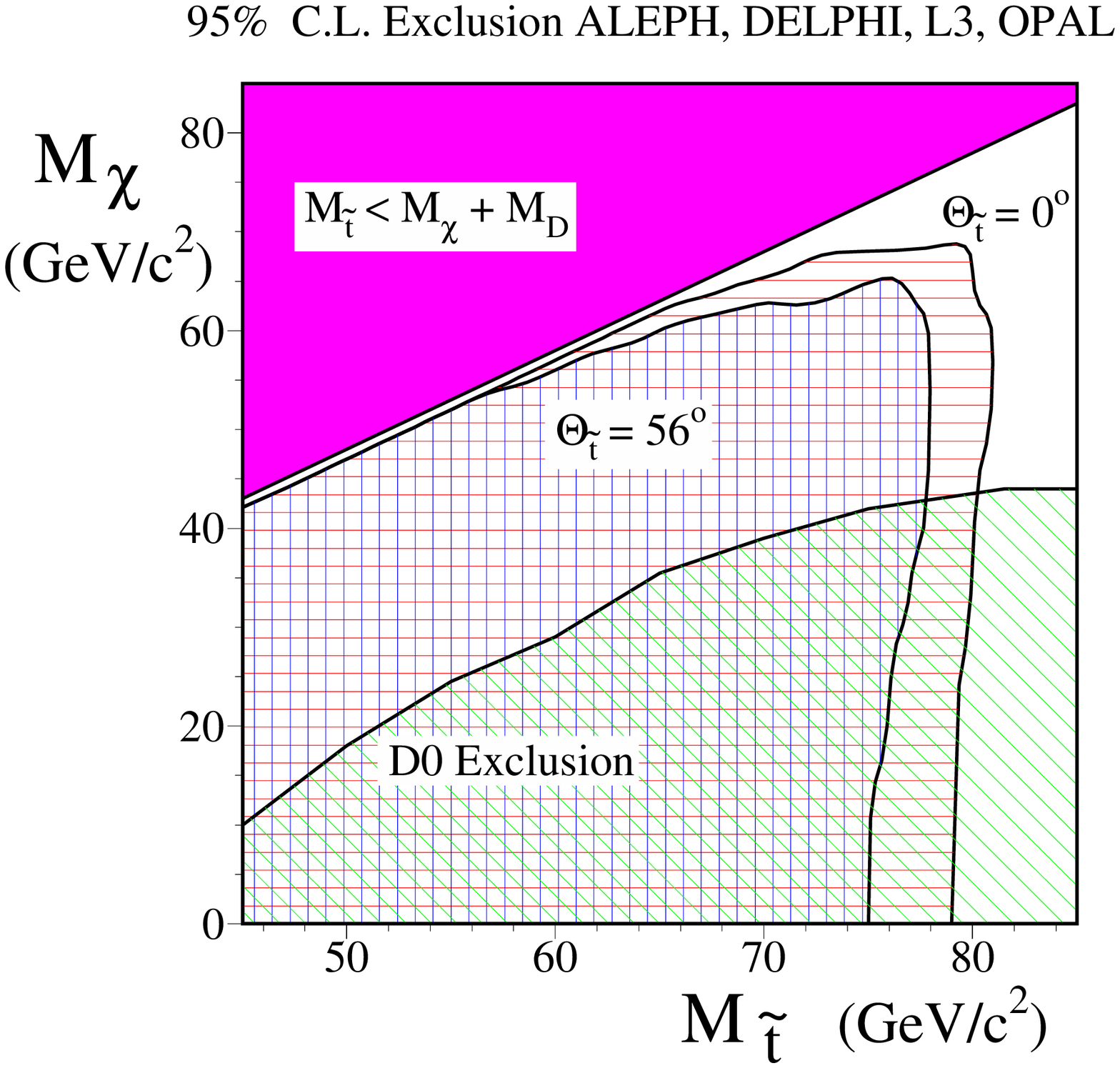}
 \vspace{-0.3cm}\caption{The excluded region in chargino-slepton and
 chargino-stop mass plane } \label{fig:slepton}
 \end{figure}

 Numerous attempts to find  superpartners at LEP II
gave no positive result thus imposing the lower bounds on their
masses~\cite{LEPSUSY}.
 They are shown on the parameter plane in Fig.\ref{fig:slepton}.

 Typical LEP II limits on the masses of
superpartners are
\begin{equation}\begin{array}{lll}
 m_{\chi^0_1} > 40 \ GeV & m_{\tilde e_{L,R}}>105\ GeV &
 m_{\tilde t}> 90\ GeV \\
 m_{\chi^\pm_1} > 100 \ GeV & m_{\tilde \mu_{L,R}}>100\ GeV &
 m_{\tilde b}> 80\ GeV \\
 & m_{\tilde \tau_{L,R}}>80\ GeV &
\end{array} \nonumber\end{equation}

\subsubsection{Experimental signatures at hadron colliders}

\begin{figure}[htb]
\begin{center}
\leavevmode \hspace*{-0.5cm}
  \epsfxsize=11cm
 \epsffile{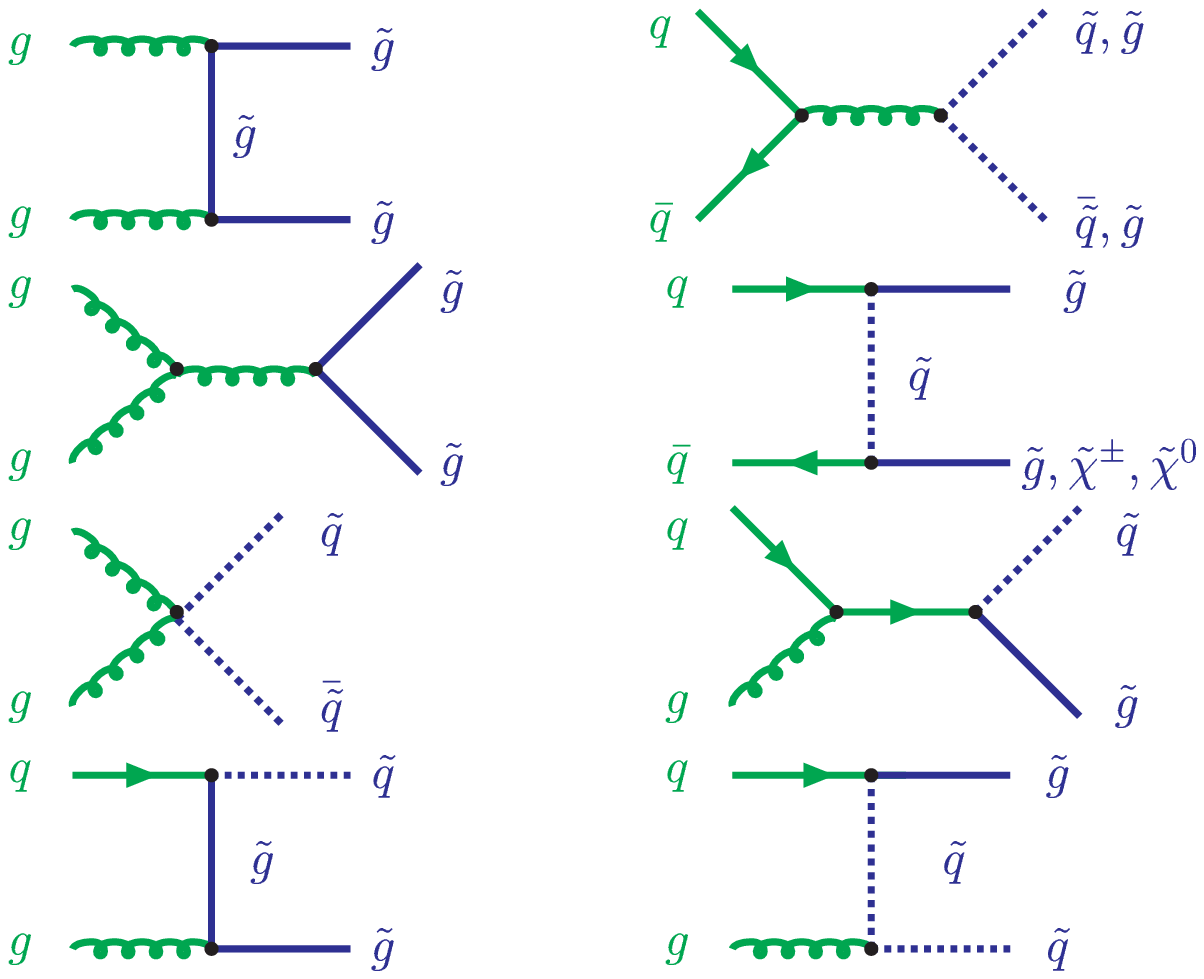}
\end{center}
\caption{Gluon fusion, $q\bar q$ scattering,
quark-gluon scattering}\label{fusion}
 \end{figure}

Experimental signatures at hadron colliders are similar to those
at $e^+e^-$ machines; however, here one has much wider
possibilities. Besides the usual annihilation channel identical to
$e^+e^-$ one with the obvious replacement of electrons by quarks
(see Fig.\ref{creation}), one has numerous processes of gluon
fusion, quark-antiquark and quark-gluon scattering (see
Fig.\ref{fusion}).

 Experimental
SUSY signatures at the Tevatron (and LHC) are
 $$\begin{array}{lll}
\mbox{\underline{Production}}&\mbox{\underline{Key Decay
Modes}}&~~ \mbox{\underline{Signatures}} \\ && \\ \bullet
~~~~~\tilde{g}\tilde{g}, \tilde{q}\tilde{q},
\tilde{g}\tilde{q}
&\left.\begin{array}{l} \tilde{g}
\to q\bar q \tilde{\chi}^0_1   \\
 ~~~~~ q\bar q' \tilde{\chi}^\pm_1  \\
 ~~~~~ g\tilde{\chi}^0_1 \end{array} \right\}
 m_{\tilde{q}}>m_{\tilde{g}} &
\begin{array}{c} \Big/ \hspace{-0.3cm E_T} + \mbox{multijets}\\
 (+\mbox{leptons}) \end{array} \\
& \left.\begin{array}{l}\tilde{q} \to q \tilde{\chi}^0_i \\
    \tilde{q} \to q' \tilde{\chi}^\pm_i \end{array} \right\}
  m_{\tilde{g}}>m_{\tilde{q}}&\\
 \bullet
~~~~~\tilde{\chi}^\pm_1\tilde{\chi}^0_2 &\tilde{\chi}^\pm_1 \to
 \tilde{\chi}^0_1 l^\pm \nu, \ \tilde{\chi}^0_2 \to
 \tilde{\chi}^0_1 ll &
 \mbox{Trilepton} + \Big/ \hspace{-0.3cm E_T} \\
&  \tilde{\chi}^\pm_1 \to \tilde{\chi}^0_1 q \bar q', \tilde{\chi}^0_2 \to
\tilde{\chi}^0_1 ll,&\mbox{Dilepton + jet} + \Big/ \hspace{-0.3cm E_T}\\
 \bullet  ~~~~~\tilde{\chi}^+_1\tilde{\chi}^-_1 &
\tilde{\chi}^+_1 \to l \tilde{\chi}^0_1 l^\pm \nu &
\mbox{Dilepton} + \Big/ \hspace{-0.3cm E_T} \\ \bullet
~~~~~\tilde{\chi}^0_i\tilde{\chi}^0_i & \tilde{\chi}^0_i \to
\tilde{\chi}^0_1 X, \tilde{\chi}^0_i  \to \tilde{\chi}^0_1 X' &
\Big/ \hspace{-0.3cm E_T} + \mbox{Dilept+(jets)+lept}\\
 \bullet  ~~~~~\tilde{t}_1\tilde{t}_1 & \tilde{t}_1 \to c
\tilde{\chi}^0_1 & \mbox{2 acollinear jets}+ \Big/ \hspace{-0.3cm
E_T} \\ & \tilde{t}_1 \to b \tilde{\chi}^\pm_1, \tilde{\chi}^\pm_1
\to \tilde{\chi}^0_1 q\bar q' &
 \mbox{single lepton} + \Big/ \hspace{-0.3cm E_T}  + b's\\
 &\tilde{t}_1 \to b \tilde{\chi}^\pm_1,\tilde{\chi}^\pm_1 \to \tilde{\chi}^0_1
l^\pm \nu ,
& \mbox{Dilepton} + \Big/ \hspace{-0.3cm E_T} + b's\\
\bullet
~~~~~\tilde{l}\tilde{l},\tilde{l}\tilde{\nu},\tilde{\nu}\tilde{\nu}
&
 \tilde{l}^\pm \to l\pm \tilde{\chi}^0_i,\tilde{l}^\pm \to \nu_l
 \tilde{\chi}^\pm_i& \mbox{Dilepton}+ \Big/ \hspace{-0.3cm E_T} \\
& \tilde{\nu} \to \nu \tilde{\chi}^0_1& \mbox{Single lept} +
\Big/ \hspace{-0.3cm E_T} +\mbox{jets} \\
&& \Big/ \hspace{-0.3cm E_T}
 \end{array}$$
Note again the characteristic missing energy and transverse
momenta events. Contrary to  $e^+e^-$ colliders, at hadron
machines the background is extremely rich and essential.

\subsubsection{The lightest superparticle}

One of the crucial questions is the properties of the lightest
superparticle. Different SUSY breaking scenarios lead to different
experimental signatures and different LSP.

$\bullet$ Gravity mediation

In this case, the LSP is the lightest neutralino
$\tilde{\chi}^0_1$, which is almost 90\% photino for a low
$\tan\beta$ solution and contains more higgsino admixture for high
$\tan\beta$. The usual signature for LSP is missing energy;
$\tilde{\chi}^0_1$ is stable and is the best candidate for the
cold dark matter in the Universe. Typical processes, where the LSP
is created, end up with jets + $\Big/ \hspace{-0.3cm}E_T$, or
leptons + $\Big/ \hspace{-0.3cm}E_T$, or both jest + leptons +
$\Big/ \hspace{-0.3cm}E_T$.

$\bullet$ Gauge mediation

In this case the LSP is the  gravitino $\tilde G$ which also leads
to missing energy. The actual question here is what the NLSP, the
next-to-lightest particle, is. There are two possibilities:

i) $\tilde{\chi}^0_1$ is the NLSP. Then the decay modes are: \
 $\tilde{\chi}^0_1 \to \gamma \tilde G, \ h \tilde G, \ Z \tilde
 G.$\
 As a result, one has two hard photons + $\Big/
 \hspace{-0.3cm}E_T$, or jets + $\Big/ \hspace{-0.3cm}E_T$.

ii) $\tilde l_R$ is the NLSP. Then the decay mode is  $\tilde l_R
\to \tau \tilde G$ and the signature is a charged lepton and the
missing energy.

$\bullet$ Anomaly mediation

In this case, one also has two possibilities:

i) $\tilde{\chi}^0_1$ is the LSP and wino-like. It is almost
degenerate with the NLSP.

ii) $\tilde \nu_L$ is the LSP. Then it appears in the decay of
chargino $\tilde \chi^+ \to \tilde \nu l$ and the signature is the
charged lepton and the missing energy.

$\bullet$ R-parity violation

In this case, the LSP  is no longer stable and decays into the SM
particles. It may be charged (or even colored) and may lead to
rare decays like neutrinoless double $\beta$-decay, etc.

\begin{figure}[ht]
 \leavevmode \hspace*{2cm}
  \epsfxsize=6.2cm \epsfysize=6cm
 \epsffile{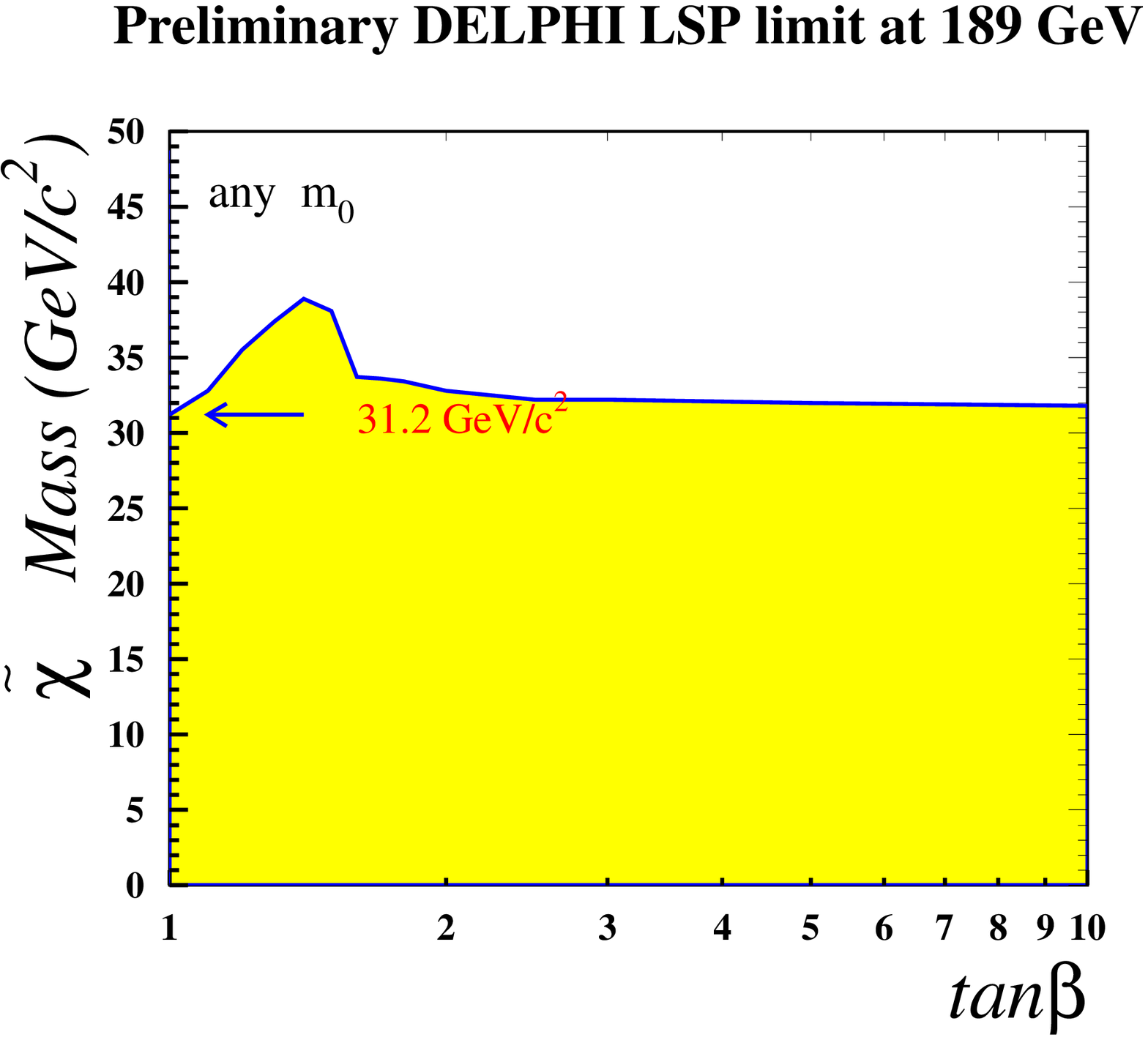}
\vspace{-5.7cm}

 \epsfxsize=6.2cm
  \hspace*{8cm}\epsffile{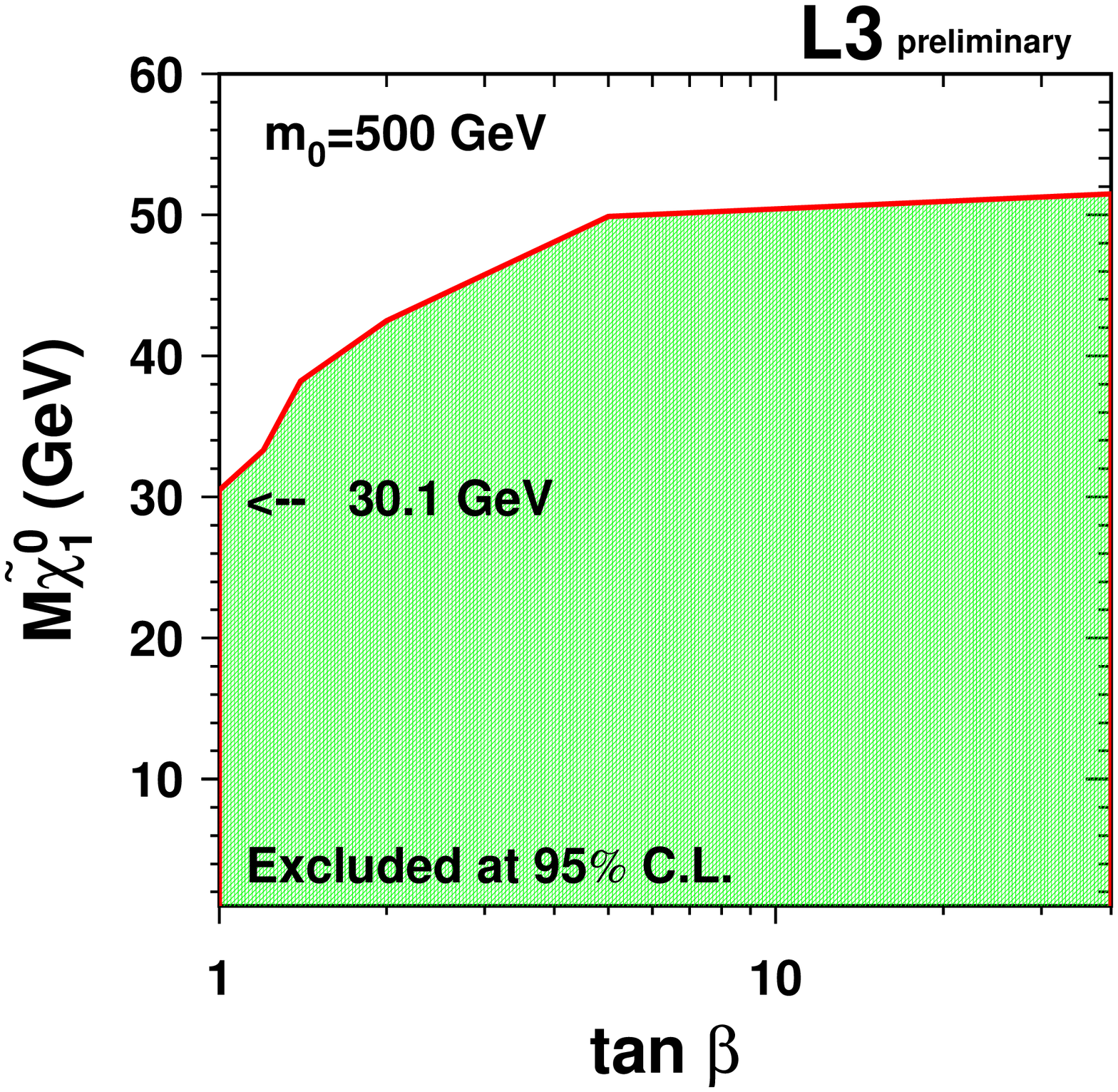}
\vspace{-0cm}
 \caption{The LSP mass limits within the MSSM ~\cite{LEPSUSY}} \label{fig:lsp}
 \end{figure}

Experimental limits on the LSP mass follow from non-observation of
the corresponding events. Modern lower limit from LEP is around 40
GeV (see Fig.\ref{fig:lsp}).

\subsection{The Higgs boson mass in the MSSM}

One of the hottest topics in the SM now is the search for the
Higgs boson. It  is also a window to a new physics. Below we
consider properties of the Higgs boson in the MSSM.

It has already been mentioned that in the MSSM the mass of the
lightest Higgs boson is predicted to be less than the $Z$-boson
mass. This is, however, the tree level result and the masses
acquire the radiative corrections. With account taken of the
one-loop radiative corrections the lightest Higgs mass is
\begin{equation}
m_h^2 \approx M_Z^2\cos^2 2\beta + \frac{3g^2 m_t^4}{16\pi^2M_W^2}
\log\frac{ \tilde{m}_{t_1}^2\tilde{m}_{t_2}^2}{m_t^4}. \label{rad}
\end{equation}
One finds that the one-loop correction is positive and increases
the mass value. Two loop  corrections have the opposite effect but
are smaller~\cite{feynhiggs}.

The Higgs mass depends mainly on the following parameters: the top
mass, the squark masses, the mixing in the stop sector and
$\tan\beta$. The maximum Higgs mass is obtained for large
$\tan\beta$, for a maximum  value of the top and squark masses and
a minimum value of the stop mixing.

\begin{figure}[ht]
\begin{center}
 \leavevmode
  \epsfxsize=8cm \epsfysize=6.3cm
 \epsffile{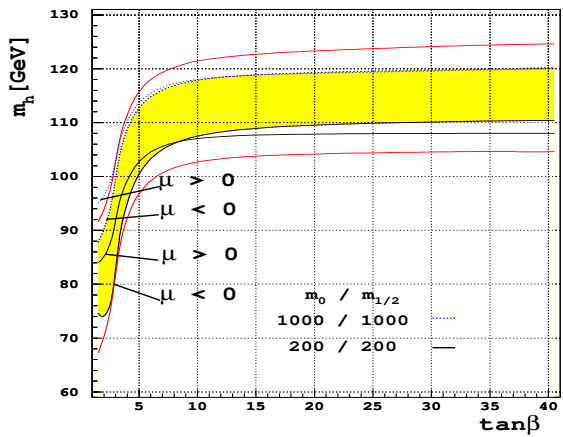}
\end{center}\vspace{-1cm}
\caption{The mass of the lightest Higgs boson in the MSSM as a
function of $\tan\beta$} \label{fig:mhtb}
\end{figure}

 The lightest
Higgs boson mass $m_h$ is shown as a function of $\tan\beta$ in
Fig.~\ref{fig:mhtb} \cite{BHGK}. The shaded band corresponds to
the uncertainty from the stop mass and stop mixing for $m_t=175$
GeV. The upper and lower lines correspond to $m_t$=170 and 180
GeV, respectively.

Combining all the uncertainties  the results  for the  Higgs mass
in the CMSSM can be summarized as follows:

$\bullet$ The low $\tan\beta$ scenario ($\tan\beta <3.3$) of the
CMSSM is excluded by the lower limit on the Higgs mass of  113.3
GeV~\cite{SM}.

$\bullet$ For the high $\tan\beta$ scenario the  Higgs mass is
found to be~\cite{BHGK}:
$$ m_h=115\pm3~ ({\rm stop m})~\pm1.5~({\rm stop mix})~\pm2~({\rm
theory})~\pm5~({\rm top m})~\rm GeV,$$
 where the errors  are the estimated standard deviations
around the central value.

One can see that the LEP came very close to SUSY  prediction  for
the Higgs mass and already ruled out low $\tan\beta$ scenario. The
next step is to be made by Tevatron. Unfortunately, the luminosity
of Tevatron at the moment is not enough to distinguish the Higgs
boson from the background. One have to wait till LHC starts
operation.

 However, these SUSY limits on the Higgs mass may not be so restricting if
non-minimal SUSY models are considered. Already in the
Next-to-Minimal model~\cite{NMSSM} the Higgs mass at low
$\tan\beta$ may be lifted by 20-30 GeV.  However, more
sophisticated models do not change the generic feature of SUSY
theories, the presence of the light Higgs boson.

\subsection{Perspectives of SUSY observation}

 With the LEP shut down, further attempts to discover supersymmetry
are connected with the Tevatron and LHC hadron colliders.

 \vspace{0.3cm}
\underline{Tevatron}
 \vspace{0.3cm}

The Fermilab Tevatron collider will define the high energy
frontier of particle physics while CERN's Large Hadron Collider is
being built. At the first stage (Run IIa), it has 2 fb$^{-1}$ of
integrated luminosity per experiment at $\sqrt{s}$ = 2 TeV. AT the
second stage (Run IIb), the luminosity is expected to reach 15
fb$^{-1}$ per experiment. However, since it is a hadron collider,
not the full energy goes into collision taken away by those quarks
in a proton that do not take part in the interaction. Any direct
search is kinematically limited to below 450 GeV.

\begin{figure}[ht]
 \leavevmode \hspace*{1cm}
  \epsfxsize=5.9cm
 \epsffile{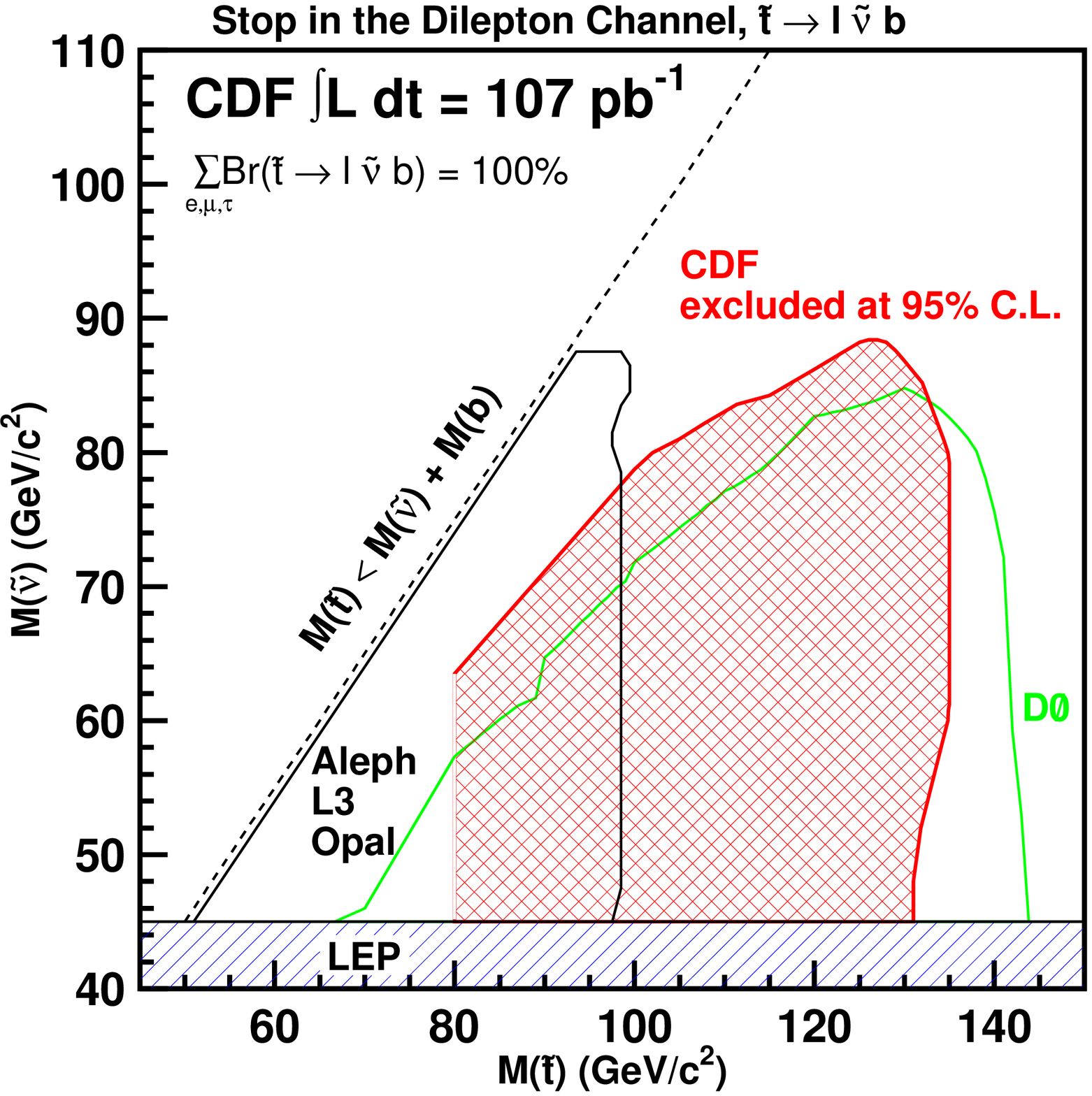}
 \vspace{-6.0cm}

 \epsfxsize=6cm 
 \hspace*{7cm}\epsffile{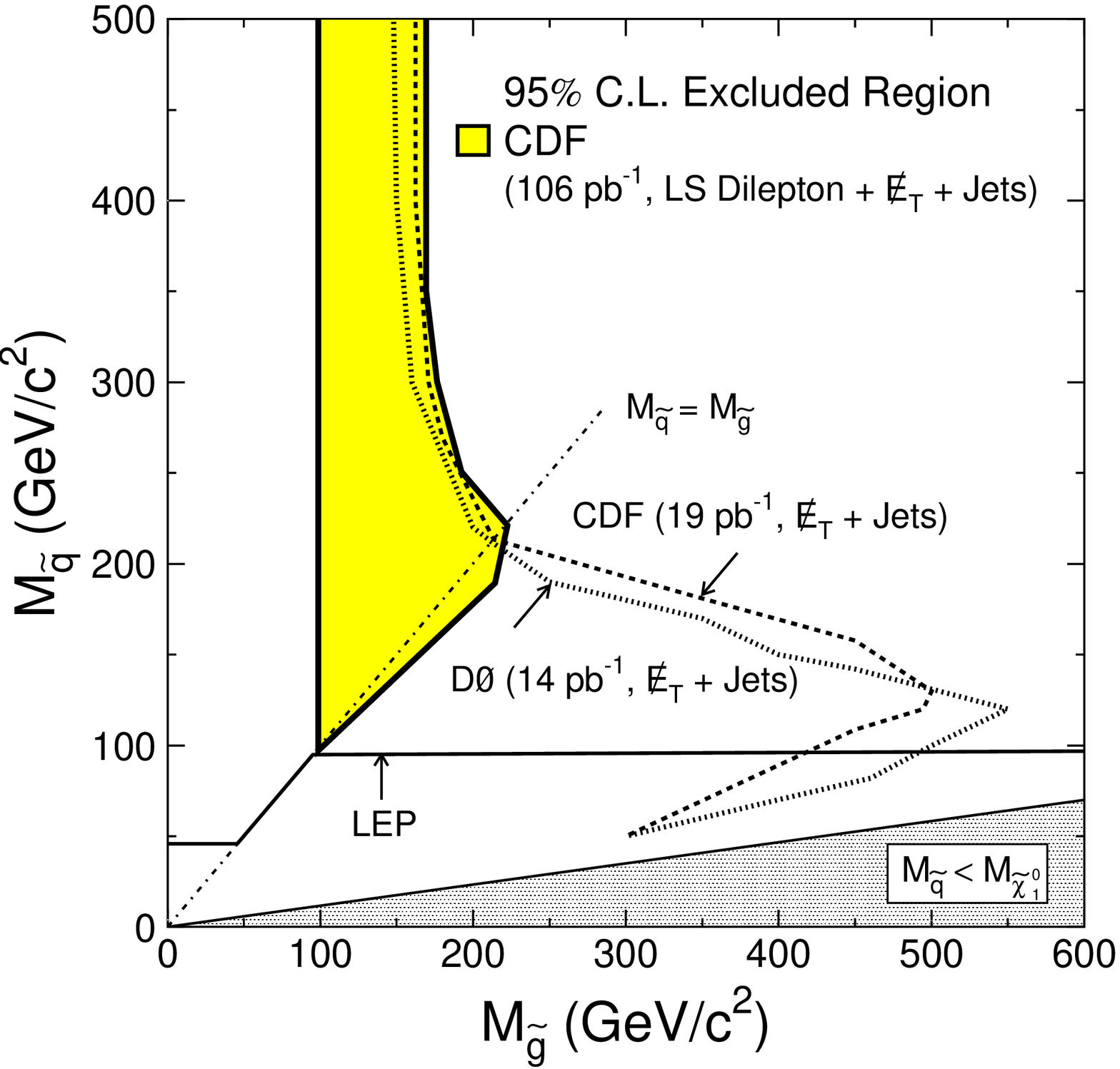}
 \caption{Exclusion plots for squarks and sneutrinos (left) and squarks and gluino
 (right) at Tevatron} \label{Tev}
\end{figure}

There exist numerous papers on SUSY searches at the
Tevatron~\cite{tevatron}-\cite{TevSUSY}. Modern exclusion areas
are shown in plots in Fig.\ref{Tev}~\cite{tevatron} for squarks,
sneutrinos, and gluino.
They impose the limits on the squark and gluino masses:\
$m_{\tilde q} \geq 300 \ GeV, \ \ m_{\tilde g} \geq 195\ GeV.$

 We show in Table \ref{T}~\cite{kamon} the discovery reach of
the Tevatron for squarks of the third generation for 20 fb$^{-1}$
of integrated luminosity.  They are still far from the expected
masses of superpartners predicted by the MSSM (see Table
\ref{t2}).

\begin{table}[htb]
\vspace{0.4cm}
\begin{center}
\begin{tabular}{ l l  l  l l }
\hline
 Decay &  Subsequent
    & Final State of
 & {Discovery Reach } \\
 ($Br$ = 100\%) & Decay  &
  $\tilde{b}_1 \bar{\tilde{b}}_1$ or $\tilde{t}_1 \bar{\tilde{t}}_1$ &
  @20 fb$^{(-1)}$ & (Run I)\\
\hline $ \tilde{b}_1\to b \tilde{\chi}^0_1$ &
    & $bb \Big/ \hspace{-0.3cm E_T}$
    & 260 \ GeV/c$^2$& (146 GeV/c$^2$ )\\
$ \tilde{t}_1\to c \tilde{\chi}^0_1$ &
    & $cc \Big/ \hspace{-0.3cm E_T}$
    & 220 \ GeV/c$^2$& (116 \ GeV/c$^2$ )\\
$\tilde{t}_1\to b l\tilde{\nu}$ & $\tilde{\nu} \to \nu
\tilde{\chi}^0_1$
    & $l^+l^- b \Big/ \hspace{-0.3cm E_T}$
    & 240 \ GeV/c$^2$ & (140 \ GeV/c$^2$ ) \\
$\tilde{t}_1\to b l\tilde{\nu}\tilde{\chi}^0_1$ &
    & $l^+l^- b \Big/ \hspace{-0.3cm E_T}$
    & - & (129 \ GeV/c$^2$ ) \\
$\tilde{t}_1\to b \tilde{\chi}^\pm_1$ & $\tilde{\chi}^\pm_1 \to
W^{(*)} \tilde{\chi}^0_1$
    & $l^+l^- b \Big/ \hspace{-0.3cm E_T}$;
    & 210 \ GeV/c$^2$ & (-)\\
$\tilde{t}_1\to b W\tilde{\chi}^0_1$ &
    &  $l^+l^- bj \Big/ \hspace{-0.3cm E_T} $
    & 190 \ GeV/c$^2$ & (-) \\
\hline
\end{tabular}
\end{center}
\caption{Discovery reaches on $M_{\tilde{b}}$ and $M_{\tilde{t}}$
expected in Run II.} \label{T}
\end{table}
Gluinos  and squarks  are pair-produced at the Tevatron. One may
have $\tilde g\tilde g, \tilde g\tilde q$, and $\tilde q\tilde q$
pairs. In most of the parameter space accessible at the Tevatron,
the left-chiral squark dominantly decays into a quark and either a
$\tilde \chi^\pm_1$ or a $\tilde \chi^0_2$. Pair-produced squarks
and gluinos have at least two large-$E_T$ jets associated with
large missing energy.  The final state with lepton(s) is possible
due to leptonic decays of the $\tilde \chi^\pm_1$ and/or $\tilde
\chi^0_2$.

We show also the discovery reach of the Tevatron in the
$m_0,m_{1/2}$ parameter plane of the MSSM in the trilepton
channel~\cite{kamon} for two values of $\tan\beta$. The trilepton
signal arises when both the lightest chargino  ($\tilde
\chi^\pm_1$) and the next-to-lightest neutralino ($\tilde
\chi^0_2$) decay leptonically in $p\bar p \to \tilde \chi^\pm_1
\tilde \chi^0_2 + X$.
\begin{figure}[htb]
\begin{center}
\vspace*{0.5cm} \leavevmode
  \epsfig{figure=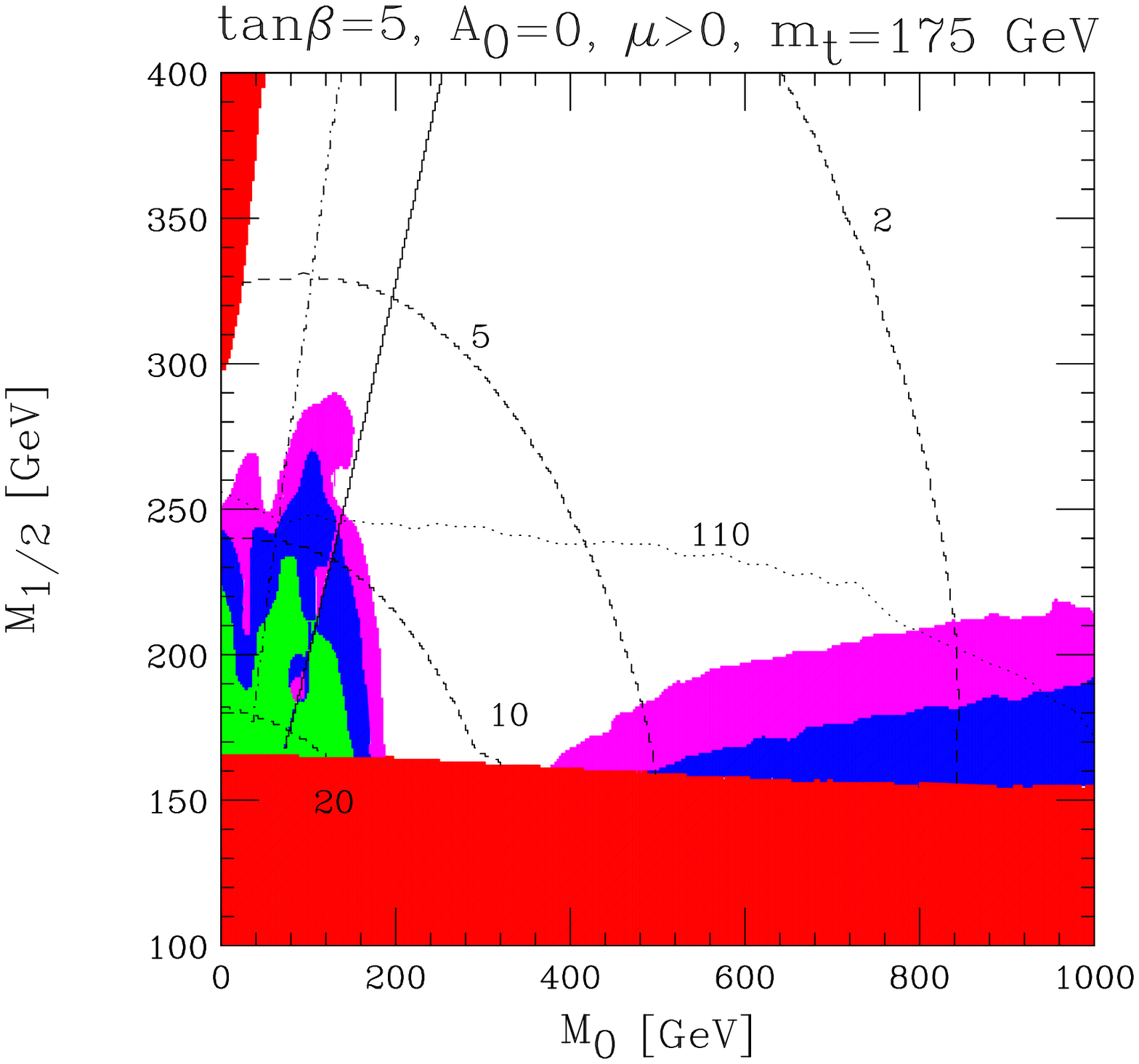,height=6cm,width=6cm,angle=90}
 \epsfig{figure=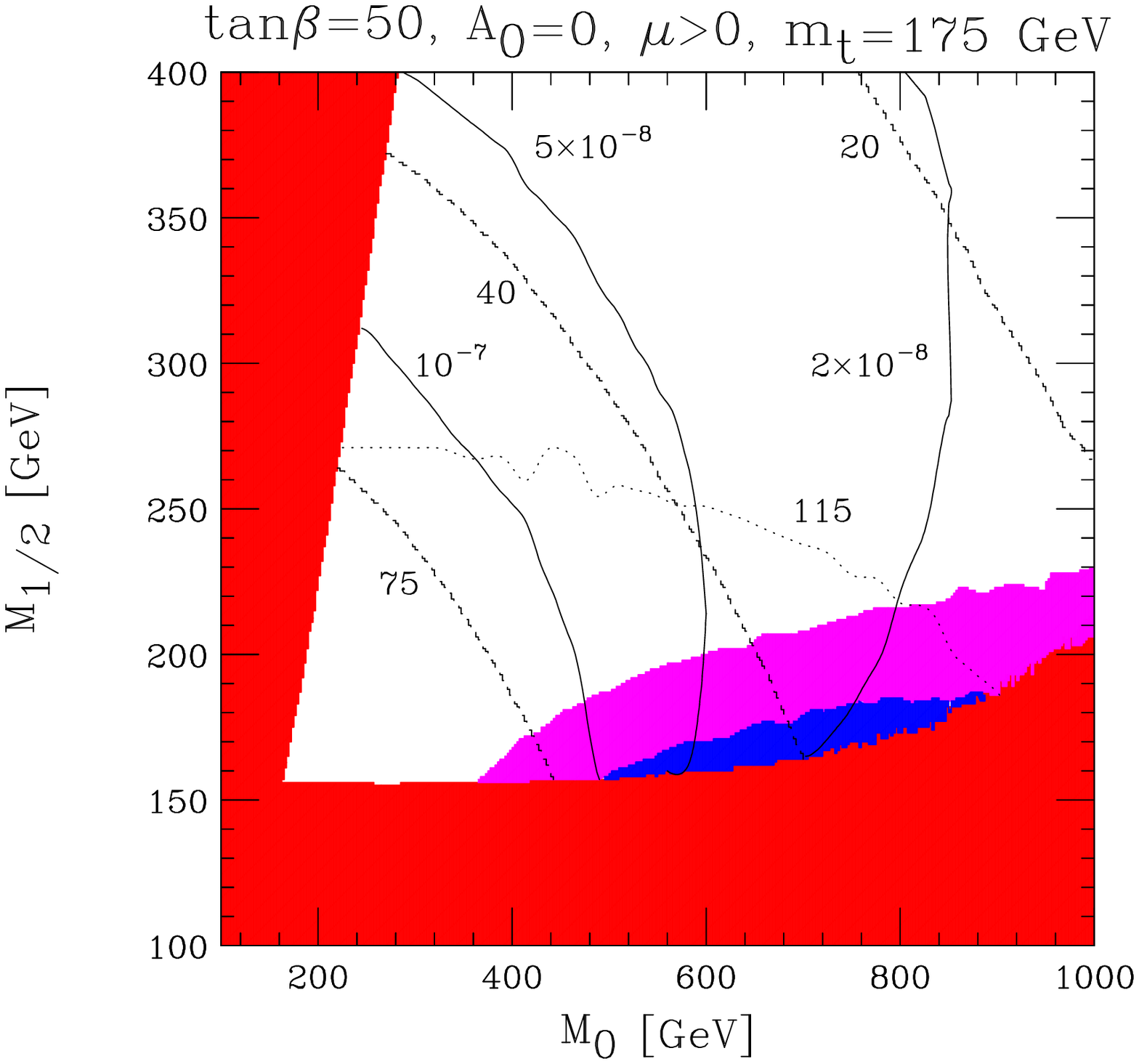,height=6cm,width=6cm,angle=90}
\end{center}
 \caption{\label{Tev2} Regions of the $m_0,m_{1/2}$ plane
where the trilepton events should be detectable at the level of
$5\sigma$ significance for $\tan\beta=5$ (left) and $\tan\beta=50$
(right)~\cite{15}. Three areas are shown for the integrated
luminosity of 30 fb$^{-1}$ (magenda), 10 fb$^{-1}$ (blue) and 2
fb$^{-1}$ (green). The large red regions are excluded. Dashed
lines represent the SUSY contribution to the muon anomalous
magnetic moment (in units of $10^{-10}$) and the dotted lines are
iso-mass contours of the lightest neutral Higgs boson.}
\end{figure}
In the trilepton channel the Tevatron  will be sensitive up to
$m_{1/2} \leq 250$ GeV if $m_0\ \leq 200$ GeV  and  up to $m_{1/2}
\leq 200$ GeV if $m_0\ \geq 500$ GeV. \vspace{0.5cm}

\underline{LHC}
 \vspace{0.3cm}

The LHC hadron collider is the ultimate machine for new physics at
the TeV scale. Its c.m. energy is planned to be 14 TeV with very
high luminosity up to a few hundred fb$^{-1}$.  The LHC is
supposed to cover A wide range of parameters of the MSSM (see
Figs. below) and discover the superpartners with the masses below
2 TeV~\cite{Kras}. This will be a crucial test for the MSSM and
the low energy supersymmetry. The LHC potential to discover
supersymmetry is widely discussed in the
literature~\cite{Kras}-\cite{LHCSUSY}.

\begin{figure}[ht]
\begin{center}
\leavevmode
  \epsfxsize=9cm 
 \epsffile{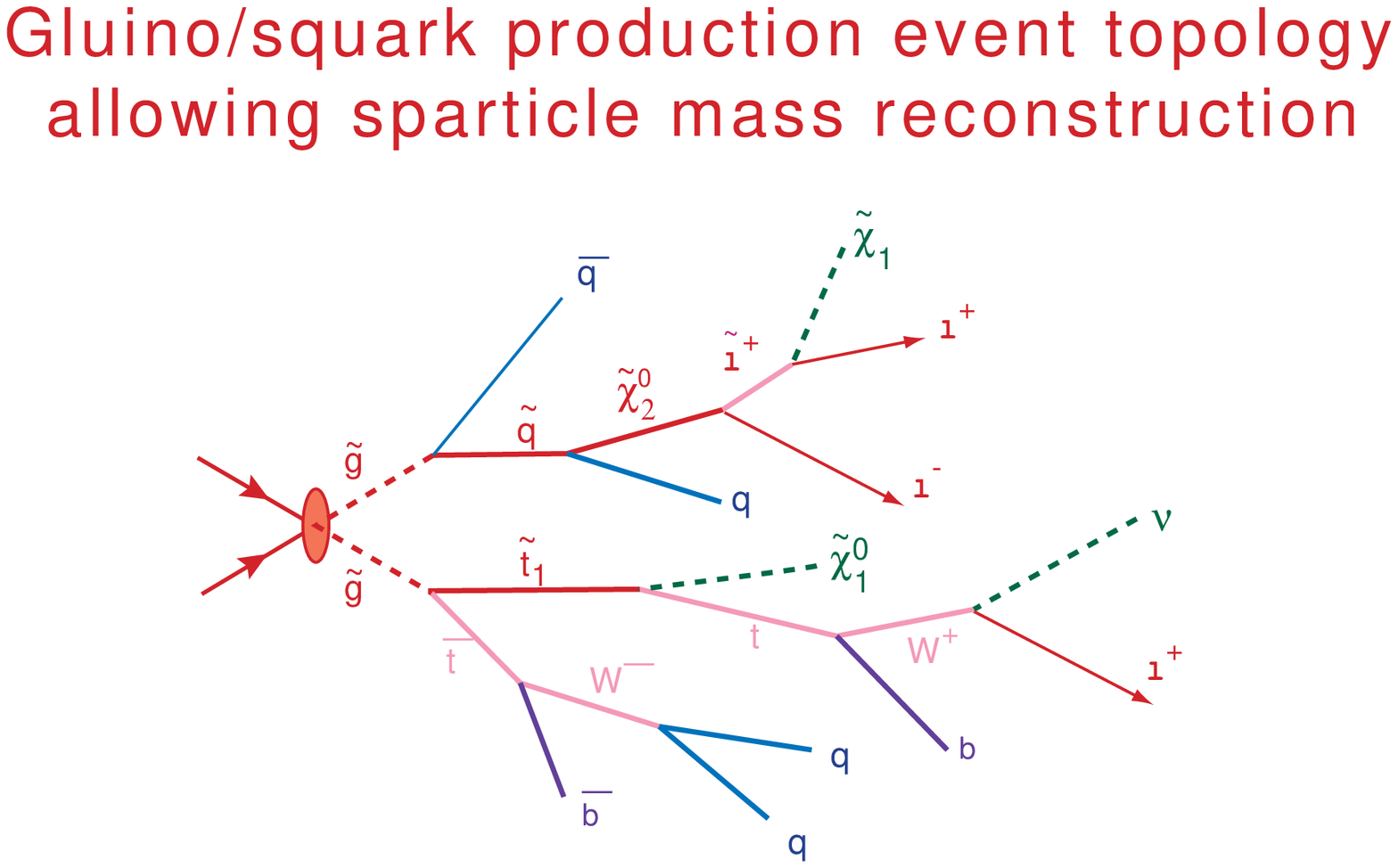}
 \vspace{-4cm}
\caption{The cascade decay of squarks into jets and neutralino
with  possible addition of multileptons} \label{cascade}
\end{center}
\end{figure}

The gluino and squark production cross sections at LHC can reach 1
pb for masses around 1 TeV. Their decays produce missing
transverse momentum from the LSP escape plus multiple jets and a
varying number of leptons from the intermediate gauginos. The main
decay mode is quarks and gluons plus the LSP. Cascade decays and
as a consequence of multilepton events are almost negligible. A
typical event with the cascade squark decay is shown in
Fig.\ref{cascade}.

\begin{figure}[ht]
\begin{center}
\leavevmode
  \epsfxsize=6.5cm
 \epsffile{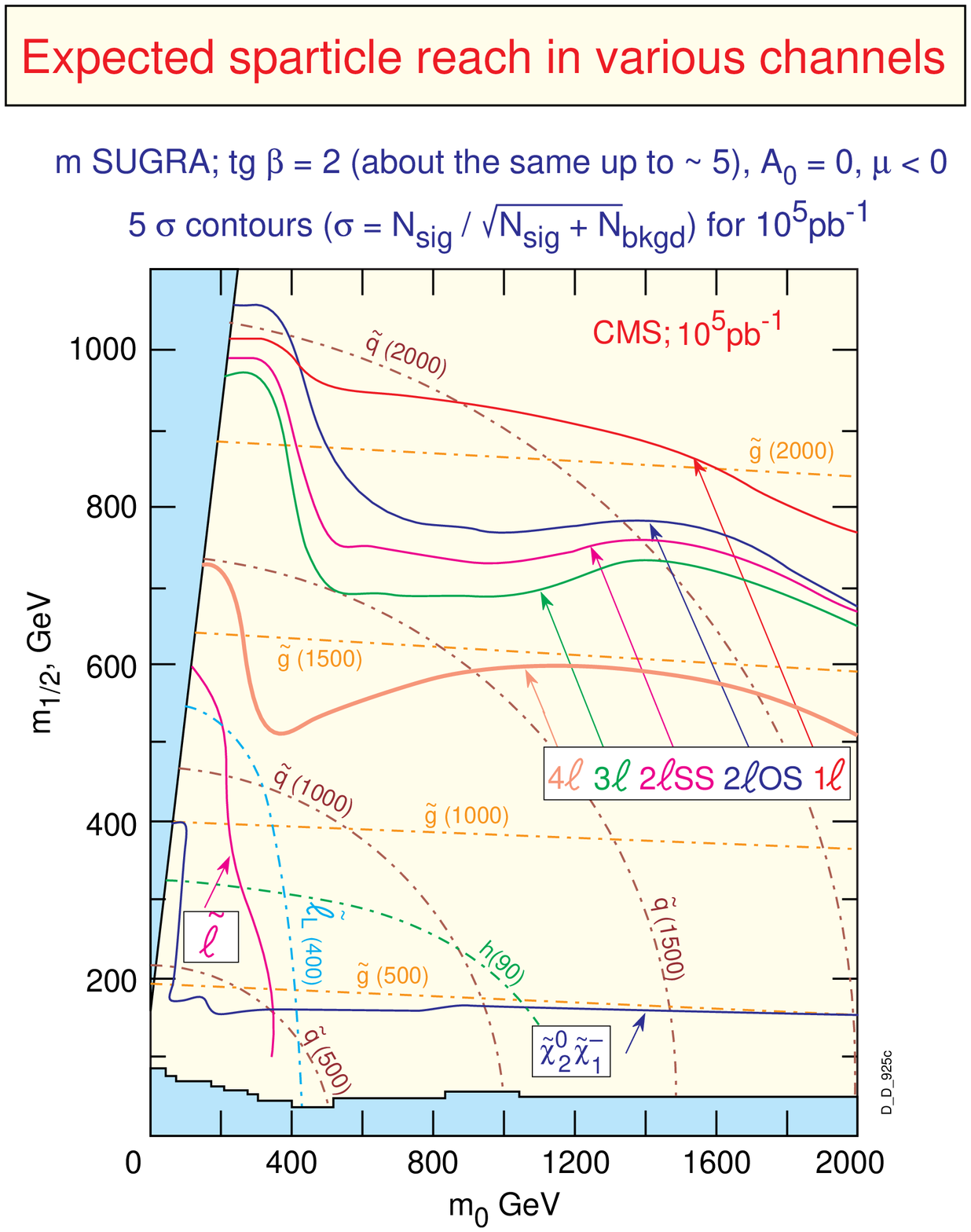}
%
 \epsfxsize=6.5cm \epsfysize=8cm
 \epsffile{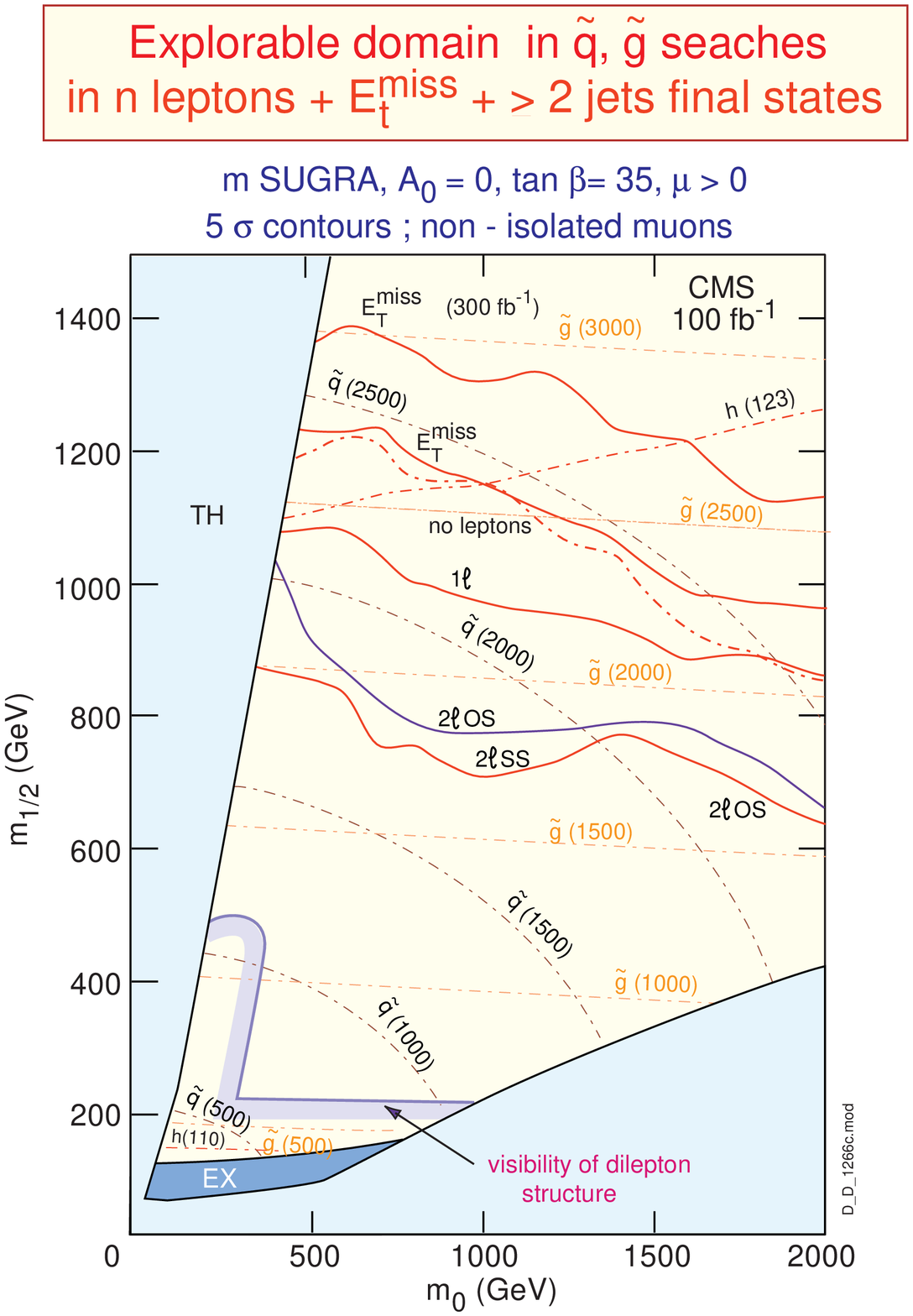}
\end{center}
\caption{Expected sparticle reach in various channels at
LHC~\cite{LHC}} \label{LHC}
\end{figure}

\begin{figure}[ht]
\begin{center}
\leavevmode
  \epsfxsize=5.5cm
 \epsffile{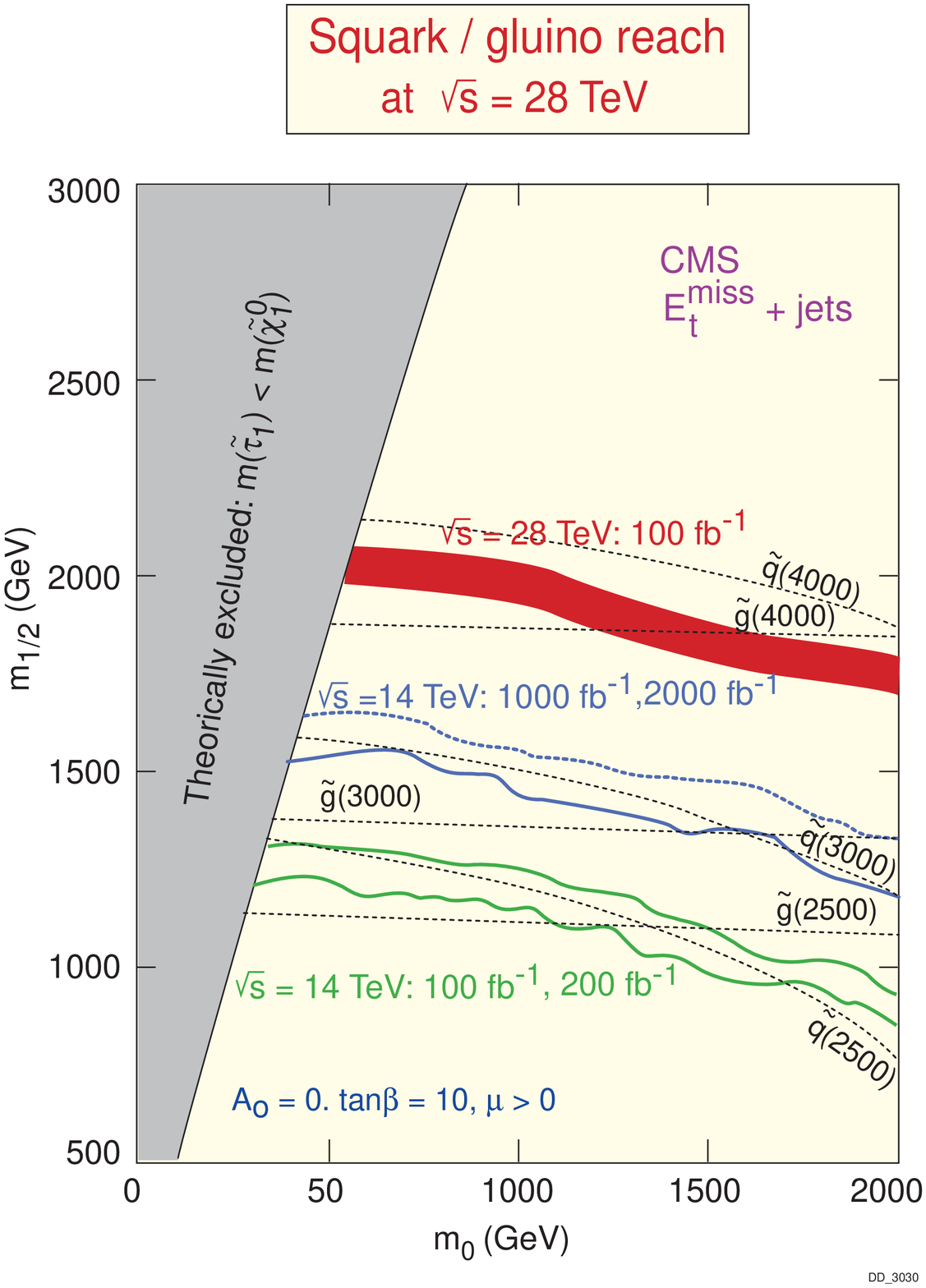}
%
 \epsfxsize=6cm \epsfysize=7cm
\hspace*{1cm} \epsffile{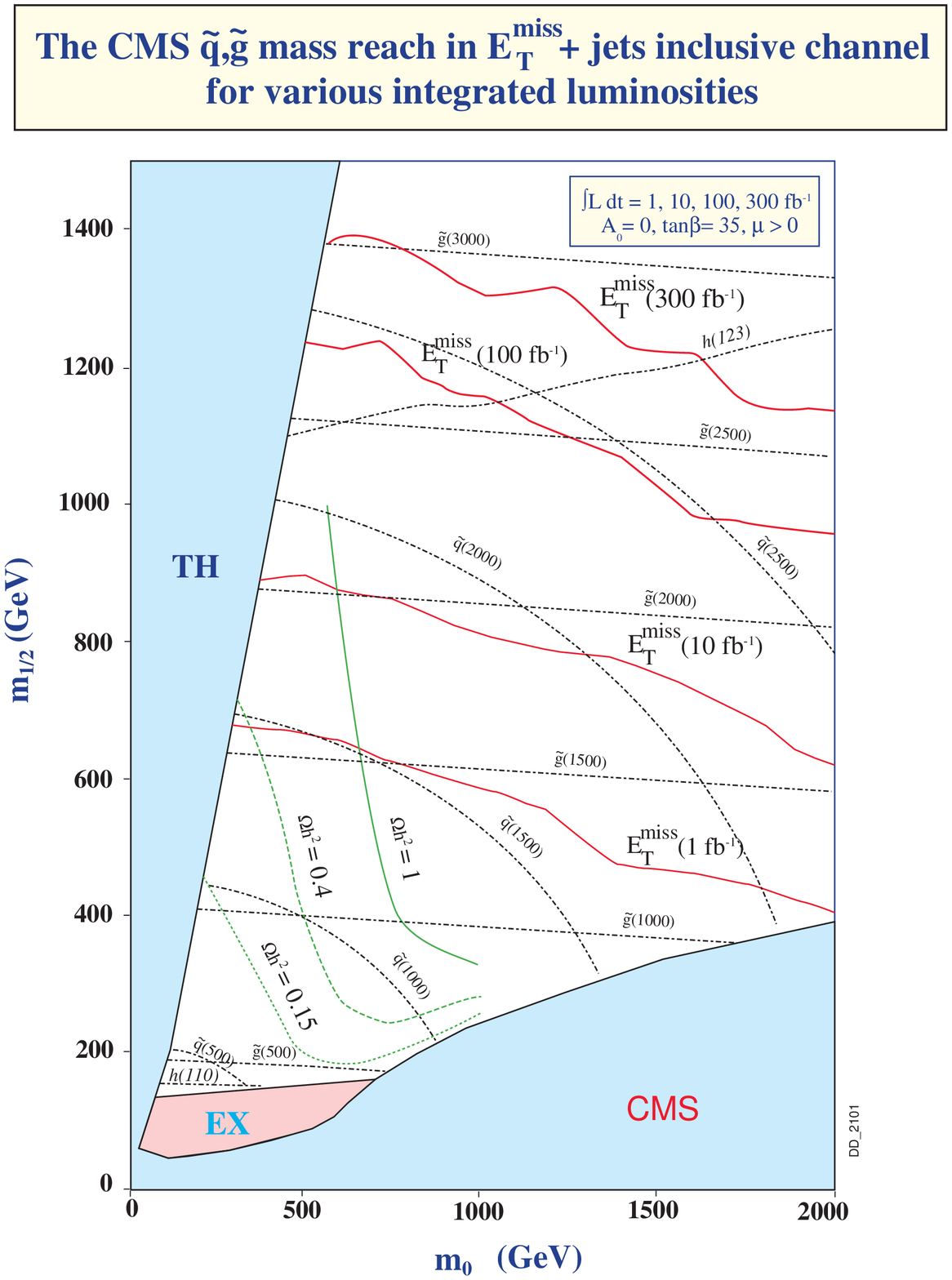}
\end{center}
\caption{Expected squarks and gluino reach at LHC for various
luminosities~\cite{LHC}} \label{LHC2}
\end{figure}

  The
LHC will be able to discover SUSY with squark and gluino masses up
to $2\div 2.5$ TeV for the luminosity $L_{tot}=100\ fb^{-1}$. The
expected discovery reach for various channels is shown in
Figs.\ref{LHC}, \ref{LHC2}. The most powerful signature for squark
and gluino detection are multijet events; however, the discovery
potential depends on relation between the LSP, squark, and gluino
masses, and decreases with the increase of the LSP mass.

\begin{figure}[ht]
\begin{center}
\leavevmode
  \epsfxsize=11.5cm
 \epsffile{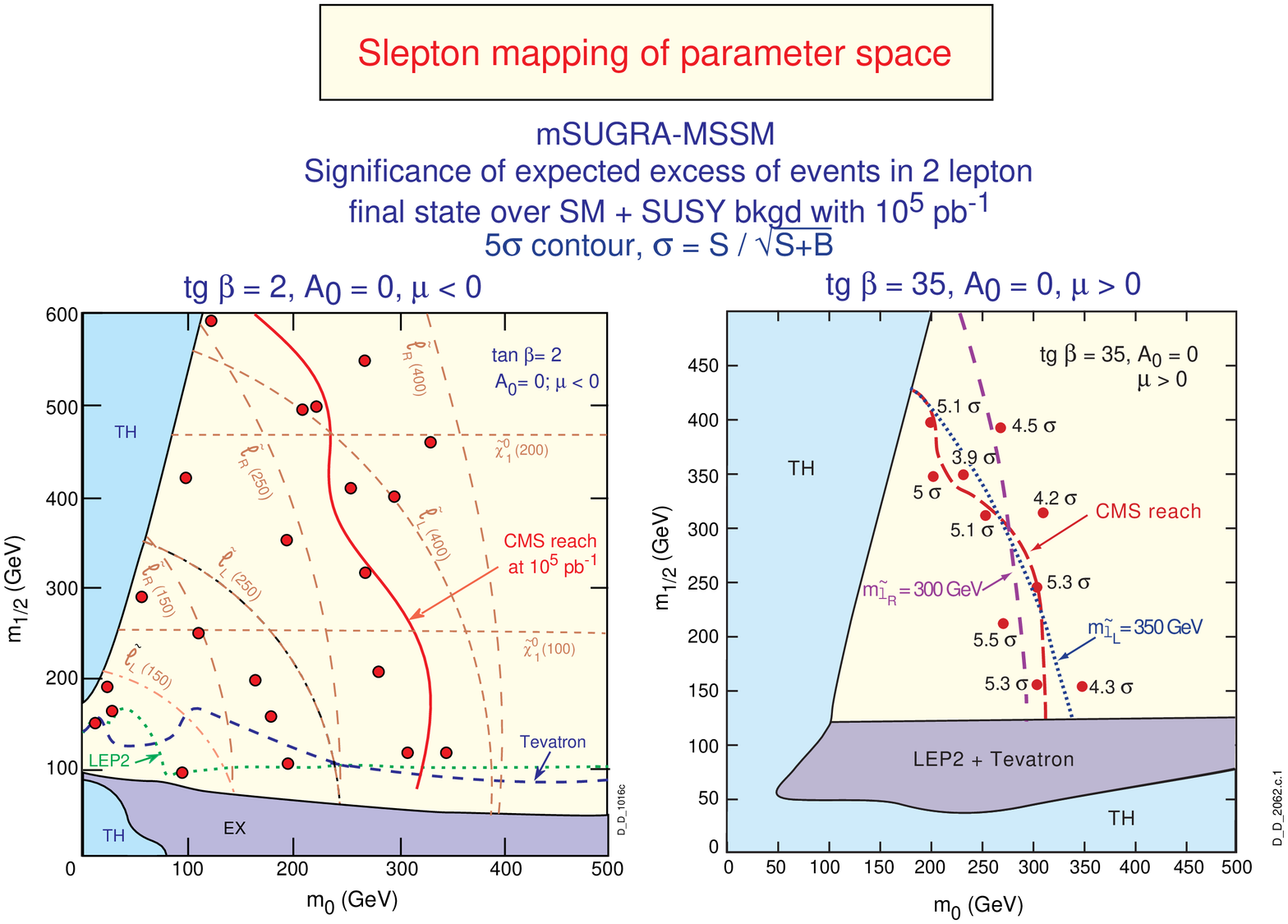}
%
\end{center}
\caption{Expected slepton reach at LHC~\cite{LHC}} \label{LHC3}
\end{figure}

Slepton pairs produced through the Drell-Yan mechanism
$pp\to\gamma^*/Z^*\to\tilde l^+\tilde l^-$ can be detected through
their leptonic decays $\tilde l \to l+\tilde \chi^0_1$. The
typical signature used for sleptons detection is the dilepton pair
with missing energy and no hadronic jets. For the luminosity
$L_{tot}=100\ fb^{-1}$ LHC will be able to discover sleptons with
the masses up to 400 GeV~\cite{Kras}.  The discovery reach for
sleptons in various channels is shown in Fig.\ref{LHC3}.

Chargino and neutralino pairs are also produced via the Drell-Yan
mechanism $pp\to \tilde \chi^\pm_1 \tilde \chi^0_2$ and may be
detected through their leptonic decays $\tilde \chi^\pm_1 \tilde
\chi^0_2 \to lll+E_T^{miss}$. So their main signature is the
isolated leptons with missing energy.
 The main background to the three
lepton channel comes from $WZ/ZZ,t\bar t, Zb\bar b$ and $b\bar b$
production. There could also be SUSY background arising from
squarks and gluino cascade decays into multileptonic modes. In the
case of light gauginos and heavy squarks and sleptons, which can
be realized in some regions of parameter space of the MSSM, the
cross sections for gaugino production may exceed those of strongly
interacting particles. Neutralinos and charginos could be detected
provided their masses are lighter than 350 GeV~\cite{Kras}.

\subsection{Conclusion}

Supersymmetry is now the most popular extension of the Standard
Model. It promises us that new physics is round the corner at a
TeV scale to be exploited at new machines of this decade. If our
expectations are correct, very soon we will face new discoveries,
the whole world of supersymmetric particles will show up and the
table of fundamental particles will be enlarged in increasing
rate.  This would be a great step in understanding the microworld.
\newpage

\section{PART II \ \  EXTRA DIMENSIONS}
\setcounter{equation} 0
\subsection{The main ideas}

Extra dimensions attracted  considerable interest in recent years
mainly due to unusual possibilities and intriguing effects even in
classical physics. (For review see, e.g. Refs.\cite{ED}. We follow
mostly Yu.Kubyshin paper.) There is no much motivation for ED in
particle physics except for the string theory paradigm. The point
is that to be consistent the string theory requires cancellation
of conformal anomaly which is possible in the critical dimension
equal to 26 for the bosonic string and 10 for the fermionic
one~\cite{string}. This way the ED come into play. Due to  the
presence of entirely new ingredients these models provide
solutions to some problems of the SM, in particular, of the
hierarchy problem in GUTs.

\begin{figure}[ht]
\leavevmode
  \epsfxsize=7cm \epsfysize=3.8cm
 \epsffile{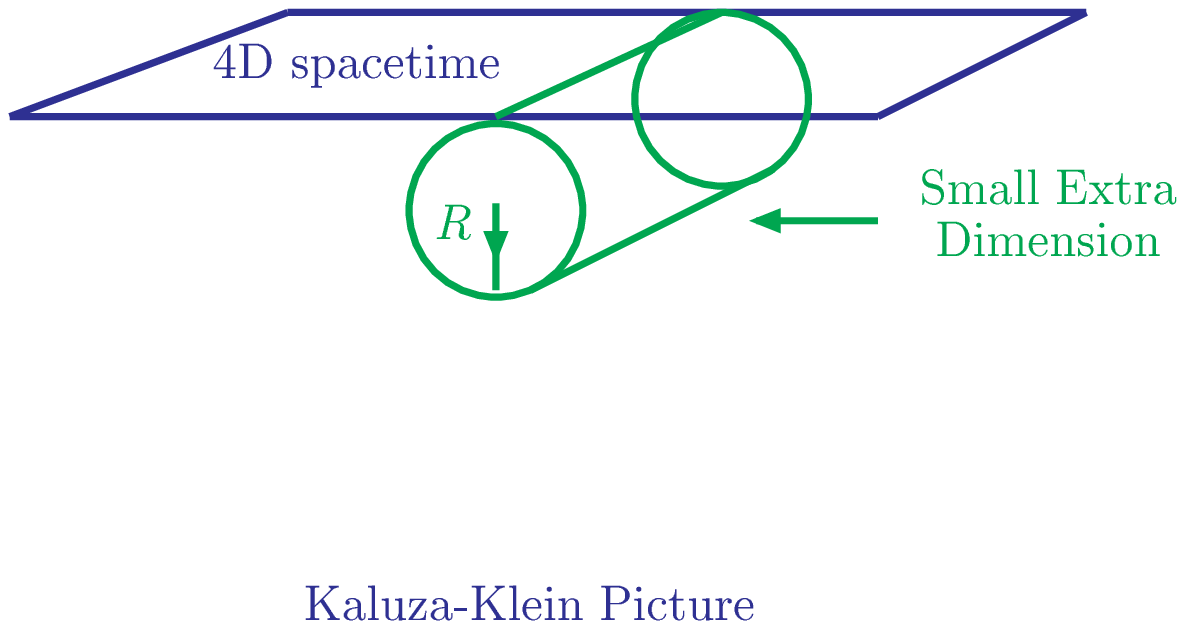}
 %
 \hspace*{1cm}
 \epsfxsize=5cm \epsfysize=5cm
 \epsffile{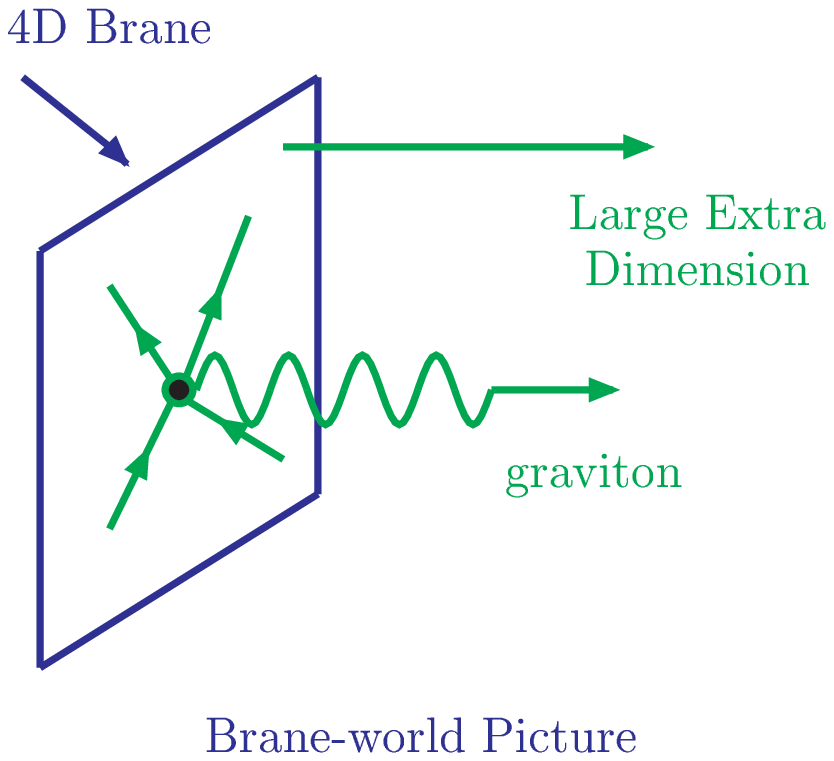}
\caption{Two possible constructions of the Extra dimensions}
\label{large}
\end{figure}

To explain why we do not see the ED, one usually refers to the
so-called Kaluza-Klein picture~\cite{KK}. It is believed that the
space-time has 3 large  spacial dimensions, and small and compact
extra ones. We do not see them because the radius of ED is too
small for the present energies, say, equal to the Planck length,
$10^{-33}$ cm. This is shown symbolically in Fig.\ref{large}.

\begin{figure}[ht]
\begin{center}
\leavevmode
  \epsfxsize=5cm 
 \epsffile{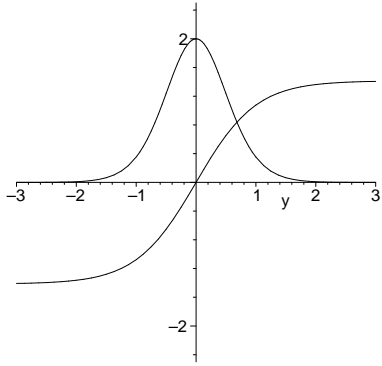}\vspace{-0.3cm}
\caption{The simplest  localization: the kink solution and its
energy density} \label{kink}
\end{center}
\end{figure}

At the same time, recently there appeared an alternative
explanation. This one is related to the so-called brane-world
picture~\cite{ADD,RS}. Here we do not assume small and compact ED,
but rather large ones, and the reason why we do not see them is
that we are "localized"  on a 4-dimensional brane in this
multidimensional space-time (see Fig.\ref{large}). This is similar
to being confined in a potential well and not being able to escape
it if the energy is not big enough. To get an example of such a
localization, consider some classical solution of the form of a
kink~\cite{kink}. Its energy is localized in the transition region
and the corresponding particle seems to be localized in this
region (see Fig.\ref{kink}).

Both the pictures  or even some combination of them may well be
realized in Nature and below we consider some consequences of such
an assumption.

\subsection{Kaluza-Klein Approach}

The KK approach is based on the hypothesis that the space-time is
a (4+d)-dimensional pseudo Euclidean space~\cite{KK-review}
 $$E_{4+d} = M_{4} \times K_{d},$$
  where $M_{4}$ is the
four-dimensional space-time and $K_{d}$ is $d$-dimensional compact
space of characteristic size (scale) $R$.  In accordance with the
direct product structure of the space-time, the metric is usually
chosen to be
\begin{equation}
ds^{2} = \hat{G}_{MN}(\hat{x}) d\hat{x}^{M} d\hat{x}^{N} = g_{\mu
\nu}(x) dx^{\mu} dx^{\nu} + \gamma_{mn}(x,y) dy^{m} dy^{n}.
\end{equation}
To interpret the theory as an effective four-dimensional one, the
field $\hat{\phi}(x,y)$ depending on both coordinates is expanded
in a Fourier series over the compact space
\begin{equation}
\hat{\phi} (x,y) = \sum_{n} \phi^{(n)}(x) Y_{n}(y), \label{KKexp}
\end{equation}
 where $Y_{n}(y)$ are orthogonal normalized
eigenfunctions of the Laplace operator $\Delta_{K_{d}}$ on the
internal space $K_{d}$,
\begin{equation}\label{Y-ef}
  \Delta_{K_{d}} Y_{n}(y) = \frac{\lambda_{n}}{R^{2}} Y_{n}(y).
\end{equation}

The coefficients $\phi^{(n)}(x)$ of the Fourier expansion
(\ref{KKexp}) are called the Kaluza-Klein modes and play the role
of fields of the effective four-dimensional theory. Their masses
are given by
\begin{equation}\label{mass}
  m_{n}^{2} = m^{2} + \frac{\lambda_{n}}{R^{2}},
\end{equation}
where $R$ is the radius of the compact dimension.

 The coupling constant $g_{(4)}$ of the 4-dimensional theory
is related to the coupling constant $g_{(4+d)}$ of the initial
(4+d)-dimensional one by
\begin{equation}\label{coupl}
  g_{(4)} = \frac{g_{(4+d)}}{V_{(d)}},
\end{equation}
$V_{(d)}\propto R^{d}$ being the volume of the space of extra
dimensions.

\subsubsection{Low scale gravity}

Consider now the Einstein $(4+d)$-dimensional gravity with the
action
\[
S_{E} = \int d^{4+d}\hat{x} \sqrt{-\hat{G}} \frac{1}{16 \pi
G_{N(4+d)}} {\cal R}^{(4+d)} [\hat{G}_{MN}],
\]
where the scalar curvature ${\cal R}^{(4+d)} [\hat{G}_{MN}]$ is
calculated using the metric $\hat{G}_{MN}$. Performing the mode
expansion and integrating over $K_{d}$ one arrives at the
four-dimensional action
\[
S_{E} = \int d^{4}x \sqrt{-g} \left\{ \frac{1}{16 \pi G_{N(4)}}
{\cal R}^{(4)} [g^{(0)}_{MN}] + \mbox{non-zero KK modes} \right\},
\]
 Similar to eq.(\ref{coupl}), the relation between the 4-dimensional and
$(4+d)$-dimensional gravitational (Newton) constants is given by
\begin{equation}\label{G-rel}
  G_{N(4)} = \frac{1}{V_{(d)}} G_{N(4+d)}.
\end{equation}
One can rewrite this relation in terms of the 4-dimensional Planck
mass $M_{Pl} = (G_{N(4)})^{-1/2} = 1.2 \cdot 10^{19}\; \mbox{GeV}$
and a fundamental mass scale of the $(4+d)$-dimensional theory $M
\equiv (G_{N(4+d)})^{-\frac{1}{d+2}}$. One gets
\begin{equation}\label{M-rel}
  M_{Pl}^{2} = V_{(d)} M^{d+2}.
\end{equation}
This formula is often referred to as the reduction formula.

Thus, the fundamental scale of a multidimensional theory becomes
$M$ rather than $M_{Pl}$. This way the problem of hierarchy is
reduced to a less severe one in extra dimensions. The scale $M$ is
not restricted by experimental value of the Newtonian constant and
may take values $\sim$ few TeV. Assuming $M=1\ TeV$ one can
rewrite eq.(\ref{M-rel}) as~\cite{ADD}
\begin{equation}\label{radcm}
  R \sim \frac{1}{M} \left( \frac{M_{Pl}}{M} \right)^{2/d} \sim
10^{\frac{30}{d}-17} \; \mbox{cm},
\end{equation}
or
\begin{equation}\label{radgev}
  R^{-1} \sim M \left( \frac{M}{M_{Pl}} \right)^{2/d} \sim
10^{-\frac{30}{d}+3} \; \mbox{GeV}.
\end{equation}

Let us analyze various cases. In the case $d=1$ it follows from
eq.(\ref{radcm})  that $R \sim 10^{13} \mbox{cm}$, i.e. the size
of extra dimensions is of the order of the solar distance. This
case is obviously excluded. For $d \geq 2$ we obtain:

\begin{tabular}{lll}
    &   & \\
for \ \ $d=2$ & $R \sim 0.1 \; \mbox{mm}$, &
     $R^{-1} \sim 10^{-3} \; \mbox{eV}$ \\
for \ \ $d=3$ \ \ & $R \sim 10^{-7} \; \mbox{cm}$, &
     $R^{-1} \sim 100 \; \mbox{eV}$ \\
  $\ldots$  \ \ &   $\ldots$  \ \ &   $\ldots$  \ \   \\
for \ \ $d=6$ \ \ & $R \sim 10^{-12} \; \mbox{cm}$, &
     $R^{-1} \sim 10 \; \mbox{MeV}$ \\
     &  &  \\
\end{tabular}

Such sizes of extra dimensions are already acceptable because no
deviations from the Newtonian gravity have been observed for
distances $r \sim 1$ mm so far (see, for example, \cite{Hoyle}).
On the other hand,  the SM  has been accurately checked already at
the scale $\sim 100$ GeV. To overcome this difficulty it is
supposed~\cite{ADD}  that  the SM fields are localized on the 4d
brane while only gravitons propagate in the bulk.

\begin{figure}[ht]
\begin{center}
\leavevmode
  \epsfxsize=8cm 
 \epsffile{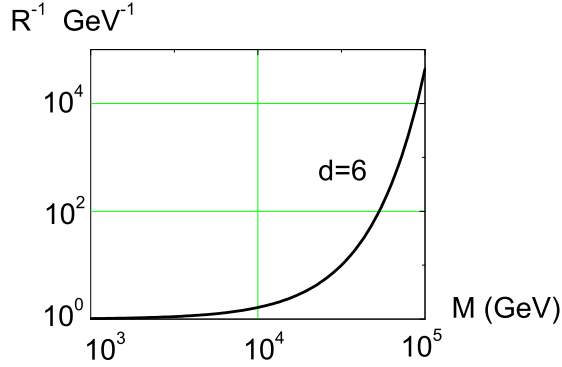}
\caption{The dependence of the radius of compactification $R$ on
the multidimensional fundamental scale $M$ for $d=6$}
\label{radius}
\end{center}
\end{figure}

At the same time, these conclusions strongly depend on the choice
of the scale M. If one takes bigger scale, the corresponding
radius decreases very fast (see Fig.\ref{radius}).

The presence of ED leads to the modification of classical gravity.
The Newton potential between two test masses $m_{1}$ and $m_{2}$,
separated by a distance $r$, is in this case equal to
\[
V(r) = G_{N(4)} m_{1}m_{2} \sum_{n} \frac{1}{r} e^{-m_{n}r} =
G_{N(4)} m_{1}m_{2} \left( \frac{1}{r} +
   \sum_{n \neq 0} \frac{1}{r} e^{-|n|r/R}  \right).
\]
The first term in the last bracket is the contribution of the
usual massless graviton (zero mode) and the second term is the
contribution of the massive gravitons. For the size $R$ large
enough (i.e. for the spacing between the modes small enough) this
sum can be replaced by the integral and one gets \cite{ADD}
\begin{eqnarray}
  V(r) & = &  G_{N(4)} \frac{m_{1}m_{2}}{r} \left[
1 + S_{d-1} \int_{1/R}^{\infty}e^{-mr} m^{d-1} dm \right] \nonumber \\
   & = &  G_{N(4)} \frac{m_{1}m_{2}}{r} \left[
1 + S_{d-1} \left( \frac{R}{r} \right)^{d} \int_{r/R}^{\infty}
e^{-z} z^{d-1} dz \right], \label{Npot}
\end{eqnarray}
 where $S_{d-1}$ is the area of the $(d-1)$-dimensional sphere
 of the unit radius. This leads to the following
behaviour of the potential at short and long distances
\begin{equation}\label{pot}
  V\approx \left\{ \begin{array}{ll} G_{N(4)} \frac{\displaystyle m_{1}
  m_{2}}{\displaystyle r}& r \gg R,\\
G_{N(4)} \frac{\displaystyle m_{1} m_{2}}{\displaystyle r} S_{d-1}
     \left( \frac{\displaystyle R}{\displaystyle r} \right)^{d} \Gamma (d) =
     G_{N(4+d)} \frac{\displaystyle m_{1} m_{2}}{\displaystyle r^{d+1}}
      S_{d-1} \Gamma (d) & r \ll R,
     \end{array} \right.
\end{equation}
Thus, one has a modification of classical gravity at small
distances which may have observational consequences.

\subsubsection{The ADD model}

The ADD model was proposed  by N. Arkani-Hamed, S. Dimopoulos and
G. Dvali in Ref. \cite{ADD}. The model includes the SM localized
on a 3-brane embedded into the $(4+d)$-dimensional space-time with
compact extra dimensions. The gravitational field is the only
field which propagates in the bulk.

 To analyze the field content of the effective
(dimensionally reduced) four-dimensional model, consider the field
$\hat{h}_{MN}(x,y)$ describing the linear deviation of the metric
around the $(4+d)$-dimensional Minkowski background $\eta_{MN}$
\begin{equation}\label{ADD:Gh}
  \hat{G}_{MN} (x,y) = \eta_{MN} +
\frac{2}{M^{1+d/2}} \hat{h}_{MN}(x,y)
\end{equation}
Let us assume, for simplicity, that the space of extra dimensions
is the $d$-dimensional torus. Performing the KK mode expansion
\begin{equation}\label{ADD-exp}
  \hat{h}_{MN} (x,y) =
\sum_{n} h^{(n)}_{MN} (x) \frac{1}{\sqrt{V_{(d)}}}
\exp(-i\frac{n_{m}y^{m}}{R}),
\end{equation}
where $V_{(d)}$ is the volume of the space of extra dimensions, we
obtain the KK tower of states $h^{(n)}_{MN} (x)$ with masses
\begin{equation}\label{ADD-mass}
m_{n} = \frac{1}{R} \sqrt{n_{1}^{2}+n_{2}^{2}+ \ldots + n_{d}^{2}}
\equiv \frac{|n|}{R},
\end{equation}
 so that the mass splitting  is $\Delta m \propto 1/R.$

The interaction of the KK modes $h^{(n)}_{MN}(x)$ with fields on
the brane is determined by the universal minimal coupling of the
$(4+d)$-dimensional theory
\[
S_{int} = \int d^{4+d}\hat{x} \sqrt{-\hat{G}} \hat{T}_{MN}
\hat{h}^{MN} (x,y),
\]
where the energy-momentum tensor of the matter  localized on the
brane at $y=0$ has the form
\[
\hat{T}_{MN}(x,y) = \delta_{M}^{\mu} \delta_{N}^{\nu} T_{\mu
\nu}(x) \delta^{(d)}(y).
\]
Using the reduction formula (\ref{M-rel}) and the KK expansion
(\ref{ADD-exp}) one obtains that
\begin{equation}
  S_{int} =   \int d^{4}x
T_{\mu \nu} \sum_{n} \frac{1}{M^{1+d/2} \sqrt{V_{(d)}}} h^{(n)\mu
\nu} (x) =  \sum_{n} \int d^{4}x \frac{1}{M_{Pl}} T^{\mu \nu}(x)
h^{(n)}_{\mu \nu}(x), \label{ADD-int1}
\end{equation}
which is the usual interaction of matter with gravity suppressed
by $M_{Pl}$.

The degrees of freedom of the four-dimensional theory, which
emerge from the multidimensional metric, include~\cite{GRW,Hew}
\begin{enumerate}
  \item the massless graviton and the massive KK gravitons $h^{(n)}_{\mu\nu}$
(spin-2 fields) with masses given by eq.(\ref{ADD-mass});
  \item $(d-1)$ KK towers of spin-1 fields which do not couple to $T_{\mu \nu}$;
  \item $(d^{2}-d-2)/2$ KK towers of real scalar fields (for $d \geq 2$), they
do not couple to $T_{\mu \nu}$ either;
\item  a KK tower of scalar fields coupled to the trace of the
energy-momentum tensor $T_{\mu}^{\mu}$, its zero mode is called
radion and describes fluctuations of the volume of extra
dimensions.
\end{enumerate}
Alternatively, one can consider the $(4+d)$-dimensional theory
with the $(4+d)$-dimensional massless graviton $\hat{h}_{MN}(x,y)$
interacting with the SM fields with couplings $\sim 1/M^{1+d/2}$.

In the 4-dimensional  picture the coupling of each individual
graviton (both massless and massive) to the SM fields is small
$\sim 1/M_{Pl}$. However, the smallness of the coupling constant
is compensated by the high multiplicity of states with the same
mass. Indeed, the number $d{\cal N}(|n|)$ of modes with the
modulus $|n|$ of the quantum number being in the interval
$(|n|,|n|+d|n|)$ is equal to
\begin{equation}\label{ADD-dN}
  d{\cal N}(|n|) = S_{d-1} |n|^{d-1}d|n| = S_{d-1} R^{d}
m^{d-1}dm \sim S_{d-1} \frac{M_{Pl}}{M^{d+2}} m^{d-1}dm,
\end{equation}
 where we used the mass formula $m=|n|/R$
and the reduction formula (\ref{M-rel}). The number of KK
gravitons $h^{(n)}$ with masses $m_{n} \leq E < M$ is equal to
\[
{\cal N}(E) \sim \int_{0}^{ER} d{\cal N}(|n|) \sim S_{d-1}
\frac{M^{2}_{Pl}}{M^{d+2}} \int_{0}^{E} m^{d-1}dm =
\frac{S_{d-1}}{d} \frac{M^{2}_{Pl}}{M^{d+2}} E^{d} \sim R^{d}
E^{d}.
\]
 One can see that for $E \gg R^{-1}$ the multiplicity of states which
can be produced is large. Hence, despite the fact that due to
eq.(\ref{ADD-int1}) the amplitude of emission of the mode $n$ is
${\cal A} \sim 1/M_{Pl}$, the total combined rate of emission of
the KK gravitons with masses $m_{n} \leq E$ is
\begin{equation}\label{ADD-rate}
  \sim \frac{1}{M_{Pl}^{2}} {\cal N}(E) \sim
\frac{E^{d}}{M^{d+2}}.
\end{equation}
 We can see that there is a considerable
enhancement of the effective coupling due to the large phase space
of KK modes or due to the large volume of the space of extra
dimensions. Because of this enhancement the cross-sections of
processes involving the production of KK gravitons may turn out to
be quite noticeable at future colliders.

\subsubsection{HEP phenomenology}

There are two types of processes at high energies in which the
effect of the KK modes of the graviton can be observed at running
or planned experiments. These are the graviton emission and
virtual graviton exchange processes~\cite{GRW}-\cite{ChK}.

We start with the graviton emission, i.e., the reactions where the
KK gravitons are created as final state particles. These particles
escape from the detector, so that a characteristic signature of
such processes is missing energy. Though the rate of production of
each individual mode is suppressed by the Planck mass, due to the
high multiplicity of KK states the magnitude of the total rate of
production is determined by the TeV scale (see
eq.(\ref{ADD-rate})). Taking eq.(\ref{ADD-dN}) into account, the
relevant differential cross section \cite{GRW} is
\begin{equation}\label{ADD-sigma}
  \frac{d^{2}\sigma}{dt dm} \sim  S_{d-1}
\frac{M^{2}_{Pl}}{M^{d+2}} m^{d-1} \frac{d \sigma_{m}}{dt} \sim
\frac{1}{M^{d+2}},
\end{equation}
 where $d \sigma_{m}/dt $ is the differential cross section of
the production of a single KK mode with mass $m$.

\begin{figure}[ht]
\begin{center}
\leavevmode
  \epsfxsize=5.1cm \epsfysize=5.1cm
 \epsffile{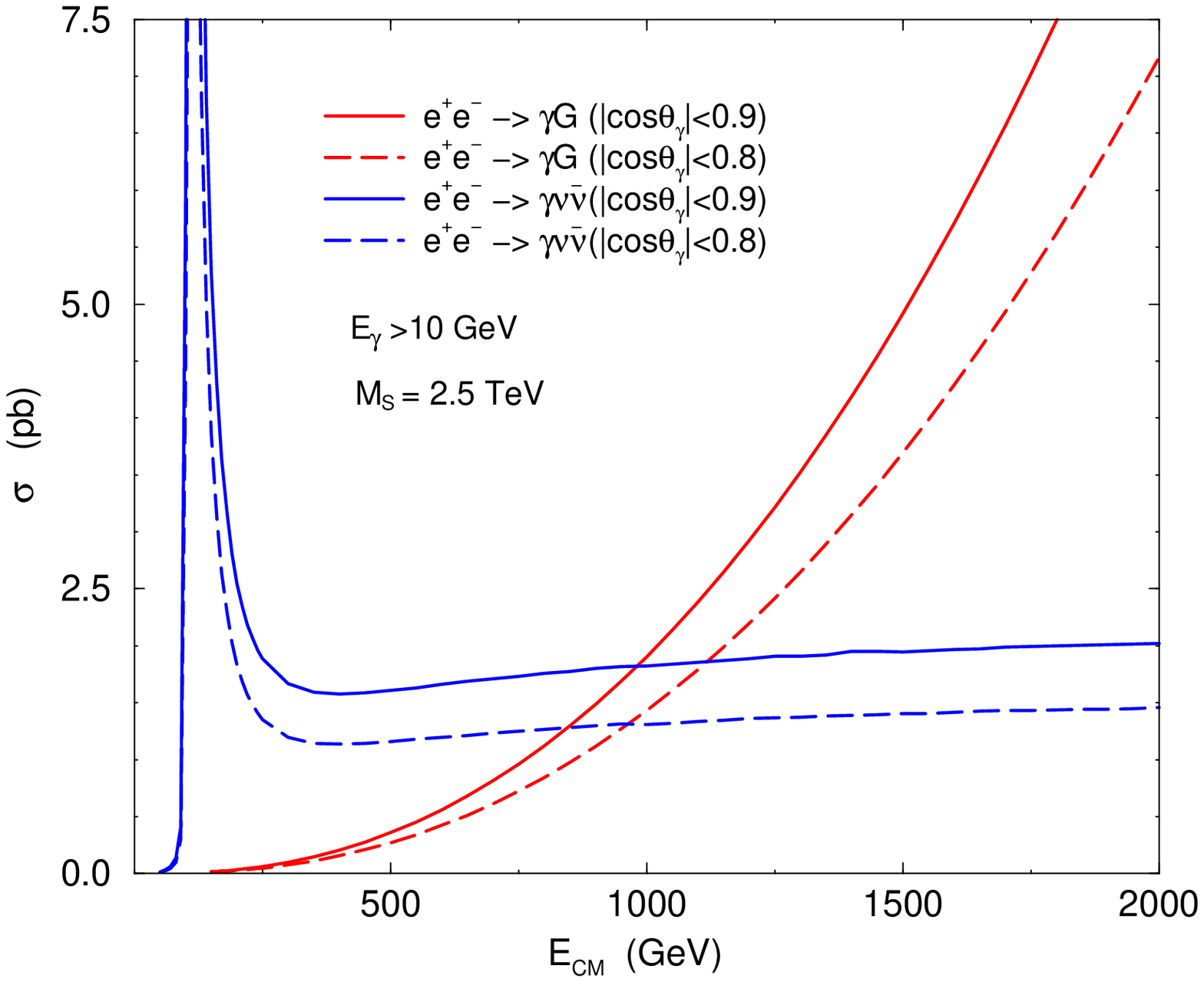}\hspace*{2cm}\epsfxsize=5.5cm
 \epsffile{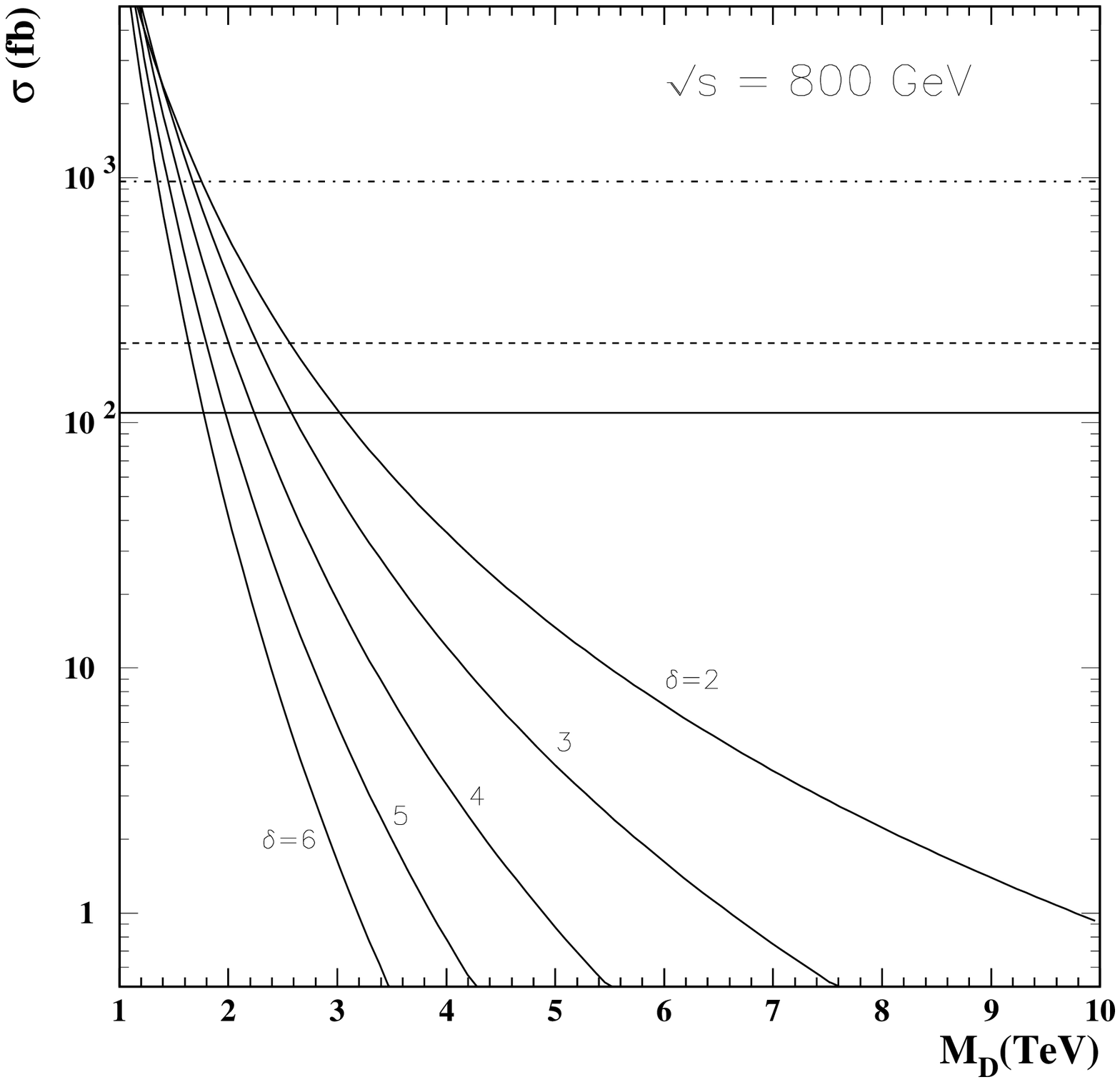}

\caption{The total cross sections for $e^{+}e^{-} \rightarrow
\gamma \nu_{i} \bar{\nu}_{i}$ ($i=e,\mu,\tau$)  and $e^{+}e^{-}
\rightarrow \gamma h$ (faster growing curves) for $d=2$ and $M =
2.5$ TeV \cite{ChK} (left) and the  cross section for $e^{+}e^{-}
\rightarrow \gamma h^{(n)}$ at $\sqrt{s}=800$ GeV as a function of
the scale $M$  for A different number  $\delta$ of extra
dimensions (right). Horizontal lines indicate the background.
\cite{TDR-Wil}} \label{fig:ADD-ee}
\end{center}
\end{figure}

At $e^{+}e^{-}$ colliders the main contribution comes from the
$e^{+}e^{-} \rightarrow \gamma h^{(n)}$ process. The main
background comes from the process $e^{+}e^{-} \rightarrow \nu
\bar{\nu} \gamma$ and can be effectively suppressed by using
polarized beams. Figure \ref{fig:ADD-ee} shows the total cross
section of the graviton production in electron-positron collisions
\cite{ChK}. To the right is  the same cross section as a function
of $M$ for $\sqrt{s} = 800 \; \mbox{GeV}$ \cite{TDR-Wil}.

Effects due to gravitons can also be observed at hadron colliders.
A characteristic process at the LHC would be $pp \rightarrow
(\mbox{jet} + \mbox{missing} \; E)$. The subprocess that gives the
largest contribution is the quark-gluon collision $qg \rightarrow
qh^{(n)}$. Other subprocesses are $q\bar{q} \rightarrow gh^{(n)}$
and $gg \rightarrow gh^{(n)}$.

\begin{figure}[ht]\hspace*{3cm}\vspace*{0.5cm}

\leavevmode
\hspace*{2cm}  \epsfxsize=5cm 
 \epsffile{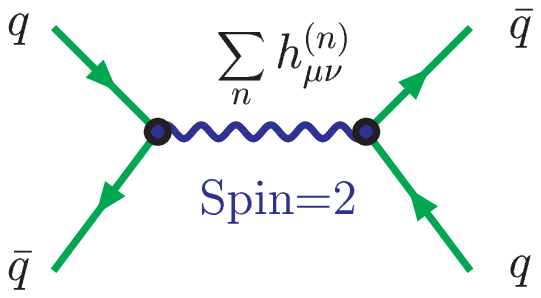}
 \vspace*{-4cm}

 \hspace*{7cm}\epsfxsize=6cm 
  \epsfig{figure=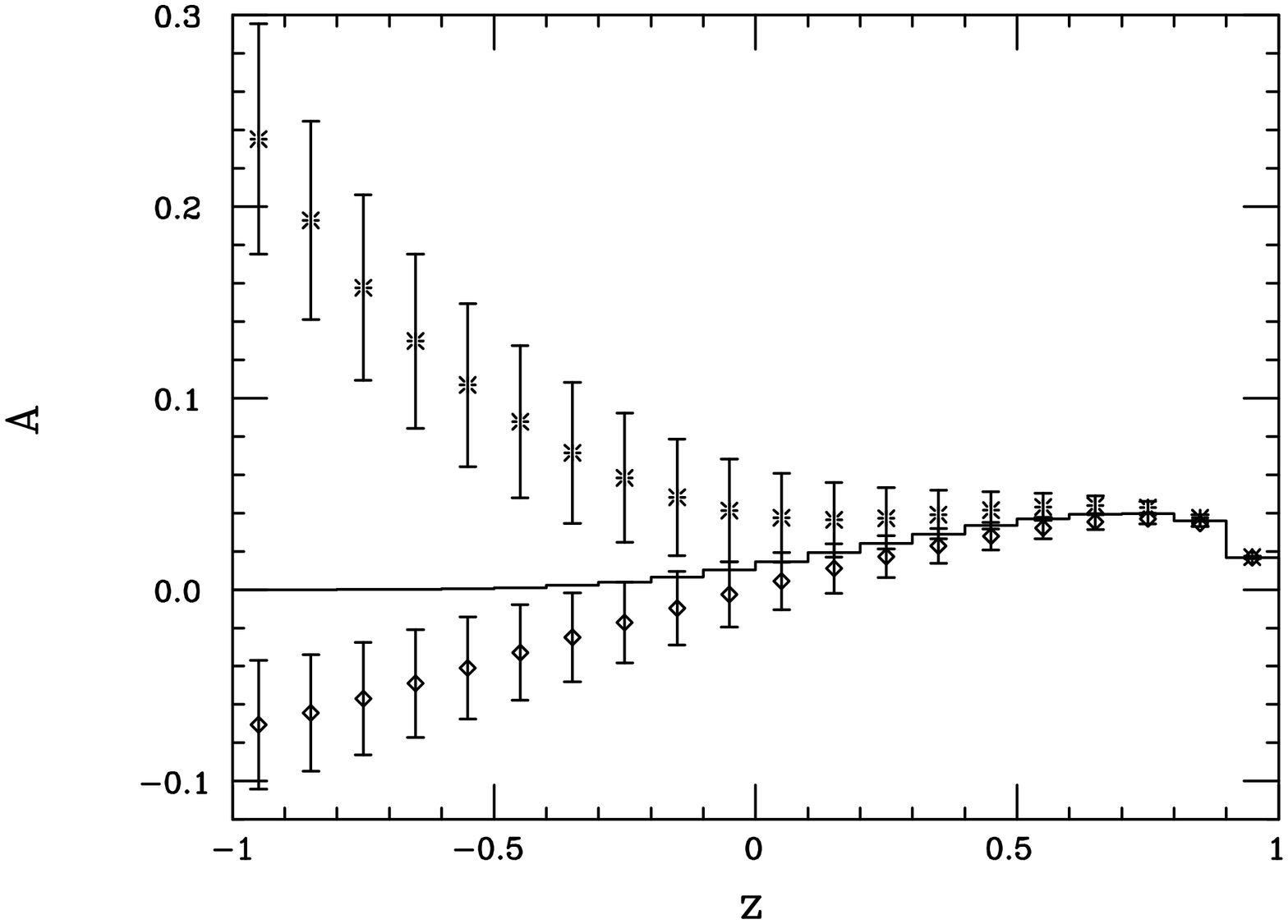,height=6cm,width=5cm,angle=-90}
\caption{The Feynman diagram for the virtual graviton exchange
(left) and deviation from the expectations of the SM (histogram)
for Bhabha scattering at a $500 \; \mbox{GeV}$ $e^{+}e^{-}$
collider for the Left-Right polarization asymmetry as a function
of $z=\cos \theta$ for $M = 1.5 \; \mbox{TeV}$ and the integrated
luminosity ${\cal L}=75 \; \mbox{fb}^{-1}$ (right) \cite{Ri99}.}
\label{grav}
\end{figure}

Processes of another type, in which the effects of extra
dimensions can be observed, are exchanges of virtual KK modes, in
particular, the virtual graviton exchanges. Contributions to the
cross section from these additional channels lead to  deviation
from the behaviour expected in the 4-dimensional model.  An
example is $e^{+}e^{-} \rightarrow f \bar{f}$ with $h^{(n)}$ being
the intermediate state (see Fig.\ref{grav}). Moreover, gravitons
can mediate processes absent in the SM at the tree-level, for
example, $e^{+}e^{-} \rightarrow HH$, $e^{+}e^{-} \rightarrow gg$.
Detection of such events with  large  cross sections may serve as
an indication of the existence of extra dimensions.

The $s$-channel amplitude of a graviton-mediated scattering
process is given by
\begin{equation}\label{amp}
{\cal A} = \frac{1}{M_{Pl}^{2}} \sum_{n} \left\{ T_{\mu \nu}
\frac{P^{\mu \nu} P^{\rho \sigma}}{s-m_{n}^{2}} T_{\rho \sigma} +
\sqrt{\frac{3(d-1)}{d+2}} \frac{T^{\mu}_{\mu}
T^{\nu}_{\nu}}{s-m_{n}^{2}} \right\},
\end{equation}
where $P_{\mu \nu}$ is a polarization factor coming from the
propagator of the massive graviton and $T_{\mu \nu}$ is the
energy-momentum tensor \cite{GRW}. It contains  a kinematic factor
\begin{eqnarray}
  {\cal S} & = &  \frac{1}{M_{Pl}^{2}}
\sum_{n}\frac{1}{s-m_{n}^{2}} \approx \frac{1}{M_{Pl}^{2}} S_{d-1}
\frac{M_{Pl}^{2}}{M^{d+2}}
\int^{\Lambda} \frac{m^{d-1}dm}{s-m^{2}}  \nonumber \\
   & = &  \frac{S_{d-1}}{2M^{4}}
\left\{ i\pi \left( \frac{s}{M^{2}} \right)^{d/2-1}  +
\sum_{k=1}^{[(d-1)/2]} c_{k} \left( \frac{s}{M^{2}} \right)^{k-1}
\left( \frac{\Lambda}{M} \right)^{d-2k} \right\}.   \label{kin}
\end{eqnarray}
Since the integrals are divergent for $d \geq 2$, the cutoff
$\Lambda$ was introduced. It sets the limit of applicability of
the effective theory. Because of the cutoff,the amplitude cannot
be calculated explicitly without the knowledge of a full
fundamental theory. Usually, in the literature it is assumed that
the amplitude is dominated by the lowest-dimensional local
operator (see \cite{GRW}).

The characteristic feature of expression (\ref{kin}) different
from the 4-dimensional model is the increase of the cross section
with energy. This is a consequence of the exchange of the infinite
tower of the KK modes. Note, however, that this result is based on
a tree-level amplitude, while the radiative corrections in this
case are power-like and may well change this behaviour.

Typical processes, in which the virtual exchange via massive
gravitons can be observed, are: (a) $e^{+}e^{-} \rightarrow \gamma
\gamma$; (b) $e^{+}e^{-} \rightarrow f \bar{f}$, for example the
Bhabha scattering $e^{+}e^{-} \rightarrow e^{+}e^{-}$ or M\"oller
scattering $e^{-}e^{-} \rightarrow e^{-}e^{-}$; (c) graviton
exchange contribution to the Drell-Yang production. A signal of
the KK graviton mediated processes is the deviation in the number
of events and in the left-right polarization asymmetry from those
predicted by the SM (see Figs.~\ref{grav}) \cite{Ri99}.

\subsection{Brane-World Models}
\subsubsection{The Randal-Sundram model}

\begin{figure}[htb]
\begin{center}
\leavevmode
  \epsfxsize=4.5cm 
 \epsffile{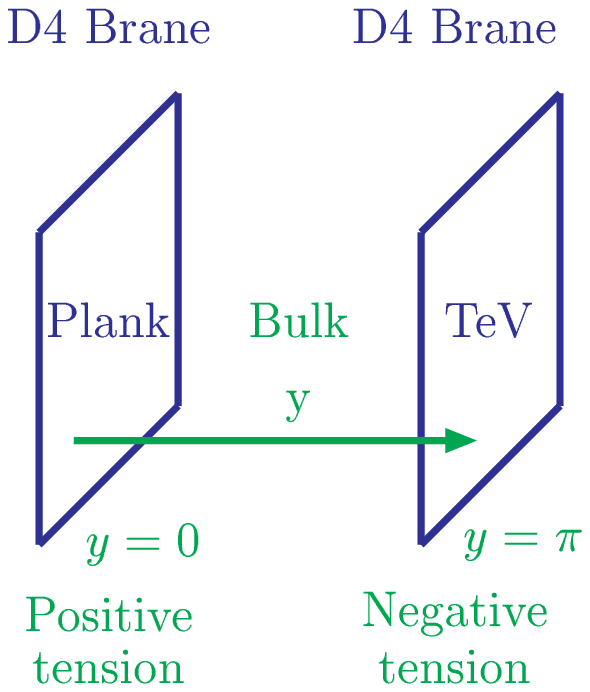}
\caption{The Randall-Sundrum construction of the extra-dimensional
space} \label{ransun}
\end{center}
\end{figure}

The RS model~\cite{RS} is a model of Einstein gravity in
five-dimensional Anti-de Sitter space-time with extra dimension
being compactified to the orbifold $S^{1}/Z_{2}$. There are two
3-branes in the model located at the fixed points $y=0$ and $y=\pi
R$ of the orbifold, where $R$ is the radius of the circle $S^{1}$.
The brane at $y=0$ is usually referred to as A Planck brane,
whereas the brane at $y=\pi R$ is called A TeV brane (see
Fig.\ref{ransun}). The SM fields are constrained to the TeV brane,
while gravity propagates in the additional dimension.

The action of the model is given by
\begin{eqnarray}
  S & = &  \int d^{4}x \int_{-\pi R}^{\pi R} dy \sqrt{-\hat{G}}
\left\{ 2 M^{3} {\cal R}^{(5)} \left[\hat{G}_{MN}\right] + \Lambda
\right\} \nonumber \\
   & +&  \int_{B_{1}} d^{4}x \sqrt{-g^{(1)}} \left( L_{1} - \tau_{1}
\right) + \int_{B_{2}} d^{4}x \sqrt{-g^{(2)}} \left( L_{2} -
\tau_{2} \right),
           \label{RS:L}
\end{eqnarray}
where ${\cal R}^{(5)}$ is the five-dimensional scalar curvature,
$M$ is a mass scale (the five-dimensional "Planck mass") and
$\Lambda$ is the cosmological constant. $L_{j}$ is a matter
Lagrangian and $\tau_{j}$ is a constant vacuum energy on brane $j$
$(j=1,2)$.

The RS  solution describes the space-time with non factorizable
geometry with the metric  given by
\begin{equation}\label{RS:RS1}
ds^{2} = e^{-2\sigma (y)} \eta_{\mu \nu} dx^{\mu} dx^{\nu} +
dy^{2}.
\end{equation}
The additional coordinate changes inside the  interval $-\pi R < y
\leq \pi R$ and the function $\sigma (y)$ in the warp factor $\exp
(-2\sigma)$ is equal to
\begin{equation}\label{RS:sigma}
  \sigma (y) = k |y|, \; \; \; (k > 0).
\end{equation}
For the solution to exist the parameters must be fine-tuned to
satisfy the relations
\[
\tau_{1} = - \tau_{2} = 24 M^{3} k, \; \; \; \Lambda = 24 M^{3}
k^{2}.
\]
Here $k$ is a dimensional parameter which was introduced for
convenience. This fine-tuning is equivalent to the usual
cosmological constant problem. If $k > 0$, then  the tension on
brane 1 is positive, whereas the tension $\tau_{2}$ on brane 2 is
negative.

For a certain choice of the gauge the most general perturbed
metric is given by
\[
ds^{2} = e^{-2k|y|} \left( \eta_{\mu \nu} + \tilde{h}_{\mu
\nu}(x,y) \right) dx^{\mu} dx^{\nu} + (1 + \phi (x) ) dy^{2}.
\]
and describes the  graviton field $\tilde{h}_{\mu \nu}(x,y)$ and
the radion field $\phi (x)$~\cite{ADM}.

As a next step the field $h_{\mu \nu}(x,y)$ is decomposed over an
appropriate system of orthogonal and normalized functions:
\begin{equation}\label{RS:decomp}
  h_{\mu \nu} (x,y) = \sum_{n=0}^{\infty} h_{\mu \nu}^{(n)}(x)
  \frac{\chi_{n}(y)}{R}.
\end{equation}
The particles localized on the branes are:\vspace{0.3cm}

\begin{minipage}[h]{7cm}
 \underline{Brane 1 (Planck):}
\begin{itemize}
  \item massless graviton $h_{\mu \nu}^{(0)}(x)$,
  \item massive KK gravitons $h_{\mu \nu}^{(n)}(x)$ with masses
  $m_{n}=\beta_{n} k e^{-\pi k R}$, where
  $\beta_{n} = 3.83, 7.02, 10.17, 13.32, \ldots$
  are the roots of the Bessel function,
  \item massless radion $\phi (x)$.
\end{itemize}
\end{minipage}\vspace{-3.7cm}

\hspace*{8cm}\begin{minipage}[h]{7cm}
 \underline{Brane 2 (TeV):}
\begin{itemize}
  \item massless graviton $h_{\mu \nu}^{(0)}(x)$,
  \item massive KK gravitons $h_{\mu \nu}^{(n)}(x)$ with masses
  $m_{n}=\beta_{n} k $,
  \item massless radion $\phi (x)$.
\end{itemize}
\end{minipage}\vspace{1cm}

The brane 2  is most interesting from the point of view of high
energy physics phenomenology.  Because of the nontrivial warp
factor $e^{-2\sigma (\pi R)}$, the Planck mass here is related to
the fundamental 5-dimensional scale $M$ by
\begin{equation}\label{RS:M-Pl}
  M_{Pl}^{2} =
e^{2k\pi R} \int_{-\pi R}^{\pi R} dy e^{-2k|y|} = \frac{M^{3}}{k}
\left( e^{2k \pi R} - 1 \right).
\end{equation}
This way one obtains the solution of the hierarchy problem. The
large value of the 4-dimensional Planck mass is explained by an
exponential wrap factor of geometrical origin, while the scale $M$
stays small.

The general form of the interaction of the fields, emerging from
the five-dimensional metric,  with the matter localized on the
branes is given by the expression:
\[
\frac{1}{2 M^{3/2}} \int_{B_{1}} d^{4}x\ h_{\mu \nu}(x,0)
T^{(1)}_{\mu \nu} + \frac{1}{2 M^{3/2}} \int_{B_{2}} d^{4}x\
h_{\mu \nu}(x,0) T^{(2)}_{\mu \nu} \sqrt{-\det \gamma_{\mu \nu}
(\pi R)}
\]
Decomposing the field $h_{\mu\nu}(x,y)$ according to
(\ref{RS:decomp}) the interaction Lagrangian can be written as
\begin{equation}
  \frac{1}{2} \int_{B_{2}} d^{4}z \left[ \frac{1}{M_{Pl}}
h^{(0)}_{\mu \nu}(z) T^{(2)\mu \nu} - \sum_{n=1}^{\infty}
\frac{w_{n}}{\Lambda_{\pi}} h^{(n)}_{\mu \nu} T^{(2)\mu \nu} -
\frac{1}{\Lambda_{\pi}\sqrt{3}}T^{(2)\mu}_{\mu} \right],
\label{RS:int2}
\end{equation}
where $\Lambda_{\pi} = M_{Pl} e^{-k \pi R} \approx \sqrt{M^{3}/k}$
and $M_{Pl}$ is given by eq.(\ref{RS:M-Pl}) .

The massless graviton, as in the standard gravity, interacts with
matter with the coupling $M_{Pl}^{-1}$. The interaction of the
massive gravitons and radion are considerably stronger: their
couplings are $\propto \Lambda_{\pi}^{-1} \sim 1 \;
\mbox{TeV}^{-1}$. If a few first massive KK gravitons have masses
$M_{n} \sim 1$TeV, then this leads to new effects which in
principle can be seen at future colliders. To have this situation,
the fundamental mass scale $M$ and the parameter $k$ are taken to
be $M \sim k \sim 1$TeV.

\subsubsection{HEP phenomenology}

With the mass of the first KK mode $M_{1} \sim 1 \; \mbox{TeV}$
direct searches for the first KK graviton $h^{(1)}$ in the
resonance production at future colliders become quite possible.
Signals of the graviton detection can be~\cite{DHR}\vspace{-0.5cm}

$\bullet$ an excess in the Drell-Yan processes \ \ \ \
$\begin{array}{l}  \\ q \bar{q}\rightarrow h^{(1)} \rightarrow
l^{+}l^{-},\\
gg \rightarrow h^{(1)} \rightarrow l^{+}l^{-}
\end{array}$

 $\bullet$ an excess in the dijet channel \hspace{1.5cm}  $
q \bar{q}, gg \rightarrow h^{(1)} \rightarrow q \bar{q},
gg.$\vspace{0.3cm}

\begin{figure}[htb]
\begin{center}
\leavevmode
  \epsfxsize=5cm 
  \epsfig{figure=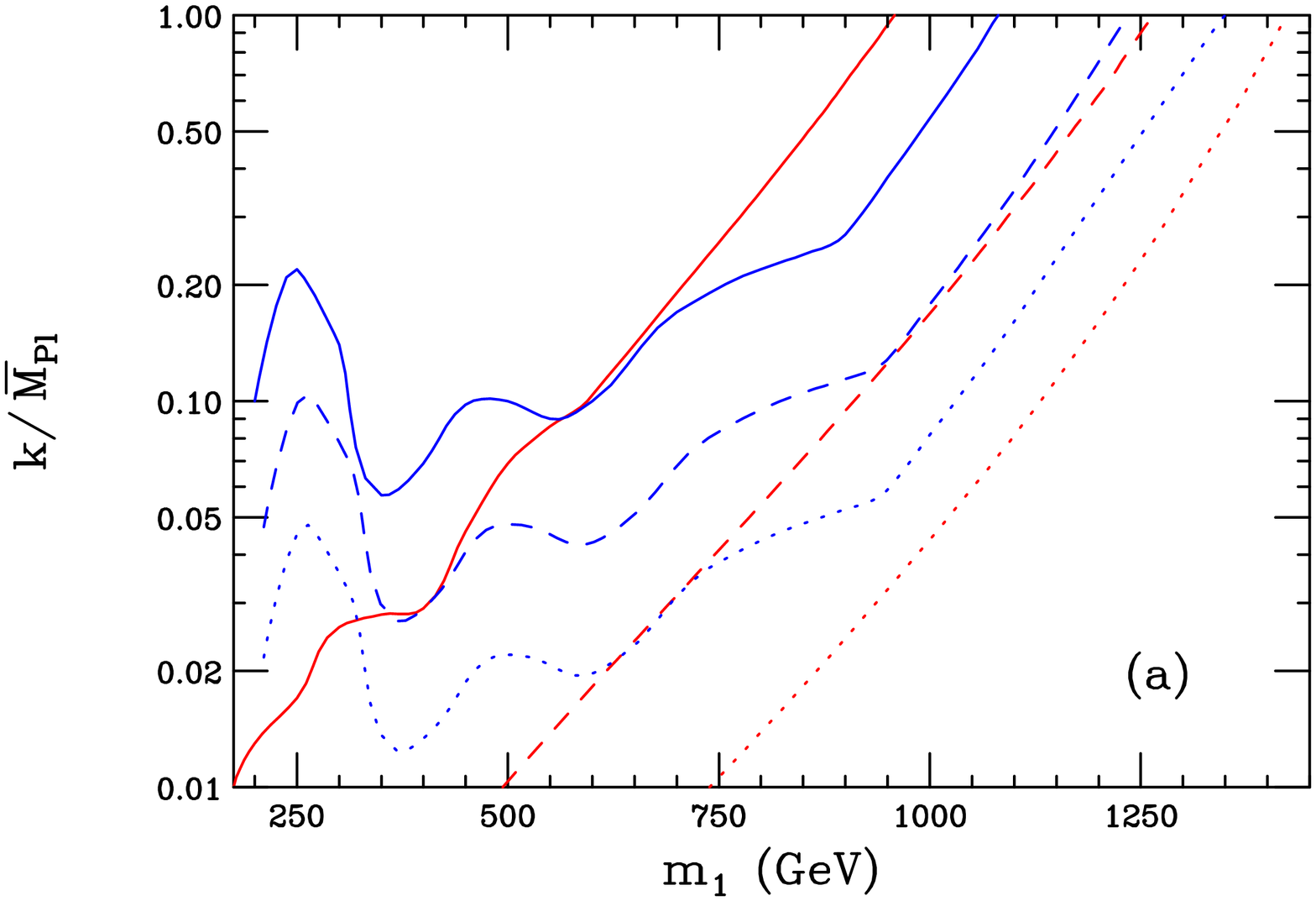,height=6cm,width=6cm,angle=90}
\epsfig{figure=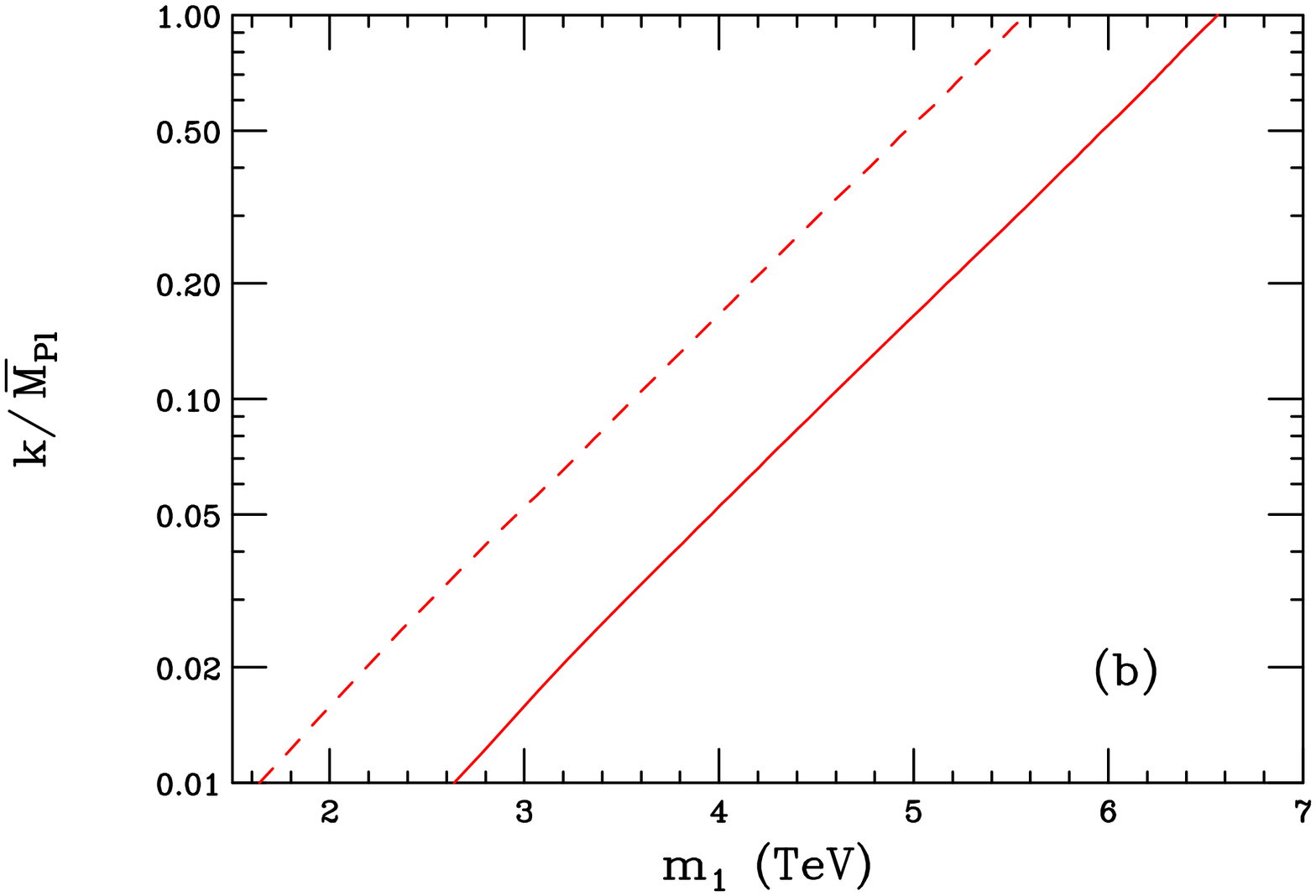,height=6cm,width=6cm,angle=90}
\caption{Exclusion region for resonance production of the first KK
graviton excitation in the Drell-Yan (corresponding to the
diagonal lines) and dijet (represented by the bumpy curves)
channels at the Tevatron (left). The solid curves represent the
results for Run I, while the dashed and dotted curves correspond
to Run II with 2, 30 fb$^{-1}$ of integrated luminosity,
respectively. The same for  the LHC (right). The dashed, solid
curves correspond to 10, 100  fb$^{-1}$ of integrated luminosity,
respectively.} \label{RS-Tev}
\end{center}
\end{figure}

\noindent The plots of the exclusion regions for the Tevatron and
LHC~\cite{DHR} are presented in Fig.~\ref{RS-Tev}.
 They show the exclusion region for resonance
production of the first KK graviton excitation in the Drell-Yan
and dijet channels. The excluded region lies above and to the left
of the curves.

\begin{figure}[htb]
\begin{center}
\leavevmode
  \epsfxsize=5cm 
  \epsfig{figure=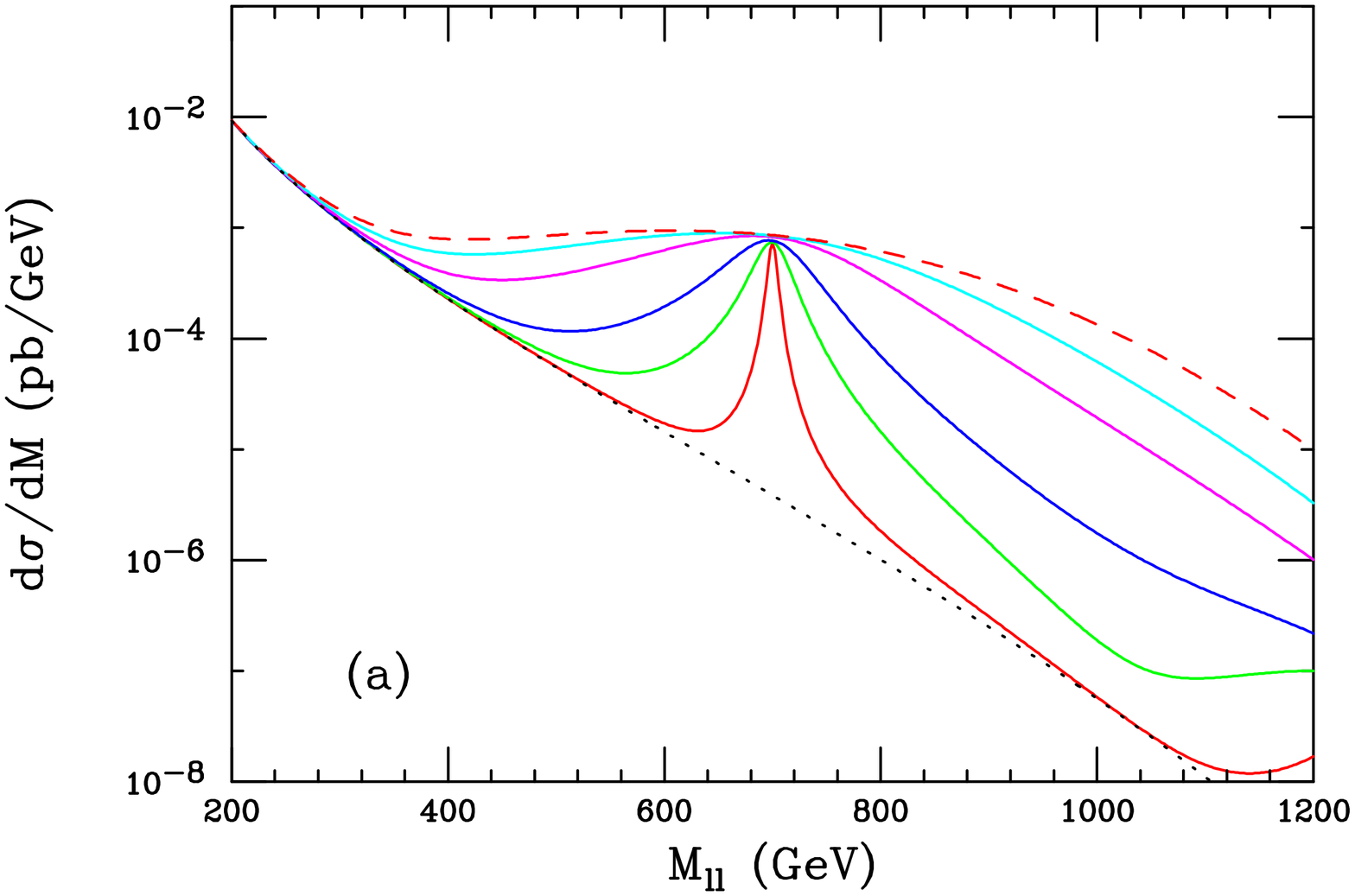,height=6cm,width=6cm,angle=90}
\epsfig{figure=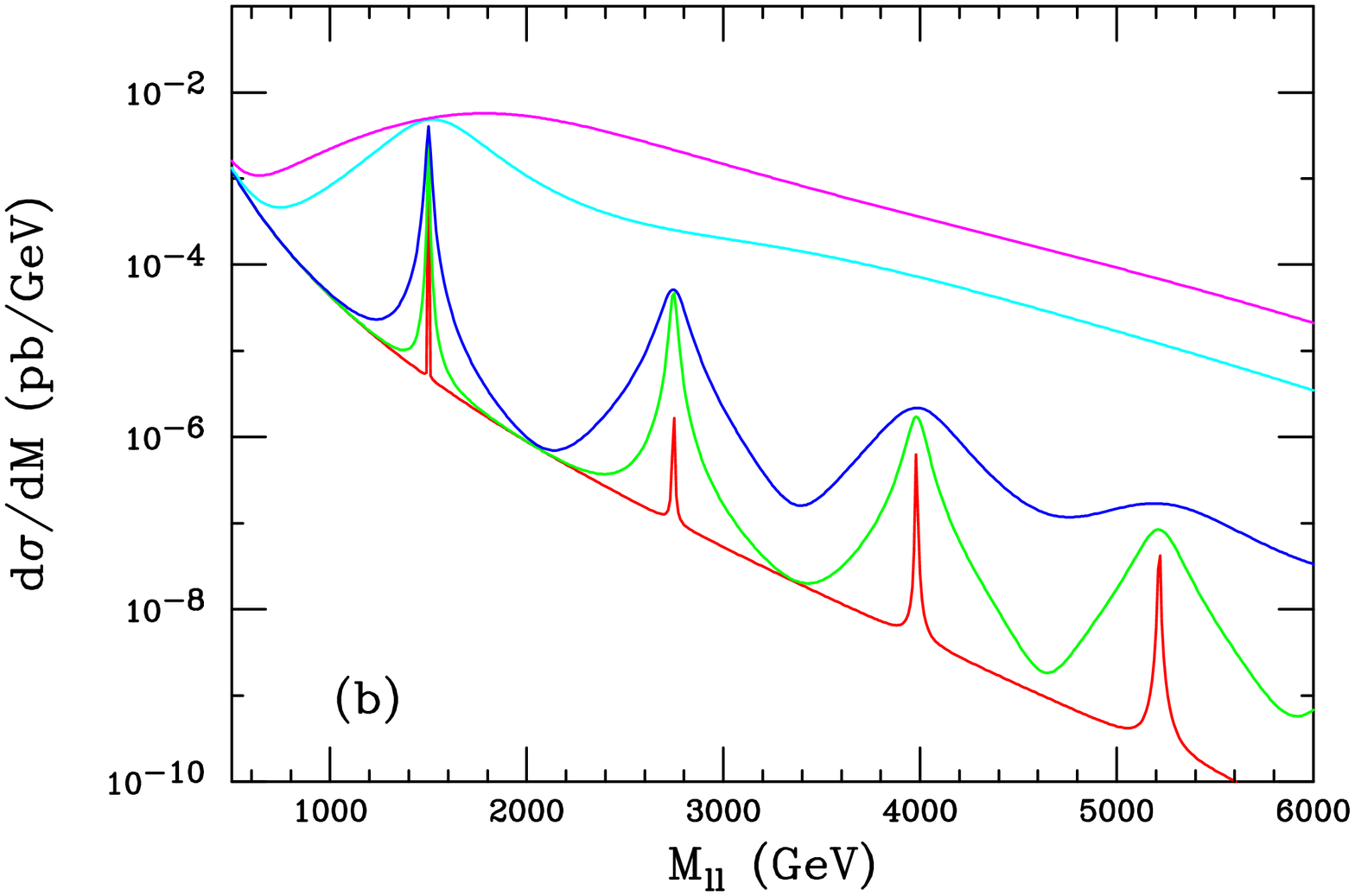,height=6cm,width=6cm,angle=90}
\caption{Drell-Yan production of the KK graviton with $M_{1}=700$
GeV at the Tevatron (left) and for the LHC (right) for
$M_{1}=1500$ GeV and its subsequent tower states \cite{DHR}.}
\label{RS:DY}
\end{center}
\end{figure}

 The next plots present the behaviour of the cross-section of the
Drell-Yan process as a function of the invariant mass of the final
leptons. It is shown for two values of $M_{1}$ for the cases of
the Tevatron and the LHC in  Fig.~\ref{RS:DY}. One can see the
characteristic peaks in the cross section for one or a series of
massive graviton modes.

\begin{figure}[htb]
\begin{center}
\leavevmode
  \epsfxsize=5cm 
  \epsfig{figure=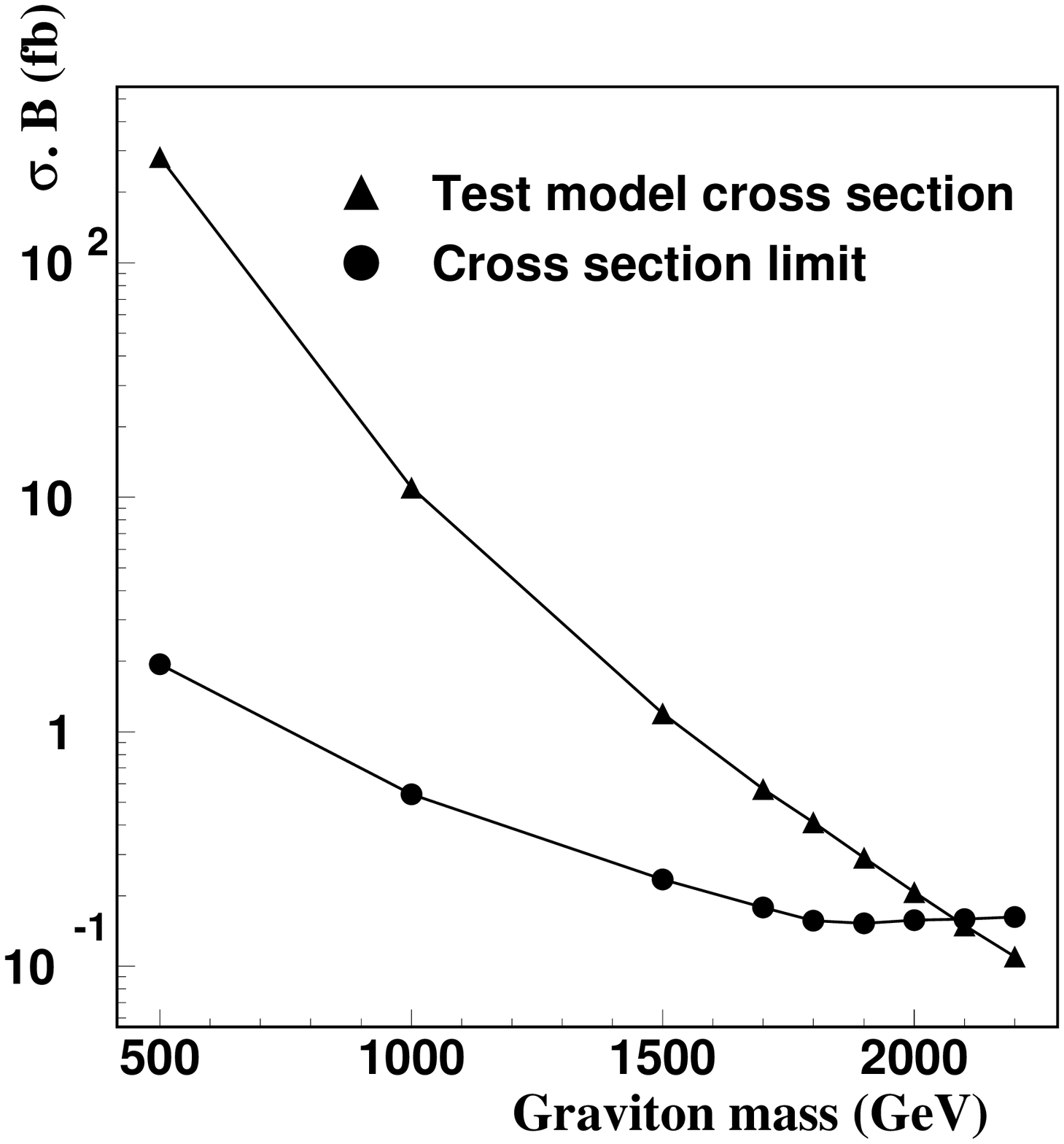,height=6cm,width=5cm,angle=00}\hspace{1cm}
\epsfig{figure=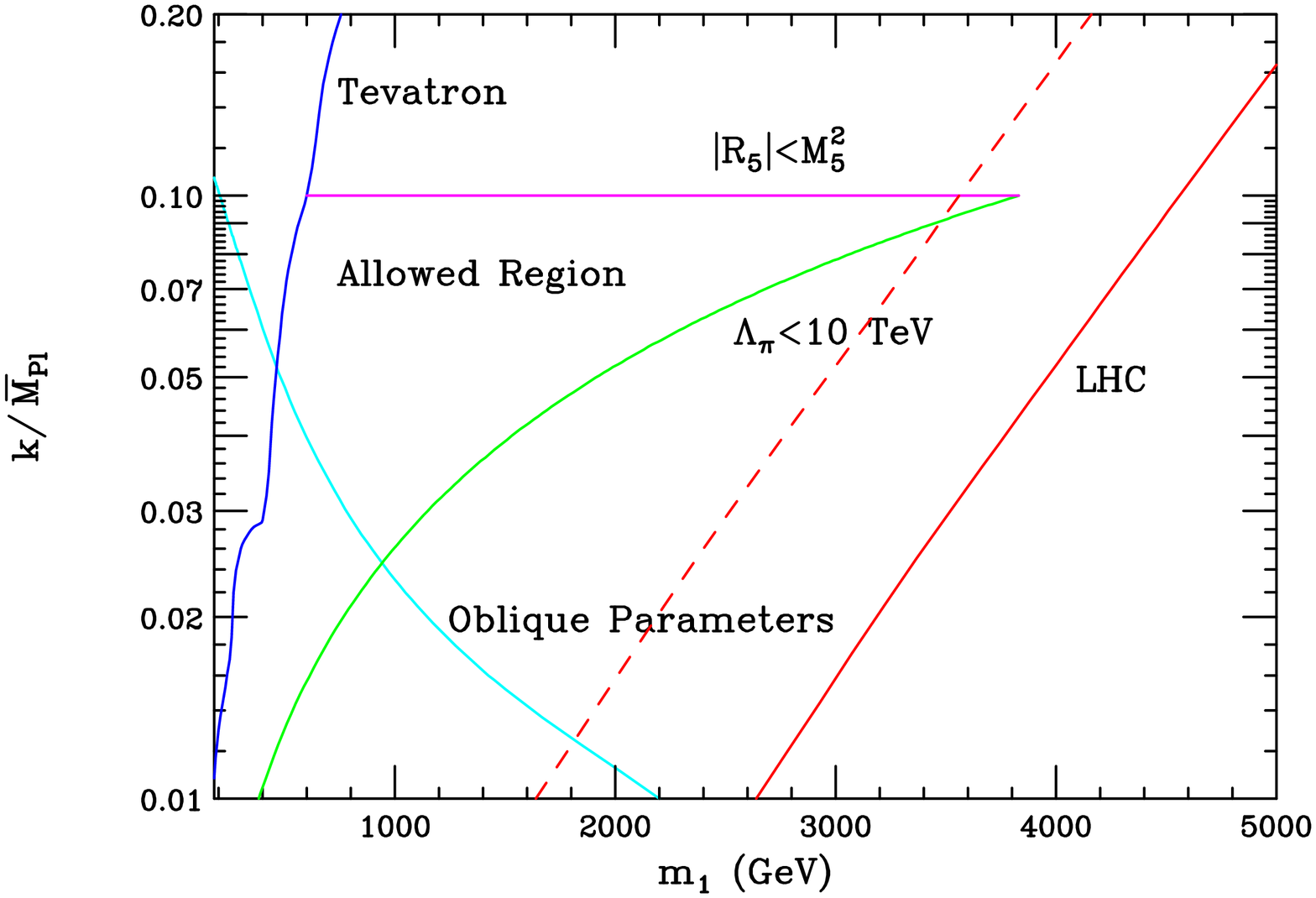,height=5.4cm,width=5.6cm,angle=90}
\caption{The cross-section times branching ratio, $\sigma \cdot
B$, for $h^{(1)} \rightarrow e^{+}e^{-}$ in the RS model and the
smallest detectable cross-section times the branching ratio,
$(\sigma \cdot B)^{min}$ \cite{AOPW} (left)
 and the summary of
experimental and theoretical constraints on the parameters $M_{1}$
and $\eta = (k/M_{Pl}) e^{k\pi R}$ (right) \cite{DHR}. The allowed
region lies as indicated. The LHC sensitivity to graviton
resonances in the Drell-Yan channel is represented by diagonal
dashed and solid curves, corresponding to 10 and 100 fb$^{-1}$ of
integrated luminosity, respectively} \label{RS:G}
\end{center}
\end{figure}

The possibility to detect  the resonance production of the first
massive graviton in the proton - proton collisions $p p
\rightarrow h^{(1)} \rightarrow e^{+}e^{-}$ at the LHC  depends on
the cross section. The main background processes are $p p
\rightarrow Z/\gamma^{*} \rightarrow e^{+}e^{-}$. The estimated
cross section of the process $h^{(1)} \rightarrow e^{+}e^{-}$ as a
function of $M_{1}$ in the RS model is shown in
Fig.~\ref{RS:G}~\cite{AOPW}. One can see that the detection might
be possible if $M_{1} \leq 2080 \; \mbox{GeV}$ .

 To be able to conclude that the observed resonance is a
graviton and not, for example, a spin-1 $Z'$ resonance or a
similar particle, it is necessary to check that it is produced by
a spin-2 intermediate state. The spin of the intermediate state
can be determined from the analysis of the angular distribution
function $f(\theta)$ of the process, where $\theta$ is the angle
between the initial and final beams. This function is
\begin{eqnarray*}
  Spin\ 0 & => &  f(\theta)=1 ,\\
  Spin\ 1 & => &  f(\theta)=1 + \cos^{2} \theta, \\
  Spin\ 2 & => & \left\{ \begin{array}{ll}
q\bar{q} \rightarrow h^{(1)} \rightarrow e^{+}e^{-} & f(\theta) =
1 - 3 \cos^{2} \theta + 4\cos^{4} \theta ,\\
gg \rightarrow h^{(1)} \rightarrow e^{+}e^{-} & f(\theta) = 1 -
\cos^{4} \theta. \end{array} \right.
\end{eqnarray*}
 The analysis, carried out in
Ref.~\cite{AOPW}, shows that angular distributions allow one to
determine the spin of the intermediate state with 90\% C.L. for
$M_{1} \leq 1720$ GeV.

As a next step it would be important to check the universality of
the coupling of the first massive graviton $h^{(1)}$ by studying
various processes, e.g. $pp \rightarrow h^{(1)} \rightarrow
l^{+}l^{-}, \; \mbox{jets}, \; \gamma \gamma, W^{+}W^{-}, HH$,
etc. If it is kinematically feasible to produce higher KK modes,
measuring the spacings of the spectrum will be another strong
indication in favour of the RS model.

The conclusion is~\cite{DHR}  that with the integrated luminosity
${\cal L} = 100 \; \mbox{fb}^{-1}$ the LHC will be able to cover
the natural region of parameters $(M_{1},\eta = (k/M_{Pl}) e^{k\pi
R})$ and, therefore, discover or exclude the RS model. This is
illustrated in the r.h.s. of  Fig.~\ref{RS:G}.

\subsection{Conclusion}

We finish  with a short summary of the main features of the ADD
and RS models.

\underline{ADD Model}.

\begin{enumerate}
\item The ADD model removes the $M_{EW}/M_{Pl}$ hierarchy, but replaces
it by the hierarchy
\[
\frac{R^{-1}}{M} \sim \left( \frac{M}{M_{Pl}} \right)^{2/d} \sim
10^{-\frac{30}{d}}.
\]
For $d=2$ this relation gives $R^{-1}/M \sim 10^{-15}$. This
hierarchy is of different type and might be easier to understand
or explain, perhaps with no need for SUSY;
\item The model predicts the modification of the Newton law at
short distances which may be checked in precision experiments;
\item For $M$ small enough high-energy physics effects,
predicted by the model, can be discovered at future collider
experiments.

\end{enumerate}

\underline{RS model}

\begin{enumerate}
\item The model solves the $M_{EW}/M_{Pl}$ hierarchy problem
without generating a new hierarchy.

\item A large part of the allowed range of parameters
of the RS model will be studied in future collider experiments
which will either discover new phenomena or exclude the most
"natural" region of its parameter space.

\item With a mechanism of radion stabilization added the model is quite
viable. In this case, cosmological scenarios, based on the RS
model, are consistent without additional fine-tuning of parameters
(except the cosmological constant problem)~\cite{Cosmoed}.
\end{enumerate}

\vspace{0.3cm} {\large \bf Acknowledgements} \vspace{0.1cm}

 I am grateful to the organizers of the School at Sant-Feliu for
providing very nice atmosphere during the school and to the
students for their interest, patience, and attention.   Financial
support from RFBR grant \# 99-02-16650,  the grant of THE Russian
Ministry of Industry, Science and Technologies \# 2339.2003.2 is
kindly acknowledged. I would like to thank the Theory Group of
KEK, where this manuscript was finished, for support and
hospitality.

\end{document}